\documentclass[10pt,journal,compsoc]{IEEEtran}
\ifCLASSOPTIONcompsoc
  \usepackage[nocompress]{cite}
\else
  \usepackage{cite}
\fi

\usepackage{booktabs} 
\usepackage[shortlabels]{enumitem}
\usepackage{float}
\usepackage{multirow}
\usepackage{balance}
\usepackage{graphicx}
\usepackage{url}
\usepackage{subfigure}
\usepackage{algorithm}
\usepackage[noend]{algpseudocode}
\usepackage{color}
\usepackage{xcolor}
\usepackage{amssymb}
\usepackage{amsmath}
\usepackage{amsfonts}

\usepackage{amsthm}
\usepackage{textcomp}
\usepackage{wrapfig}
\usepackage[normalem]{ulem}
\usepackage{setspace}
\usepackage{nccmath}
\usepackage[justification=centering]{caption}
\usepackage[shortlabels]{enumitem}
\setenumerate[1]{itemsep=0pt,partopsep=0pt,parsep=\parskip,topsep=0pt}
\setitemize[1]{itemsep=0pt,partopsep=0pt,parsep=\parskip,topsep=0pt}
\setdescription{itemsep=0pt,partopsep=0pt,parsep=\parskip,topsep=0pt}

\newtheorem{definition}{\bf Definition}

\newtheorem{example}{\bf Example}

\usepackage{makecell}

\usepackage{xargs}   
\usepackage{xcolor}  
\usepackage[colorinlistoftodos,prependcaption,textsize=tiny]{todonotes}
\newcommandx{\rone}[2][1=]{\todo[linecolor=blue,backgroundcolor=blue!25,bordercolor=blue,#1]{#2}}
\newcommandx{\rtwo}[2][1=]{\todo[linecolor=orange,backgroundcolor=orange!25,bordercolor=orange,#1]{#2}}
\newcommandx{\rthree}[2][1=]{\todo[linecolor=olive,backgroundcolor=olive!25,bordercolor=olive,#1]{#2}}
\newtheoremstyle{exampstyle}
  {.2em} 
  {.2em} 
  {\itshape} 
  {} 
  {\bfseries} 
  {.} 
  {.5em} 
  {} 
\theoremstyle{exampstyle} \newtheorem{strategy}{Strategy}
\theoremstyle{exampstyle}
\newtheorem*{theorem*}{Problem}

\newcommand\ICQ{\textsc{ICQ}}
\newcommand\SICQ{\textsc{S-ICQ}}
\newcommand\CICQ{\textsc{C-ICQ}}
\newcommand\EICQ{\textsc{E-ICQ}}
\newcommand\RICQ{\textsc{R-ICQ}}

\ifCLASSINFOpdf
\else
\fi
\begin{document}
%
\title{Contact Tracing over Uncertain Indoor Positioning Data (Extended Version)}
%
%
%
%

\author{Tiantian Liu,
        Huan Li,~\IEEEmembership{Member,~IEEE},
        Hua Lu,~\IEEEmembership{Senior~Member,~IEEE},
        Muhammad Aamir Cheema,~\IEEEmembership{Senior~Member,~IEEE},
        and Harry Kai-Ho Chan,~\IEEEmembership{Member,~IEEE}
\IEEEcompsocitemizethanks{\IEEEcompsocthanksitem T.~Liu and H.~Lu are with the Department
of People and Technology, Roskilde University, Denmark.
E-mail: \{tliu, luhua\}@ruc.dk
\IEEEcompsocthanksitem H.~Li is with the College of Computer Science and Technology, Zhejiang University, China. E-mail: lihuan.cs@zju.edu.cn
\IEEEcompsocthanksitem M.~A.~Cheema is with the Faculty of Information Technology, Monash University, Australia.
E-mail: aamir.cheema@monash.edu
\IEEEcompsocthanksitem H.~K.-H. Chan is with the Information School, University of Sheffield, United Kingdom.
E-mail: h.k.chan@sheffield.ac.uk}
}

\IEEEtitleabstractindextext{%
\begin{abstract}
Pandemics often cause dramatic losses of human lives and impact our societies in many aspects such as public health, tourism, and economy. 
To contain the spread of an epidemic like COVID-19, efficient and effective contact tracing is important, especially in indoor venues where the risk of infection is higher.
In this work, we formulate and study a novel query called Indoor Contact Query (\ICQ{}) over raw, uncertain indoor positioning data that digitalizes people's movements indoors. 
Given a query object $o$, e.g., a person confirmed to be a virus carrier, an \ICQ{} analyzes uncertain indoor positioning data to find objects that most likely had close contact with $o$ for a long period of time.
To process \ICQ{}, we propose a set of techniques.
First, we design an enhanced indoor graph model to organize different types of data necessary for \ICQ{}. 
Second, for indoor moving objects, we devise methods to determine uncertain regions and to derive positioning samples missing in the raw data.
Third, we propose a query processing framework with a close contact determination method, a search algorithm, and the acceleration strategies.
We conduct extensive experiments on synthetic and real datasets to evaluate our proposals.
The results demonstrate the efficiency and effectiveness of our proposals.
\end{abstract}

}

\maketitle

\IEEEdisplaynontitleabstractindextext

%
\IEEEpeerreviewmaketitle

\section{Introduction}
\label{sec:intro}

\IEEEPARstart{I}{n} the last 20 years, many pandemics broke out all over the world. 
For example, SARS spread to 20+ countries, sickening over 8,000 people and killing nearly 800 before it vanished in 2003~\cite{health-disease}.
Swine flu (H1N1) killed around 600 thousand people worldwide and still circulates in the human population today~\cite{health-disease}.
As of the end of July 2022, the most recent COVID-19 pandemic has caused more than 561 million infection cases globally, including over 6 million deaths since 2019\footnote{\url{https://covid19.who.int}}. 
These pandemics have caused dramatic losses of human lives and impacted our societies in many aspects such as public health, tourism, and economy.


To contain the spread of such an epidemic, several measures have been proposed and practised, e.g., wearing masks, keeping social distance, and vaccination.
Apart from these measures, it is of high importance to conduct efficient (i.e., timely) and effective contact tracing, e.g., finding likely infected people who had close contact with a person whose infection was already confirmed.
This is especially relevant in many indoor venues such as railway stations and airports where people travel from and to many different cities, and malls that accommodate many people when restrictions are loosened or lifted.
All such indoor venue scenarios call for efficient and effective contact tracing.
For example, if one person who visited a mall is confirmed to be infected, finding others in close contact with this person is necessary for timely appropriate measures like testing and quarantine to break the chain of infection.
Otherwise, the failure of contact tracing for such an indoor venue would no doubt allow the virus to spread further and infect more people.

In such a case mentioned above, a simple way is to treat all people in the indoor venue at that time as close contact. However, this is very ineffective and brings about a lot of false alarms and unnecessary inconvenience to people. 
Moreover, it is hard to scale to trace close contacts with many infected people in a building.
Thus, we desire to propose an efficient way to analyze the indoor movement of people and identify those that are likely to be close contacts.

In this work, we study a new type of query called Indoor Contact Query (\ICQ{}) over indoor positioning data that digitalizes peoples' indoor movements. Given a query object $o$, e.g., a person confirmed to be infected with coronavirus, an \ICQ{} analyzes historical indoor positioning data to find those objects (other people) that most likely had close contact with object $o$ for a sufficiently long period of time. 

In addition to epidemics, \ICQ{} can have many other uses. For example, in criminal investigations, law enforcement agencies can pinpoint potential suspects by identifying individuals who likely had close contact with the victim. Also, \ICQ{} can be utilized as a tool for identifying associates of a specific individual --- if they have had prolonged close contact, it is likely they know each other.

Processing \ICQ{} is challenged by several substantial technical difficulties. 
First, due to the limitations of indoor positioning~\cite{baniukevic2011improving}, the quality of raw indoor positioning data is often lower than that of outdoor GPS data, making it difficult to accurately determine indoor uncertainty regions at sampling times that are not present in the raw positioning data.
Second, the indoor environment is complex with special entities such as walls and doors, making it difficult to model and process indoor data.
Third, \ICQ{} involves a number of constraints, such as distance and contact duration. Such constraints render the query processing more complex, and it is non-trivial to form an appropriate and easy-to-compute formulation of close contact for \ICQ{}.

Existing solutions fall short in solving \ICQ{}.
First of all, the proximity-based contact tracing solutions~\cite{tedeschi2021iotrace,rivest2020pact,ng2022epidemic, bay2020bluetrace, li2021vcontact, shrestha2021bluetooth, di2020bluetooth} find a pair of close contact devices if they perceive each other for a certain period.
However, these methods need to install apps on devices, and they do not consider spatial information and thus fail to customize close contact criteria.
There also exist location-based contact tracing methods~\cite{xiong2020react, eusuf2020web, kato2021pct, zhang2021efficient, chao2021efficient, ami2021computation, hasan2021covid, li2022towards} that find spatially close people based on their spatial distances in the historical times.
However, these methods mainly focus on outdoor spaces and neglect the indoor topology in our problem setting.
Besides, there are studies using indoor positioning data for distance-based analysis~\cite{lu2011spatio,li2018search,li2018finding, chan2022continuous}, but they either ignore the uncertainty of the positioning data~\cite{lu2011spatio}, or fail to analyze the relationship between two individual moving objects~\cite{li2018search, li2018finding}, or focus on a different, online early warning problem~\cite{chan2022continuous}.

To enable efficient and effective \ICQ{} answering, we propose a complete set of data management techniques in this paper. 
First, we design an enhanced graph model to accommodate heterogeneous indoor data, namely the indoor space topology and distances, indoor moving objects and their raw positioning data, and objects' sampled trajectories. All these types of data are necessary for processing \ICQ{}. 

Second, since the query is over uncertain indoor positioning data, we need to analyze the uncertain locations for objects when the data contains no positioning records for them at particular times.
To this end, we devise a method to determine the indoor uncertainty regions at timestamps that are ``unseen'' in the raw positioning data. 
Also, we introduce a method to derive samples based on two consecutive indoor uncertainty regions with respect to the contextual positioning records before and after a particular sampling time of interest.

Third, we design a framework to process \ICQ{}, which searches for all objects in close contact with the query object $o$ by calling the proposed \emph{constrained} search algorithm \CICQ{}. 
\CICQ{} is built on a method that determines the close contact by checking instant contacts during the period of interest. 
We further develop several strategies to avoid unnecessary computations and speed up querying.

Last but not least, we conduct extensive experiments to evaluate the proposed techniques on synthetic and real datasets. The results show that our sampling based contact tracing is effective at searching for real close contact objects, and \CICQ{} is efficient at finding them.

In summary, we make the following major contributions.
\begin{itemize}
    \item First, we formulate Indoor Contact Query (\ICQ{}) for contact tracing over historical uncertain indoor positioning data.
    \item Second, to adapt to the complex indoor environment and the low quality indoor positioning data, we build technical foundations of \ICQ{}, including an enhanced indoor graph model, the determination of indoor uncertainty regions for moving objects, and the generation of derived samples from uncertain positioning data. 
    \item Third, we propose an \ICQ{} processing framework, together with a close contact determination method, a search algorithm, and several strategies that accelerate the complex \ICQ{} processing. 
    \item Fourth, we conduct extensive experiments to evaluate our proposals on synthetic and real datasets under various settings.
\end{itemize}

The rest of the paper is organized as follows. Section~\ref{sec:pre} presents the data, problem, and solution overview.
Section~\ref{sec:foundation} lays the technical foundations of \ICQ{}.
Section~\ref{sec:algorithm} details \ICQ{} processing.
Section~\ref{sec:experiment} reports on the extensive experiments. 
Section~\ref{sec:related} reviews related work. Section~\ref{sec:conclusion} concludes the paper and discusses future directions.

\section{Preliminaries}
\label{sec:pre}

Table~\ref{tab:notations} lists the notations frequently used in this paper.

\begin{table}[htbp]
    \caption{Notations}\label{tab:notations}
    \centering
    \begin{tabular}{|l|l|}
    \hline
    \textbf{Symbol} & \textbf{Meaning}\\
    \hline\hline
    $o$, $O$ & indoor object, object set\\
    \hline
    $\Psi_o$, $\Psi^s_o$ & $o$'s raw trajectory and sampled trajectory\\
    \hline
    $l$ & indoor location \\
    \hline
    $\mathit{dist}(l_i, l_j)$ & contact distance between $l_i$ and $l_j$ \\
    \hline
    $t^s_w$ & a sampling time \\
    \hline
    $s = (l, \rho)$ & a sample at $l$ with probability $\rho$ \\
    \hline
    $\mathsf{P}(o, o', t^s_w)$ & contact probability between $o$ and $o'$ at $t^s_w$ \\
    \hline
    \end{tabular}
\end{table}

\subsection{Raw and Sampled Indoor Trajectories}
\label{ssec:data}

\begin{figure*}[ht]
    \centering
    \begin{minipage}[t]{0.23\textwidth}
        \centering
        \includegraphics[width=\textwidth]{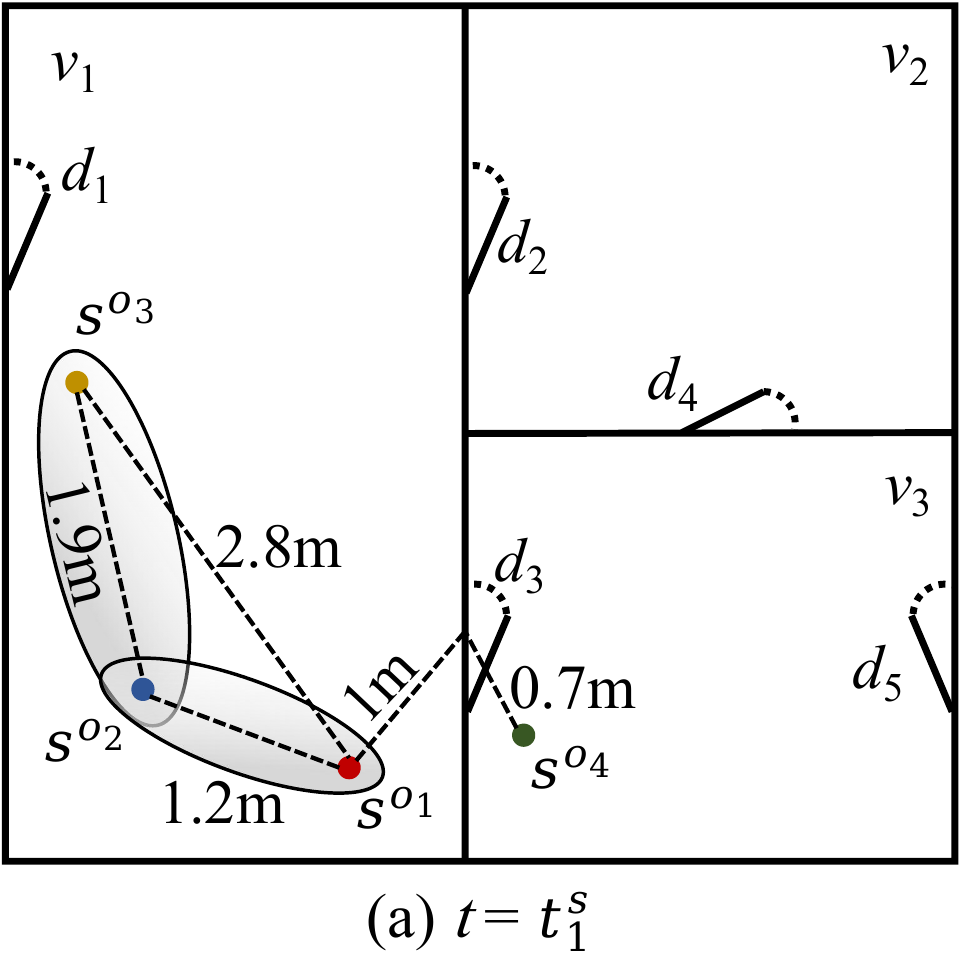}
        \label{fig:example_a}
    \end{minipage}
    \begin{minipage}[t]{0.23\textwidth}
        \centering
        \includegraphics[width=\textwidth]{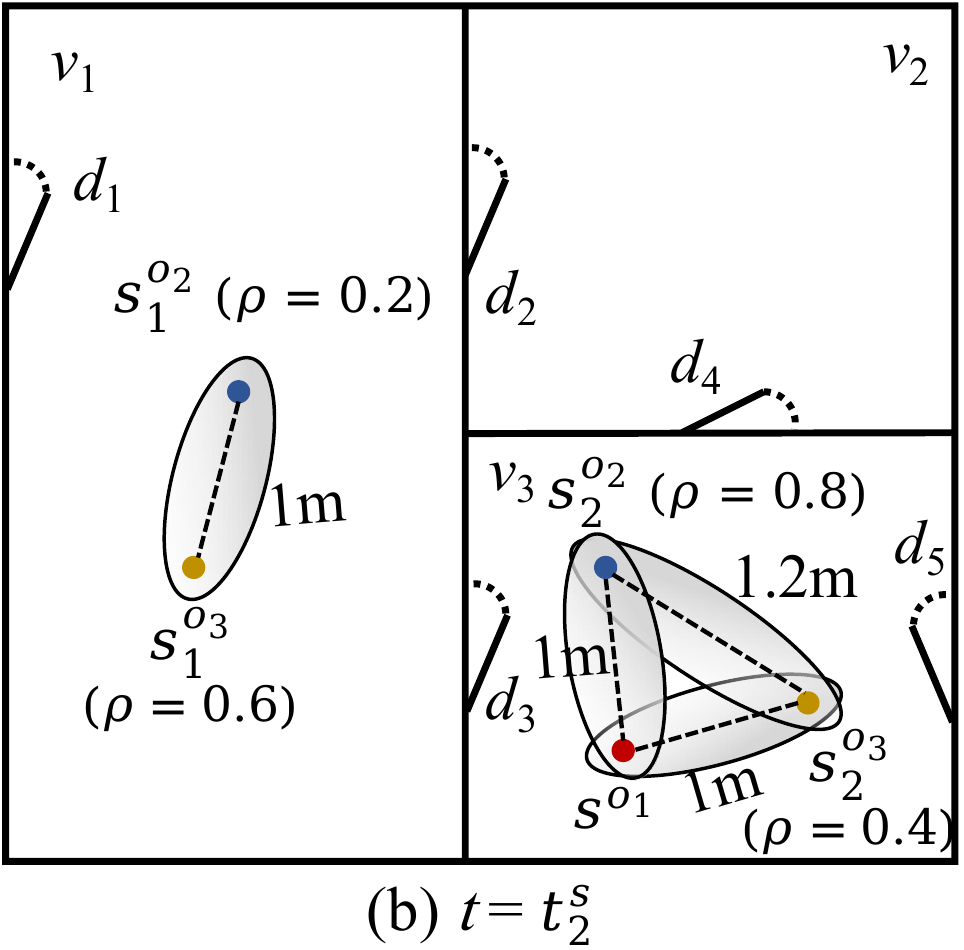}
        \label{fig:example_b}
    \end{minipage}
    \begin{minipage}[t]{0.23\textwidth}
        \centering
        \includegraphics[width=\textwidth]{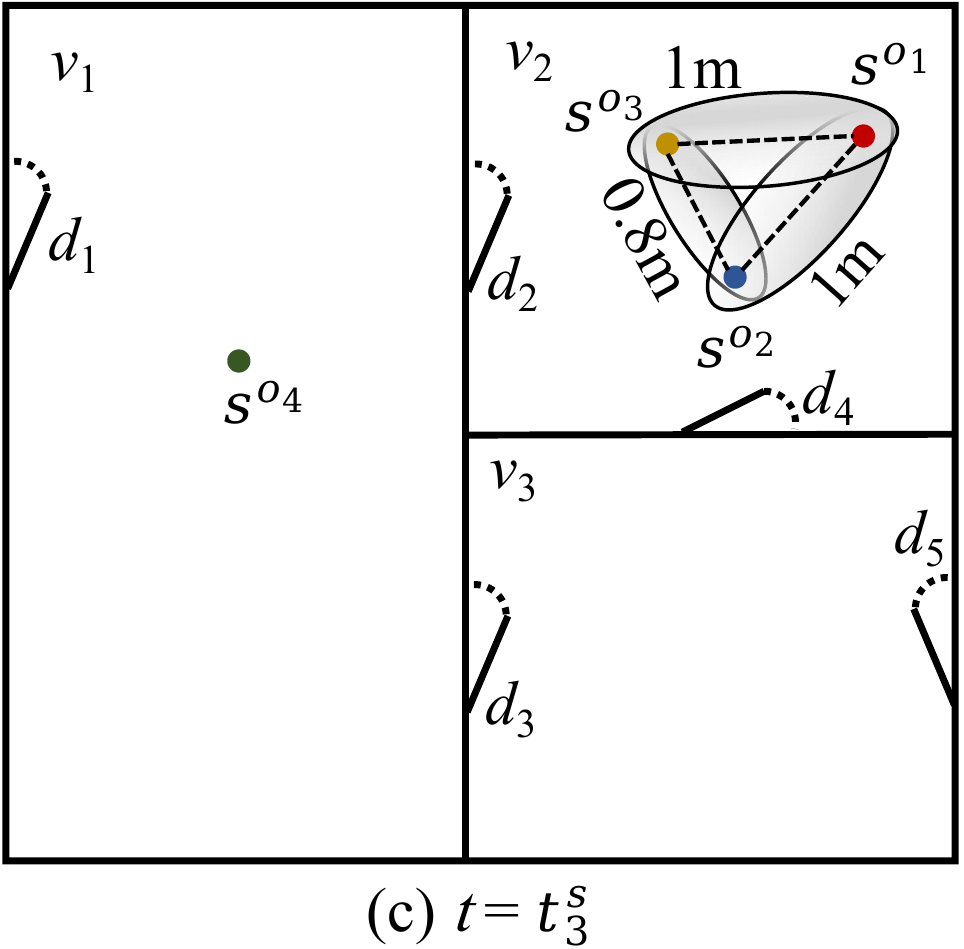}
        \label{fig:example_c}
    \end{minipage}
    \begin{minipage}[t]{0.23\textwidth}
        \centering
        \includegraphics[width=\textwidth]{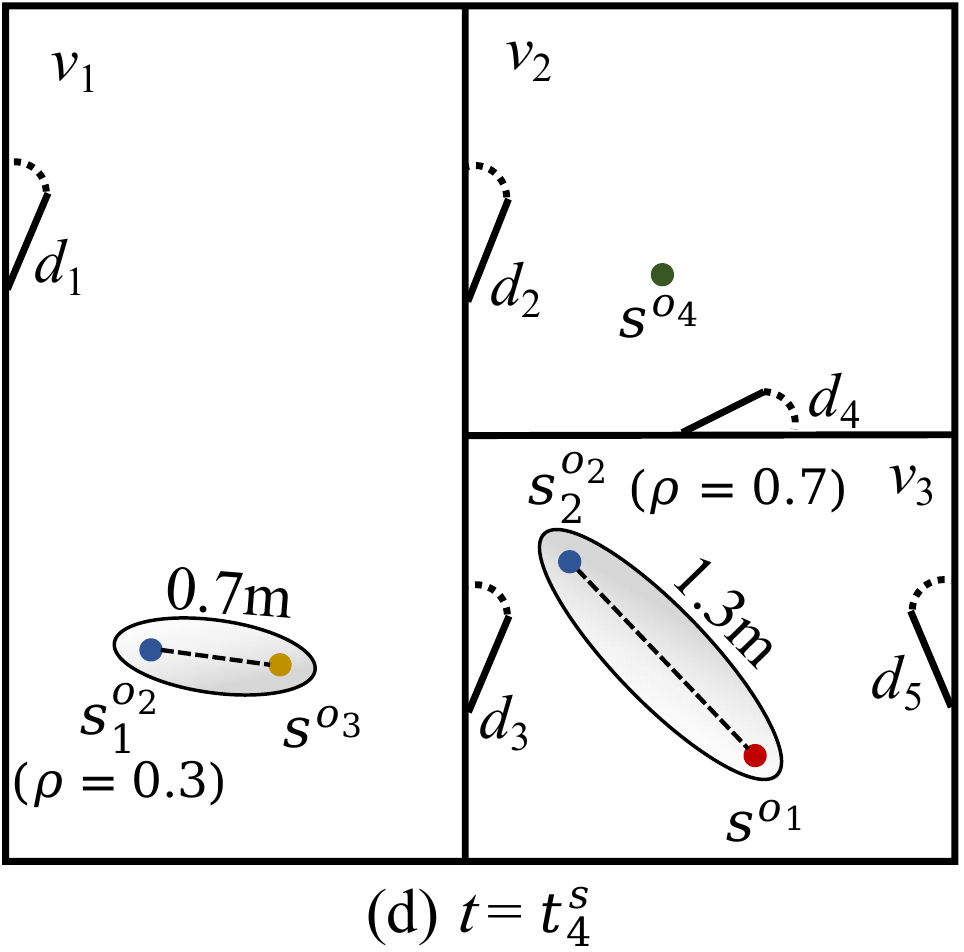}
        \label{fig:example_d}
    \end{minipage}
    \caption{Indoor objects at four consecutive sampling times (the restriction distance $\delta$ = $2$ m).} \label{fig:example_contact}
\end{figure*}
In our setting, an indoor positioning system aperiodically reports an indoor moving object $o$'s positioning record as $\psi = (l, t, et)$, which means that object $o$ was found at a location $l$ at time $t$ and the result is expired at a later time $et$.
A location $l$ is captured as $(x,y,f)$, meaning a point $(x,y)$ on a floor $f$.
Each object $o$'s \textbf{raw trajectory} $\Psi_o$ is a time-ordered sequence of all its positioning records.
In such a raw trajectory $\Psi_o$, two consecutive positioning records $\psi_i = (l_i, t_i, et_i)$ and $\psi_{i+1} = (l_{i+1}, t_{i+1}, et_{i+1})$ satisfy that $et_i < t_{i+1}$.
%
Due to the discrete nature of indoor positioning, the records of an indoor trajectory $\Psi_o$ most often do not fully disclose the object whereabouts during $\Psi_o$'s lifespan.

To enable the comparison of two objects' trajectories throughout a common part of their lifespans, we generate the \textbf{sampled trajectory} of each object as follows.
First, we set a unified sampling time interval $\Delta t$, and sample the object locations at each sampling time $t^s_w = t_o + w \cdot \Delta t$ where $t_o$ is the global origin time and $w$ is a non-negative integer.
Subsequently, the location sample at each sampling time $t^s_w$ is generated in two cases.
If $t^s_w$ falls within the time range $[t_i, et_i]$ of a record $\psi_i$,
a sample $s = (l_i, 1)$ is generated, meaning that $o$ is located at $l_i$ with the probability of 1. 
Such a sample is directly obtained from the raw positioning record and thus is called an \textbf{original sample}.
Otherwise, the two positioning records before and after the sampling time $t^s_w$ are obtained, and a set $S_w$ of samples in the form of $s_i = (l_i, \rho_i)$ are derived based on the two obtained positioning records.
Such a sample is called a \textbf{derived sample}.
Each sampled location $l_i$ is associated with a probability $\rho_i$ and we have $\sum_{s_i \in S_w}s_i.\rho_i = 1$.
How to obtain the location samples based on the contextual positioning records will be detailed in Section~\ref{ssec:sampling}.

As a result, a \textbf{sampled trajectory} $\Psi^s_o$ is obtained as the time-ordered sequence of the sample sets at all sampling times, formally $\langle (S_1, t^s_1), \ldots, (S_W, t^s_W) \rangle$.
In this way, all objects' location samples are aligned in the time dimension and the comparison of objects is performed only at the sampling times.
Note that an object's sampled trajectory is only generated within the lifespan of its raw trajectory.

\subsection{Problem Formulation}
\label{ssec:problem_formulation}

In real applications, it is interesting to find out those moving objects that had very close contact with a particular object for a sufficiently long period of time, equivalently a sufficient number of consecutive sampling times from the perspective of sampled trajectories.
In this paper, we propose a self-contained set of contact tracing techniques that allow us to efficiently find close contacts to a given object by using uncertain indoor positioning data.

We first define the \textit{contact distance} as follows.

\begin{definition}[Contact Distance]\label{definition:contact_distance}
  The \textit{contact distance} between two indoor locations $l_i$ and $l_j$ is computed as
  \begin{equation}
    \begin{medsize}
    \mathit{dist}(l_i, l_j) =
    \begin{cases}
      ||l_i, l_j||_E, &\text{if $l_i$ and $l_j$ fall in the same partition}; \\
      \infty, & \text{otherwise}.
    \end{cases}
    \end{medsize}
  \end{equation}
  where $||\cdot, \cdot||_E$ computes the Euclidean distance.
\end{definition}

In the definition above, a \textit{partition} refers to a basic topological unit of the indoor space, e.g., rooms, staircases, and elevators.
Two persons are not considered contact if they are in different partitions\footnote{In certain situations, objects located near a door but in separate partitions, such as $s^{o_1}$ and $s^{o_4}$ in Fig.~\ref{fig:example_contact}, may be considered in contact. Our method can be easily adapted to cover this case by linking an object near a door within a certain distance to both adjacent partitions in the proposed enhanced indoor graph model in Section~\ref{ssec:data_organization}.}.
For example, in Fig.~\ref{fig:example_contact} (a), the contact distance between $s^{o_2}$ and $s^{o_3}$ is the Euclidean distance $1.9$ meters, while the contact distance between $s^{o_2}$ and $s^{o_4}$ is infinite because they are not in the same partition.
Given the distance threshold $\delta$ in contact restriction, two objects' \textit{contact probability} at a sampling time $t^s_w$ is defined as follows.

\begin{definition}[Contact Probability]\label{definition:contact_probability}
  Given two indoor objects $o_i$ and $o_j$, we obtain their sample sets at time $t^s_w$ as $S_{i,w}$ and $S_{j,w}$. The \textit{contact probability} is computed as
  \begin{equation}\label{equation:contact_probability}
  \mathsf{P}(o_i, o_j, t^s_w) = \sum_{(s_i, \rho_i) \in S_{i,w}, (s_j, \rho_j) \in S_{j,w}} c(l_i, l_j) \rho_i \rho_j,
  \end{equation}
  where $c(l_i, l_j)$ is 1 if the contact distance violates the restriction and 0 otherwise.
\end{definition}

\begin{figure}[h]
    \centering
    \includegraphics[width=\columnwidth]{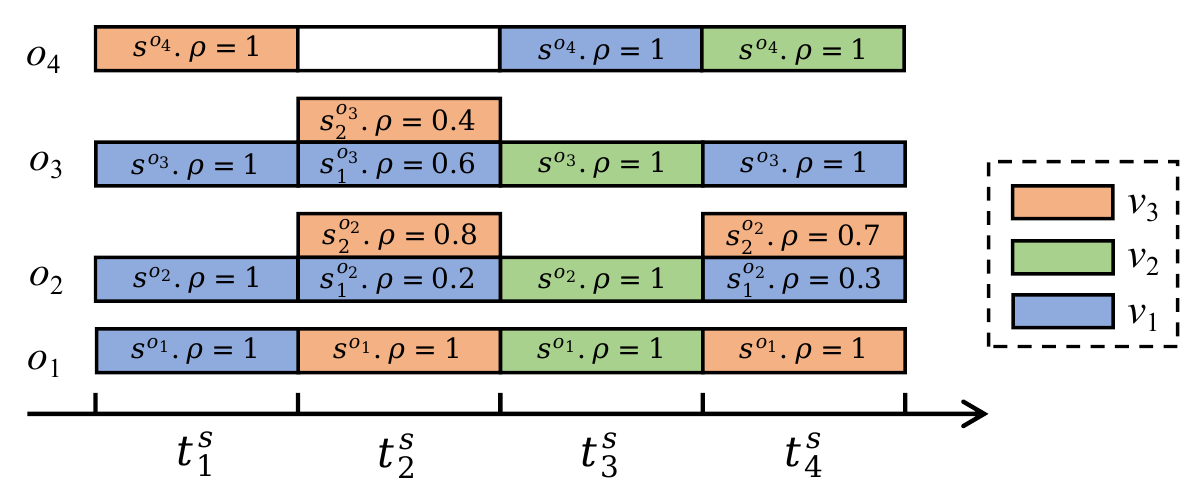}
    \caption{Temporal view of objects.}
    \label{fig:temporal_view}
\end{figure}

We say two objects are \textbf{instant contact} with each other at time $t$ if their contact probability is higher than a specified threshold $\eta$.

\begin{example}
  Take a look at Fig.~\ref{fig:example_contact} (b) and Fig.~\ref{fig:temporal_view}. We calculate the contact probability $\mathsf{P}(o_2, o_3, t^s_2)$ as follows. At time $t^s_2$, the object $o_2$ has two samples, one in the partition $v_1$ with a probability of 0.2 and the other in $v_3$ with a probability of 0.8; the object $o_3$ has a sample in $v_1$ with a probability of 0.6 and another in $v_3$ of a probability of 0.4. In both $v_1$ and $v_3$, the two samples of $o_2$ and $o_3$ are within a distance less than the restriction distance $2$ m. As a result, $\mathsf{P}(o_2, o_3, t^s_2) = 0.2 \times 0.6 + 0.8 \times 0.4 = 0.44$.
  
\end{example}

Finally, our research problem is formulated as follows.

\begin{theorem*}[Indoor Contact Query, \ICQ{}]\label{problem}
    Given a query object $o$, a time interval $T$, a distance constraint $\delta$, a contact probability threshold $\eta$, and a contact number $k$, an \textit{indoor contact query} $\ICQ{}(o, T, \delta, \eta, k)$ returns all moving objects having instant contacts with $o$ (i.e., their contact probability is no less than $\eta$) for at least $k$ consecutive sampling times within $T$. In particular, the contact number $k$ helps determine the time period during which two objects are judged to be close contacts.
\end{theorem*}


\begin{example}
  Figs.~\ref{fig:example_contact} and~\ref{fig:temporal_view} exemplifies the indoor objects at four consecutive sampling times.
  Given a query $\ICQ{}(o_2, [t^s_1, t^s_4], 2, 0.5, 3)$, the close contact between $o_2$ and $o_i$ at each sampling time is determined by the threshold $\eta = 0.5$.
  Referring to $o_3$, the contact probabilities between $o_2$ and $o_3$ at sampling times $t^s_1$ to $t^s_4$ are $[1, 0.44, 1, 0.3]$.
  As the contact probabilities at $t^s_2$ and $t^s_4$ are less than the threshold $0.5$, $o_3$ is not returned. 
  In contrast, the contact probabilities between $o_2$ and $o_1$ are $[1, 0.8, 1, 0.7]$ from $t^s_1$ to $t^s_4$, which all exceed $\eta = 0.5$.
  As there are $4$ ($>k=3$) consecutive instant contacts, $o_1$ is determined as a close contact of $o_2$ and thus returned.
  Likewise, $o_3$ is returned due to its four consecutive instant contacts with $o_2$.
\end{example}


\subsection{Solution Overview}

\begin{figure}[htbp]
    \centering
    \includegraphics[width=\columnwidth]{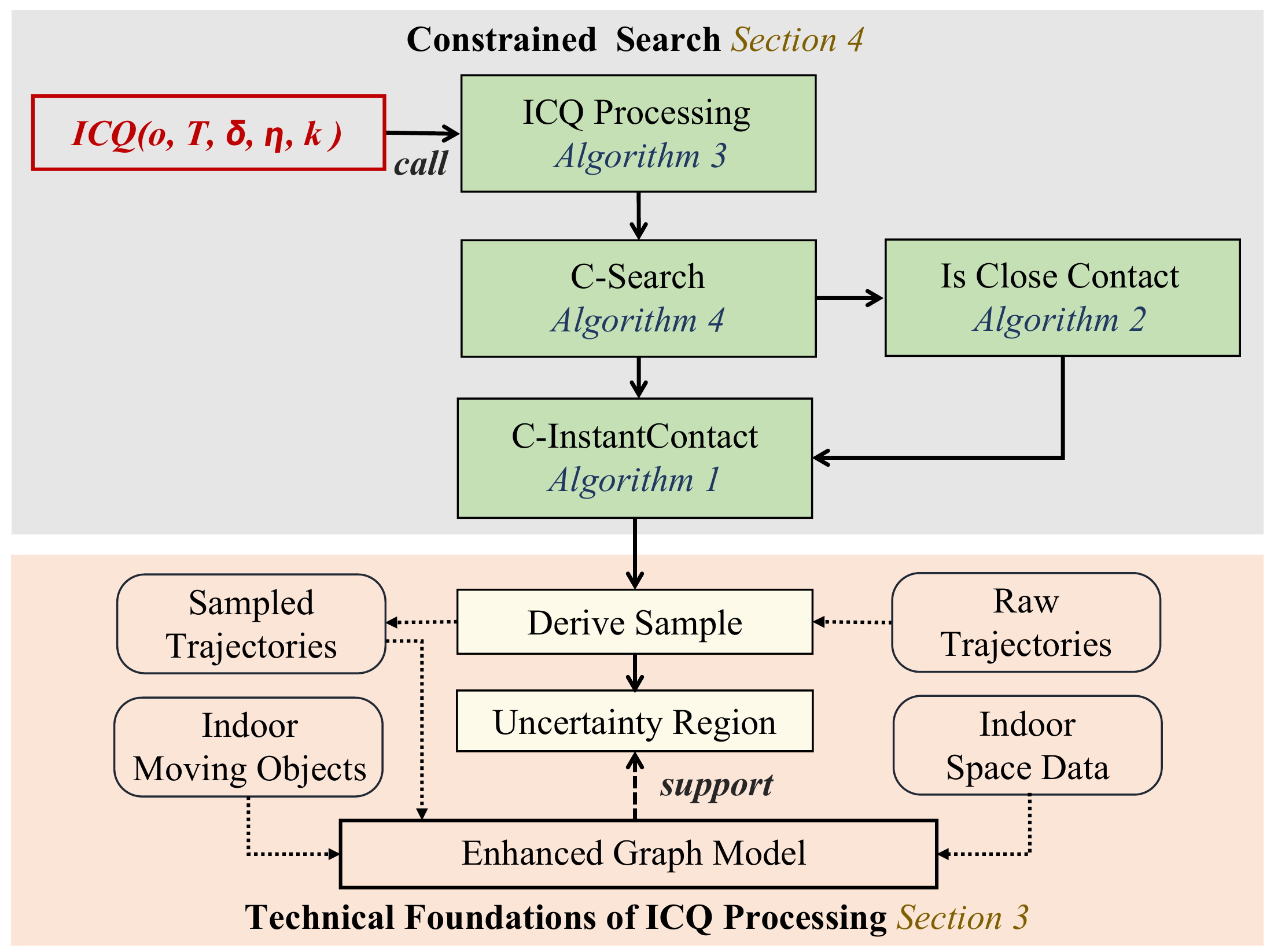}
    \caption{The solution for indoor contact query.}
    \label{fig:framework}
\end{figure}

Fig.~\ref{fig:framework} shows our solution for \ICQ{}.
The bottom layer lays the technical foundations of \ICQ{} processing, as to be detailed in Section~\ref{sec:foundation}.
In particular, an enhanced graph model is constructed to accommodate the indoor space data, indoor moving objects, and sampled trajectories for relevant computations.
On top of the enhanced graph model, it first finds a moving object's uncertainty region, i.e., the object's possible locations at a sampling time. 
Based on the uncertainty regions, it derives samples which in turn are maintained in the enhanced graph model for subsequent use.

To process a query instance $\ICQ{}(o, T, \delta, \eta, k)$, an overall framework is proposed as Algorithm~\ref{alg:ICQ}.
It finds all objects in close contact with the query object $o$ by calling Algorithm~\ref{alg:constrained_search} (the constrained search), which employs a set of strategies for acceleration.
Then, Algorithm~\ref{alg:constrained_search} calls Algorithm~\ref{alg:constrained_instant_contact} to determine the instant close contact and further help set the start timestamp. It also calls Algorithm~\ref{alg:contact} for close contact determination when processing a candidate object. Algorithm~\ref{alg:contact} calls Algorithm~\ref{alg:constrained_instant_contact} to compute the contact probability at an instant timestamp.
We present the close contact determination in Section~\ref{ssec:close_contact} and the \ICQ{} search in Section~\ref{ssec:ICQ_processing}.

It is noteworthy that Algorithms~\ref{alg:constrained_instant_contact} calls a \textit{Derive Sample} function when computing contact probabilities over object samples, whose details can be found in Appendix~\ref{sec:appendixB}.

\section{Foundations of Query Processing}
\label{sec:foundation}

This section lays the technical foundations of \ICQ{} processing. Section~\ref{ssec:data_organization} introduces an enhanced graph model to organize indoor data for relevant computations.
Section~\ref{ssec:uncertainty_region} determines the indoor uncertainty regions at sampling times that are ``unseen'' in its raw positioning records. 
Section~\ref{ssec:sampling} describes the generation of derived samples based on the two indoor uncertainty regions with respect to the contextual positioning records before and after.
Section~\ref{ssec:data_preprocessing} introduces the indoor data preprocessing.


\subsection{Enhanced Graph Model}
\label{ssec:data_organization}

\begin{figure}[!htbp]
    \centering
    \includegraphics[width=0.9\columnwidth]{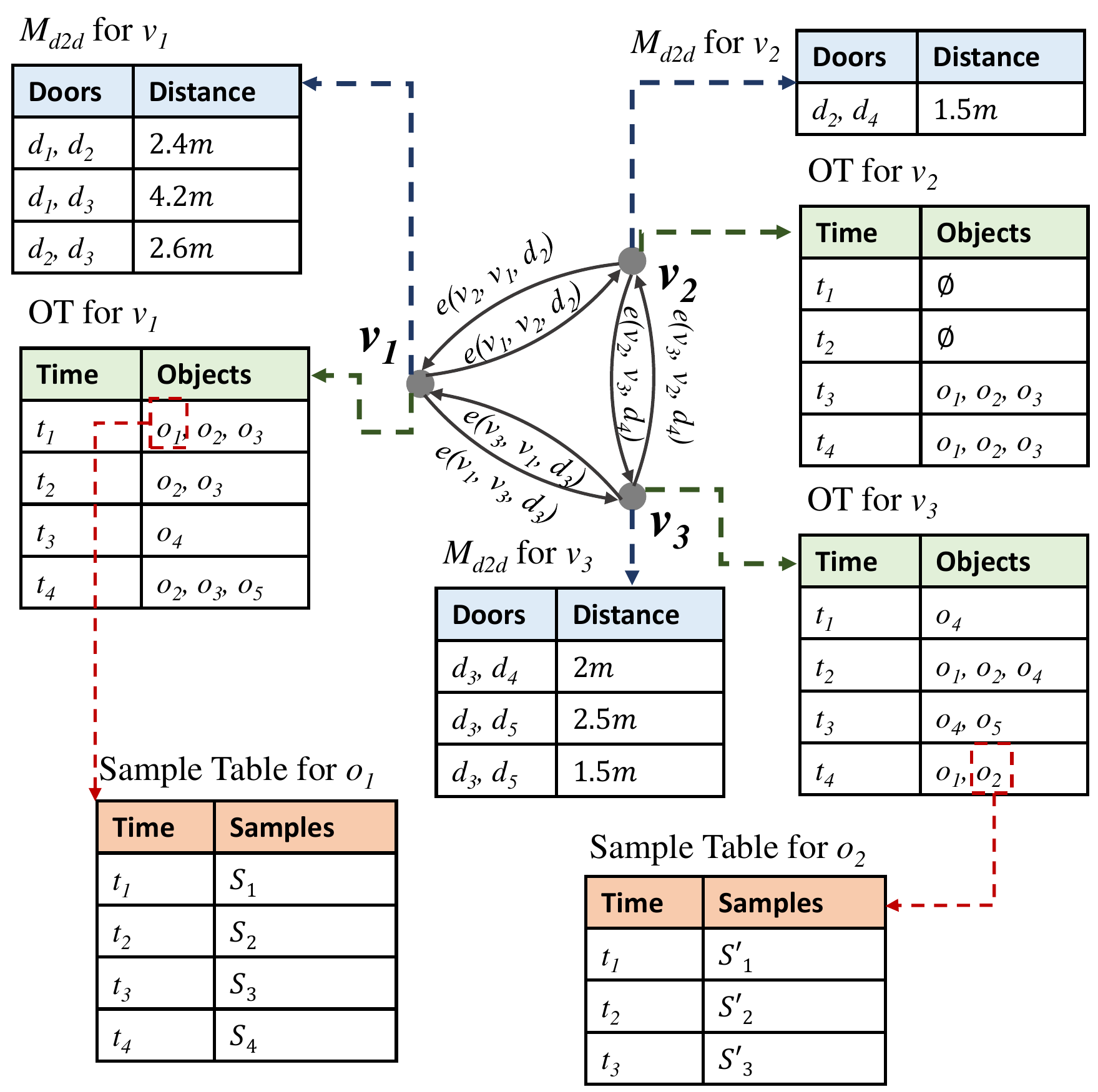}
    \caption{The example of enhanced indoor graph model.}
    \label{fig:graph}
\end{figure}

We design an enhanced graph model to organize three kinds of data, namely the indoor space information, indoor moving objects, and their sampled trajectories.
The enhanced graph model is depicted in Fig.~\ref{fig:graph}.

Following a previous work~\cite{lu2012foundation}, the core part of the model captures the indoor space topology and distances by a directed and labeled graph $G(V, E, L_V)$ where
\begin{enumerate}
    \item $V$ is the set of vertices, each for an indoor partition.\footnote{Indoor partitions can be extracted from indoor geometric data stored in digital files~\cite{boysen2014journey, boysen2014constructing}. }
    \item $E$ is the set of directed edges. An edge $e(v_i, v_j, d_k) \in E$ means one can enter a partition $v_j$ from a partition $v_i$ via a door $d_k$.
    \item $L_V$ is the set of vertex labels. Each label is a three-tuple ($v_i$, $\uparrow M_{d2d}$, $\uparrow \mathit{OT}$) where $v_i$ refers to the associated vertex, $\uparrow M_{d2d}$ is a pointer to the matrix $M_{d2d}$ that stores the intra-partition distance between each pair of doors within $v_i$, and $\uparrow \mathit{OT}$ is a pointer to the object table that maintains the bucket of objects at each sampling time.
\end{enumerate}

For each partition $v_i$'s object table $\mathit{OT}^{v_i}$, $\mathit{OT}^{v_i}[t^s]$ maintains all objects that have at least a sample inside $v_i$ at time $t^s$.
Each object $o$ is linked to its corresponding sample table that maintains its set of samples at each sampling time.


In addition to the enhanced graph model, all raw trajectories are stored in a hashtable with object IDs as the keys.
Moreover, raw trajectories are indexed by a B-tree based on the positioning records' time attributes.
This facilitates obtaining an object's contextual positioning records for a sampling time.


%


\subsection{Indoor Uncertainty Region Determination}
\label{ssec:uncertainty_region}

Given an object $o$ and a sampling timestamp $t^s$ unseen in the raw trajectory, we obtain a positioning record $\psi = (l, t, et)$ prior to $t^s$.
%
In a free space, $o$'s possible location at time $t^s$ is constrained by a circle $\bigcirc(l, r)$ centered at $l$ with radius $r= v_\mathit{max} \cdot (t^s - et)$, where $v_\mathit{max}$ is the maximum object speed.
However, such a circular uncertainty region is imprecise in the presence of indoor topology.

\begin{figure}[htbp]
    \centering
    \subfigure[Precise $\mathit{UR}_I$]{
    \begin{minipage}[t]{0.45\columnwidth}
    \centering
    \includegraphics[width=\columnwidth]{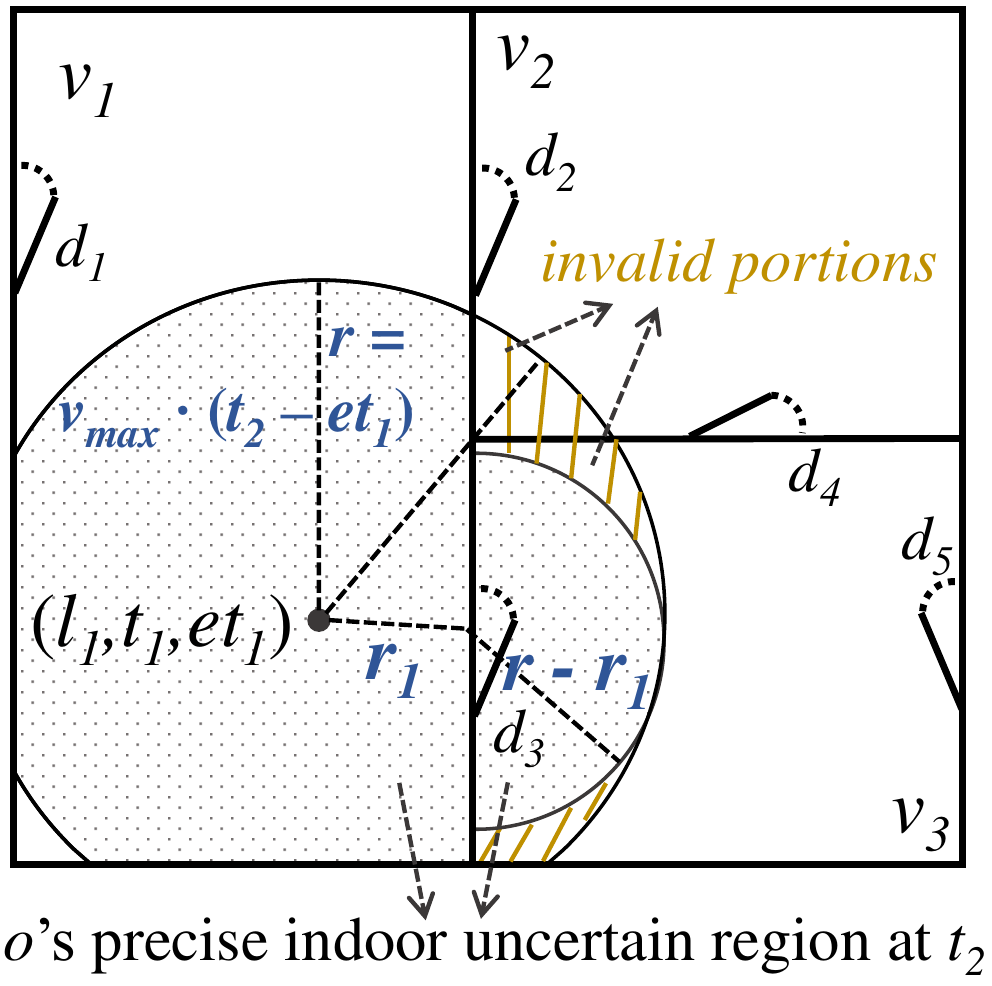}
    \end{minipage}
    }
    \subfigure[Intersection of Two $\mathit{UR}_I$s]{
    \begin{minipage}[t]{0.45\columnwidth}
    \centering
    \includegraphics[width=\columnwidth]{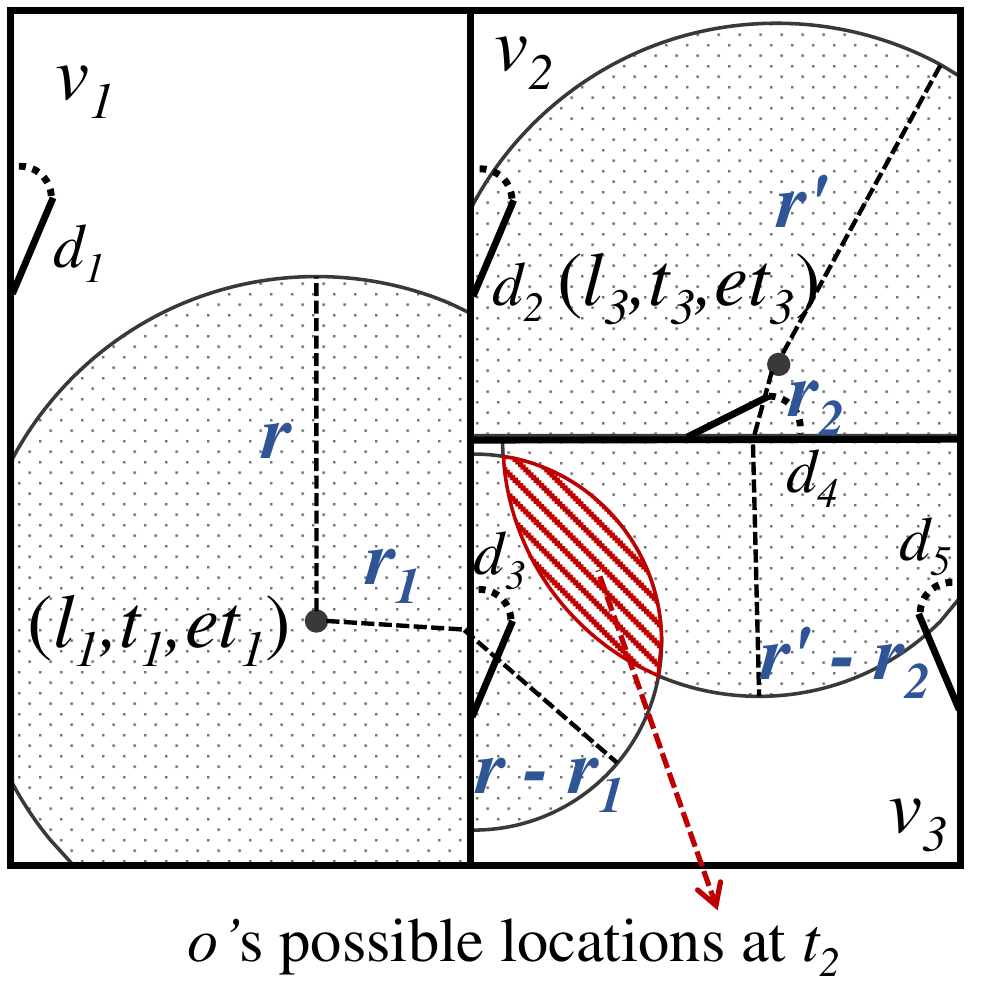}
    \end{minipage}
    }
    \centering
    \caption{The example of the object $o$'s uncertainty regions.}
    \label{fig:uncertainty}
\end{figure}

Referring to the example in Fig.~\ref{fig:uncertainty}(a), $o$'s possible location at time $t_2$ cannot be at the portion of $\bigcirc(l_1, r)$ inside the indoor partition $v_2$ as $o$ cannot go through the wall to reach $v_2$ within the distance $r = v_{max} \cdot (t_2 - et_1)$.
Indeed, $o$'s possible location at time $t_2$ is precisely constrained by an \textit{indoor uncertainty region}~\cite{li2018search} $\mathit{UR}_I(l_1, r)$ that consists of all indoor portions within the indoor distance $r$ from $l_1$.
As illustrated in Fig.~\ref{fig:uncertainty}(a), $\mathit{UR}_I(l_1, r)$ consists of the shaded part in partition $v_1$ and the shaded sector centered at the door $d_3$ with radius $r - r_1$ in partition $v_3$, where $r_1$ is the distance from $l_1$ to $d_3$.
Such indoor portions can be found via an indoor range query, which is formalized in 
\textsc{FindIUR} in Appendix~\ref{sec:appendixA}.

\subsection{Generation of Derived Samples}
\label{ssec:sampling}

Given an object $o$'s raw positioning record $\psi_{\dashv} = (l_{\dashv}, t_{\dashv}, et_{\dashv})$ prior to the unseen sampling timestamp $t^s$, $o$'s possible location at $t^s$ is constrained by $\mathit{UR}_I(l_{\dashv}, v_\mathit{max} \cdot (t^s - et_{\dashv}))$.
Likewise, $o$'s possible location at $t^s$ should also be constrained by $\mathit{UR}_I(l_{\vdash}, v_\mathit{max} \cdot (t_{\vdash} - t^s))$ with respect to $o$'s positioning record $\psi_{\vdash} = (l_{\vdash}, t_{\vdash}, et_{\vdash})$ immediately after $t^s$. 
Taking into account the two contextual positioning records $\psi_{\dashv}$ and $\psi_{\vdash}$, $o$'s possible location can only be within the intersection of $\mathit{UR}_I(l_{\dashv}, v_\mathit{max} \cdot (t^s - et_{\dashv}))$ and $\mathit{UR}_I(l_{\vdash}, v_\mathit{max} \cdot (t_{\vdash} - t^s))$.

Referring to Fig.~\ref{fig:uncertainty}(b), the red shaded region depicts the intersection of the two indoor uncertainty regions with respect to the contextual records reported at $t_1$ and $t_3$.
%
%
Generally, the intersection of two indoor uncertainty regions is an arbitrary geometry that is difficult to compute.
Therefore, we adopt a sample-based approach to approximate the intersection with a set of lattice points contained by the intersection.
The approach is formalized in \textsc{Derive} in Appendix~\ref{sec:appendixB}.

\subsection{Data Preprocessing}
\label{ssec:data_preprocessing}
We do not allow for $k'$ or more consecutive derived sample sets in a trajectory, because that would mean an object absent throughout the $k'$ or more sampling times may still be considered as close contact due to the derived sample sets. 



%
%

Therefore, we preprocess each object's raw trajectory as follows.
We go through the sampling times and check if each of them has a corresponding original sample in the raw trajectory.
If no original sample is seen for $k'$ consecutive sampling times, the raw trajectory is split immediately after the raw positioning record that provides the latest original sample.
When another original sample is seen at a subsequent sampling time, a new raw trajectory is started from the corresponding raw positioning record.
In this way, a raw trajectory may be split into multiple raw trajectories, over each of which at most $k'-1$ consecutive derived sample sets are generated.



\begin{figure}[!htbp]
    \centering
    \includegraphics[width=0.9\columnwidth]{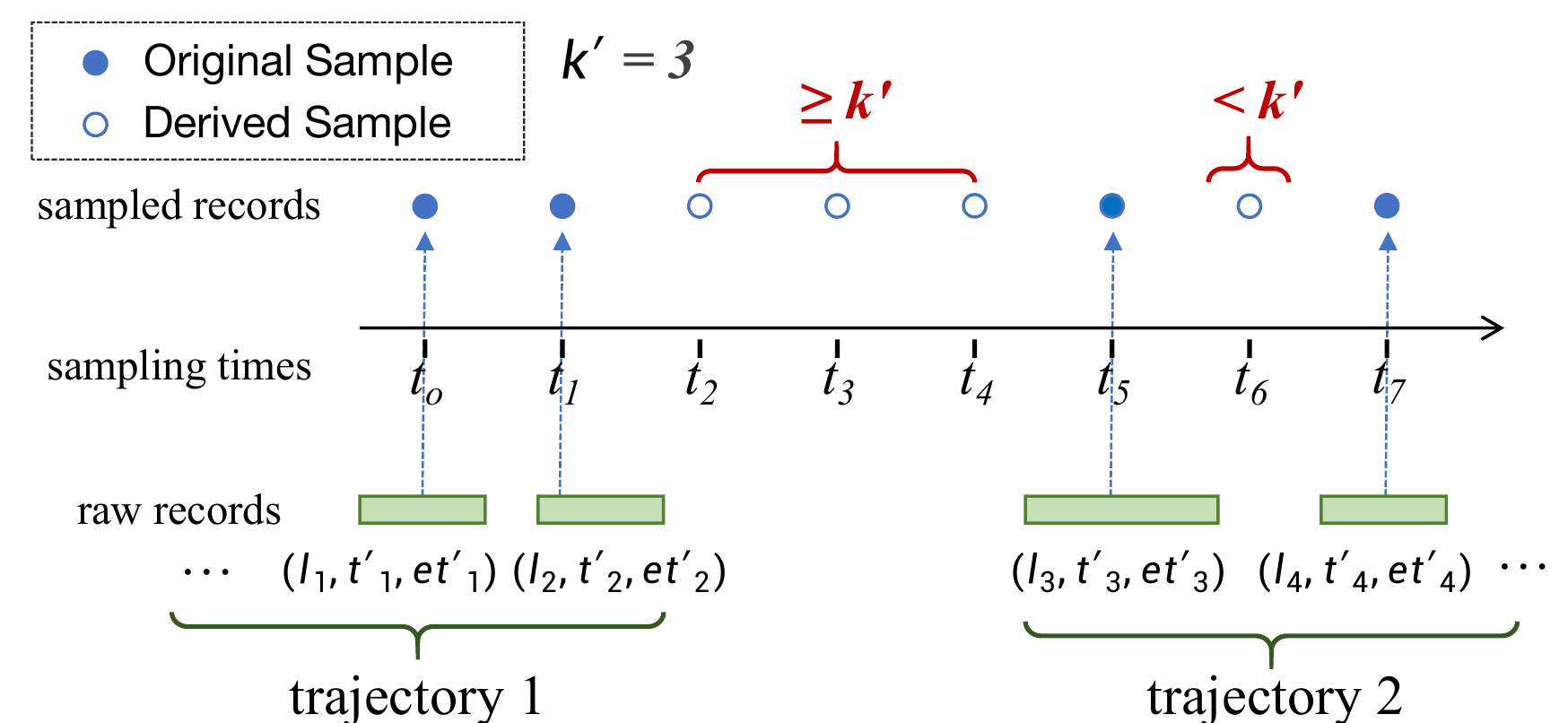}
    \caption{An example of splitting a raw trajectory $(k' = 3)$.}
    \label{fig:example_lifespan}
\end{figure}

\begin{example}
    Referring to Fig.~\ref{fig:example_lifespan}, the original samples are not seen from the raw positioning records at sampling times $t_2$ to $t_4$. Suppose $k'=3$. Then the current raw trajectory is split after the record $(l_2, t'_2, et'_2)$ that provides the latest original sample at the sampling time $t_1$. Subsequently, a new trajectory is started from the record $(l_3, t'_3, et'_3)$. This new trajectory will not be split between $(l_3, t'_3, et'_3)$ and $(l_4, t'_4, et'_4)$ because only one derived sample set is needed.
\end{example}

Our query is processed over the preprocessed trajectories. The preprocessing brings up an important property that facilitates query processing.
That is, an object must have an original sample within $k'$ consecutive sampling times.
Based on this, we can prioritize the instant contact determination with an original sample of the candidate object while avoiding the step-by-step generation of derived sample sets. 
This is implemented in line~9 of Algorithm~\ref{alg:constrained_search}, to be detailed in Section~\ref{ssec:ICQ_processing}.

\section{Query Processing Algorithms}
\label{sec:algorithm}

This section presents the \ICQ{} query processing algorithms.
Section~\ref{ssec:close_contact} introduces the method of identifying all close contacts with the query object $o$.
Section~\ref{ssec:ICQ_processing} details the \ICQ{} processing utilizing a constrained search method.
Section~\ref{ssec:complexity} analyzes the time and space complexity of the proposed solution.

\subsection{Close Contact Determination}
\label{ssec:close_contact}

To find all close contacts with the given object $o$, we must determine if an object $o_i$, in the same partition with $o$, has close contact with $o$ for $k$ consecutive times. To achieve this, we must first devise a method to determine whether two objects are in instant contact.
A simple approach would be to sequentially traverse all pairs of samples of these two objects and compute the instant contact probability. If the probability is above the threshold, the two objects are considered instant contact. The method for sequential instant contact determination is described in Algorithm~\ref{alg:instant_contact} in Appendix~\ref{sec:appendixC}.
However, this method is inefficient in time and space (see our analysis in Section~\ref{ssec:complexity}). In the following, we propose three strategies to speed-up the determination.

\begin{strategy}[Reverse Early Termination]\label{strategy:reverse_early_termination}
  At the level of indoor partitions, an object's sample set $S$ is merged as a new set, represented as $\{ (v, \sum_{\pi_l(s_i) \in v} \rho_i) \}$, each tuple in which is called a \textit{merged sample}.
  Suppose $v$ and $v'$ are two different partitions. For object $o$'s each merged sample $(v, \rho_v)$ and object $o'$'s each merged sample $(v', \rho_{v'})$, the corresponding \textit{non-contact probability} is computed as $\rho_v \rho_{v'}$.
  The overall non-contact probability between $o$ and $o'$ is accumulated over all the pairs of different partitions.
  Once the accumulated non-contact probability reaches or exceeds $1 - \eta$, the contact probability must be lower than $\eta$ and the two corresponding objects are definitely not in close contact.
\end{strategy}

\begin{strategy}[Object Sample Pruning]\label{strategy:sample_pruning}
  A pair of two object samples in different partitions is directly pruned as their distance is $\infty$ (see Equation~\ref{definition:contact_distance}) and larger than the distance constraint $\delta$.
\end{strategy}

\begin{strategy}[Early Termination]\label{strategy:early_termination}
  For two objects and their corresponding sample sets, their contact probability is accumulated by iteratively processing each pair of the two objects' samples. Given the threshold $\eta$, two objects are definitely in instant contact once their current accumulated probability reaches or exceeds $\eta$.
\end{strategy}

With these accelerate strategies, we formalize the constrained instant contact determination in Algorithm~\ref{alg:constrained_instant_contact}.
First, it uses a caching mechanism such that two objects' instant contact result is directly fetched if the result has been computed and cached (line~1).
Then, it prepares the candidate object $o'$'s sample set $S'$ by either calling \textsc{Derive} or retrieving from the enhanced graph model (lines~2--3).
Next, it merges the samples for $o$ and $o'$, respectively, in lines~4--5.
Lines~6--12 implement Strategy~\ref{strategy:reverse_early_termination}, in which the overall non-contact probability $\mathsf{NP}$ is accumulated and $\mathit{false}$ is returned directly if $\mathsf{NP}$ exceeds $1-\eta$.
Lines~14--16 implement Strategy~\ref{strategy:sample_pruning} such that only a pair of object samples in the common partition $v^{*}$ is considered in computing concrete instant contact probability.
Lines~19--20 implement Strategy~\ref{strategy:early_termination}. After processing each object sample pair, the accumulated contact probability is checked for a potential quick determination.
Note that the result is cached before it is returned (see lines~12, 20, and 22).

\begin{algorithm}
      \caption{\textsc{C-InstantContact} (query object $o$, candidate object $o'$, current time $t^s$, distance constraint $\delta$, contact probability threshold $\eta$)} \label{alg:constrained_instant_contact}
      \begin{algorithmic}[1]
        \State return the cached result if it exists
        \If {$t^s$ is not seen in $\Psi^s_{o'}$}
                $S' \gets$ \textsc{Derive}$(o', t^s)$
            \Else
                ~obtain $(S', t^s)$ from $\Psi^s_{o'}$
        \EndIf
        \State merge samples in $S$ as $\{ (v, \sum_{\pi_l(s_i) \in v} \rho_i) \}$
        \State merge samples in $S'$ as $\{ (v', \sum_{\pi_l(s'_i) \in v'} \rho'_i) \}$
        \State $\mathsf{NP} \gets 0$
        \For {each merged sample $(v, \rho_v)$}
          \For {each merged sample $(v', \rho_{v'})$ having $v \neq v'$}
            \State $\mathsf{NP} \gets \mathsf{NP} + \rho_v \cdot \rho_{v'}$
            \If {$\mathsf{NP} > (1 - \eta)$} \Comment{Strategy~\ref{strategy:reverse_early_termination}}
                \State update the latest non-contact time of $o'$ as $t^s$
                \State cache the result; \textbf{return} $\mathit{false}$
            \EndIf
          \EndFor
        \EndFor
        \State $\mathsf{P} \gets 0$
        \For {each partition $v^{*}$ of $S$ and $S'$} \Comment{Strategy~\ref{strategy:sample_pruning}}
            \For {each sample $(l, \rho)$ in $S$ falling in $v^{*}$}
                \For {each sample $(l', \rho')$ in $S'$ falling in $v^{*}$}
                  \If {$\mathit{dist}(l, l') < \delta$}
                      \State $\mathsf{P} \gets \mathsf{P} + \rho \cdot \rho_i$
                      \If {$\mathsf{P} \geq \eta$} \Comment{Strategy~\ref{strategy:early_termination}}
                          \State cache the result; \textbf{return} $\mathit{true}$
                      \EndIf
                  \EndIf
                \EndFor
            \EndFor
        \EndFor
        \State update the latest non-contact time of $o'$ as $t^s$
        \State cache the result; \textbf{return} $\mathit{false}$
    \end{algorithmic}
\end{algorithm}

Then, we can further check whether $o'$ is a close contact of $o$. \textsc{IsCloseContact} is formalized in Algorithm~\ref{alg:contact}. It sequentially scans the sampling times from $t^s$ to $t^s + (k-1) \Delta t$.
At each scanned time $t^q$ (initilized in line~1), the algorithm calls algorithm \textsc{C-InstantContact} to determine if the query object $o$ and the candidate object $o'$ are instant contact.
If the two objects have no instant contact at a scanned time $t^q$, $\mathit{false}$ is returned (lines~3--4).
Otherwise, $\mathit{true}$ is returned to indicate that the two objects are in close contact at all scanned times (line~6).
\begin{algorithm}
    \caption{\textsc{IsCloseContact} (query object $o$, candidate object $o'$, current time $t^s$, distance constraint $\delta$, contact probability threshold $\eta$, integer $k$)} \label{alg:contact}
    \begin{algorithmic}[1]
        \State $t^q \gets t^s$
        \While {$t^q \leq t^s + (k-1) \Delta t$}
            \If {!$\textsc{C-InstantContact}(o, o', t^q, \delta, \eta)$}
                \State\textbf{return} $\mathit{false}$
            \EndIf
            \State $t^q \gets t^q + \Delta t$
        \EndWhile
        \State \textbf{return} $\mathit{true}$
    \end{algorithmic}
\end{algorithm}

\subsection{\ICQ{} Processing}
\label{ssec:ICQ_processing}

We proceed to present the overall framework for \ICQ{} processing. Given a query \ICQ{}$(o, T, \delta, \eta, k)$, we search for each possible timestamp and find out the objects that are in contact with the query object $o$. 
The main steps are as follows.
First, we determine the period [$t^s_s$, $t^s_e$] when $o$ appears in the space.
Second, we obtain all objects present during [$t^s_s$, $t^s_e$].
Third, for each such object, we check if it had close contact with $o$ for $k$ consecutive times by 1) generating samples at unseen timestamps and then 2) calculating the instant contact probability at each sampled timestamp to determine the instant contact.
Finally, we return all objects that had close contact with $o$.
The overall process is formalized in Algorithm~\ref{alg:ICQ}.

\begin{algorithm}
    \caption{\textsc{\ICQ{}-Processing} (query object $o$, time period $T$, distance constraint $\delta$, contact probability threshold $\eta$, contact number $k$)} \label{alg:ICQ}
    \begin{algorithmic}[1]
        \State $t^s_s \gets \max(\min(T), \min(\{ \psi.t \mid \psi \in \Psi_o \}))$
        \State $t^s_e \gets \min(\max(T), \max(\{ \psi.et \mid \psi \in \Psi_o \}))$
        \State $t^s \gets t^s_s$
        \While {$t^s \le t^s_e$}
            \For {each object $o'$ that appeared within $[t^s_s, t^s_e]$}
              \If {$t^s$ is seen in $\Psi_{o'}$}
                  \State obtain record $(l_i, t_i, et_i) \in \Psi_{o'}$, $t^s \in [t_i, et_i]$ 
                  \State add $(\{ (l_i, 1) \}, t^s)$ to $\Psi^s_{o'}$
              \Else
                  \If{$o'$ is query object $o$}
                      \State $S \gets$ \textsc{Derive}$(o', t^s)$; add $(S, t^s)$ to $\Psi^s_{o'}$
                  \EndIf
              \EndIf
            \EndFor
            \State $t^s \gets t^s + \Delta t$
        \EndWhile
        \State \textbf{return} \textsc{C-Search}($o$, $[t^s_s, t^s_e]$, $\delta$, $\eta$, $k$)
    \end{algorithmic}
\end{algorithm}

Initially, the earliest and latest sampling times to process are obtained (lines~1--2).
In particular, the earliest sampling time $t^s_s$ is the older timestamp of the start of the query time interval $T$ and the start of object $o$'s lifespan.
The latest sampling time $t^s_e$ is obtained similarly.

Afterwards, the framework prepares the sampled trajectory $\Psi^s_{o'}$ for each object $o'$ that appeared within $[t^s_s, t^s_e]$ (lines~4--12).
Specifically, if a sampling time $t^s$ is seen in $o'$'s raw trajectory, the reported location $l_i$ and a probability of $1$ form the unique sample at $t^s$ (lines~6--8).
Otherwise, the Derive Sample function $\textsc{Derive}(o, t^s)$ is called to get the samples at $t^s$.
To avoid heavy computations, here we only derive samples for the query object $o$ since its samples will certainly be used in search (lines~10--11).
Other objects' sampling will be conducted during the concrete search.
All generated samples will be maintained in the enhanced graph model (see Section~\ref{ssec:data_organization}) for subsequent retrieval.
Finally, the algorithm calls an efficient search method to find those objects that are in close contact with $o$ during $[t^s_s, t^s_e]$ (line~13).\footnote{Within Algorithm~\ref{alg:ICQ}, the efficient search method could be replaced by a naive sequential search method that determines the instance contacts at each sampling time one by one. It is time-consuming and so we give its details in Algorithm~\ref{alg:sequential_search} ($\textsc{S-Search}$) in Appendix~\ref{sec:appendixC}.}
The search method employs constrained search,
which avoids unnecessary
computations of concrete instant contact probabilities using Strategy~\ref{strategy:skipping}.

\begin{strategy}[Time Skipping]\label{strategy:skipping}
  Given a sampling time $t^s$, if the candidate object does not have instant contact with the query object, all close sampling times within $[t^s-(k-1)\Delta t, t^s+(k-1)\Delta t]$ are skipped for the candidate object.
\end{strategy}

The constrained search (Algorithm~\ref{alg:constrained_search}) works as follows. 
First, the result set $\mathit{result}$ is initialized (line~1).
Then, it initialize each object's latest non-contact time (line~2).
Subsequently, the algorithm processes each sampling time $t^s$ sequentially (lines~3--23).
For each current time $t^s$, the sample set $S$ is retrieved from the enhanced graph model and a set $O_\mathit{visited}$ is used to record all visited objects in this iteration (line~4).
For each sample $s (l, \rho)$ in $S$, the host partition $v$ is obtained (lines~5--6).
The partition $v$ will be skipped (line~7) if it has been visited (see line~8).
Next, the algorithm checks each candidate object $o_i$ that has samples\footnote{According to the preprocessing described in Section~\ref{ssec:data_organization}, a candidate object must have a certain sample generated by Algorithm~\ref{alg:ICQ} within $(k-1)\Delta t$. Therefore, the object will certainly be processed if it is in close contact to $o$ for $k$ consecutive sampling times.} in $v$ at $t^s$ (line~9).
If $o_i$ is the query object $o$, or it has been processed, it will be skipped (line~10).
Otherwise, $o_i$ is added to the visited set (line~11).
The algorithm obtains $o_i$'s last sampling time $t^s_\mathit{ln}$ when $o_i$ has no instant contact with $o$ (line~12).
Such a non-contact time is initialized as $-\infty$ for all objects (line~2) and updated when determining the instant contact (see lines~11 and 21 in Algorithm~\ref{alg:constrained_instant_contact}).
For the candidate object $o_i$ that currently has sample(s) in the same partition as $o$ (see line~9), its instant contact with $o$ is determined by calling \textsc{C-InstantContact}.
After the determination, two cases are differentiated (lines~13--15).
If $o_i$ contacts with $o$ at $t^q$, the end check time $t^s_\mathit{ec}$ is updated as $t^q$, meaning that the sequential checking of close contact should be performed until $t^s_\mathit{ec}$. Otherwise, $t^s_\mathit{ec}$ is set to $t^s - \Delta t$ since there is no need to check until the non-contact time $t^s$.
Accordingly, the starting check time $t^s_\mathit{sc}$ is also determined (line~16).
In particular, $t^s_\mathit{sc}$ must be the latest time among the start search time $t^s$, the sampling time before $t^s_\mathit{ec}$ for $(k-1)\Delta t$, and the sampling time next to the last non-contact time $t^s_{ln}$.
The sequential scanning from $t^s_\mathit{sc}$ to $t^s_\mathit{ec}$ (lines~18--22) is performed only if their time difference exceeds $(k-1)\Delta t$ (line~17).
Once the current time $t^s$ is processed, the algorithm moves to the next sampling time $t^s + \Delta t$ (line 23). It returns the result when all sampling times have been processed (line 24).

\begin{algorithm}
    \caption{\textsc{C-Search} (query object $o$, time period $[t^s_s, t^s_e]$, distance constraint $\delta$, contact probability threshold $\eta$, contact number $k$)}\label{alg:constrained_search}
    \begin{algorithmic}[1]
        \State $\mathit{result} \gets \varnothing$
        \State initialize each object's latest non-contact time as $-\infty$
        \While {$t^s \leq t^s_e$}
            \State obtain $(S, t^s)$ from $\Psi^s_o$; $O_\mathit{visited} \gets \varnothing$
            \For {each $s (l, \rho) \in S$}
                \State $v \gets \mathit{host}(l)$
                \If {$v$ has been visited}
                    \textbf{continue}
                \EndIf
                \State marked $v$ as visited
                \For {each $o_i \in \mathit{OT}^v[t^s]$}
                    \If {$o_i = o$ or $o_i \in O_\mathit{visited}$}
                        \textbf{continue}
                    \EndIf
                    \State add $o_i$ to $O_\mathit{visited}$
                    \State  $t^s_\mathit{ln} \gets$ $o_i$'s latest non-contact time
                    \If {\textsc{C-InstantContact}$(o, o_i, t^s, \delta, \eta)$}
                        \State $t^s_{ec} \gets t^s$
                    \Else
                        ~$t^s_{ec} \gets t^s - \Delta t$
                    \EndIf
                    \State $t^s_{sc} \gets \max(t^s_s, t^s_{ec} - (k - 1)\Delta t, t^s_\mathit{ln} + \Delta t)$
                    \If {$t^s_{ec} - t^s_{sc} > (k -1) \Delta t$}\Comment{Strategy~\ref{strategy:skipping}}
                        \State $t^q \gets t^s_{sc}$
                        \While {$t^q + (k - 1)\Delta t \leq t^s_{ec}$}
                            \If {\textsc{IsCloseContact}($o$, $o_i$, $t^q$, $\delta$, $\eta$, $k$)}
                                \State add $o_i$ to $\mathit{result}$; \textbf{break}
                            \EndIf
                            \State $t^q \gets t^q + \Delta t$
                        \EndWhile
                    \EndIf
                \EndFor
            \EndFor
            \State $t^s \gets t^s + \Delta t$
        \EndWhile
        \State \textbf{return} $\mathit{result}$
    \end{algorithmic}
\end{algorithm}

\subsection{Complexity Analysis}
\label{ssec:complexity}

Let $\mathtt{o}$ be the average number of objects falling into the partition of the query object at a time point, and $\mathtt{v}$ be the average number of partitions that an object's samples may fall into at a time point.
Generally speaking, we have $\mathtt{o} \ll |O|$ where $|O|$ is the total number of objects, and $\mathtt{v} \ll |V|$ where $|V|$ is the total number of partitions.
Let $\mathtt{s}_a$ be the average size of an object' sample sets in a partition at a time point.
Let $\mathtt{s}_m$ be the maximal size among all objects' sample sets.

\begin{figure}[htbp]
    \centering
    \begin{minipage}[t]{0.45\columnwidth}
        \centering
        \includegraphics[width=\textwidth]{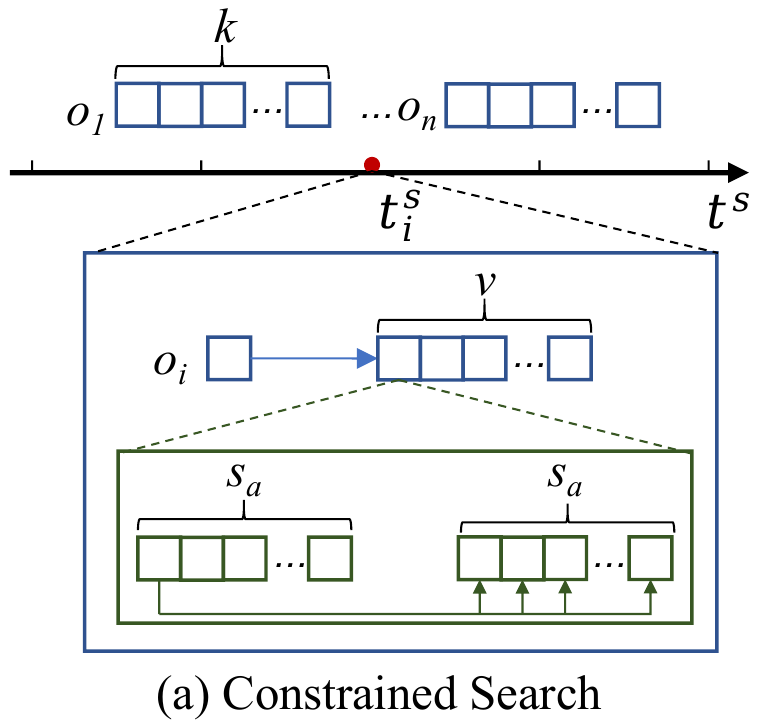}
        \label{fig:constrained}
    \end{minipage}
    \begin{minipage}[t]{0.45\columnwidth}
        \centering
        \includegraphics[width=\textwidth]{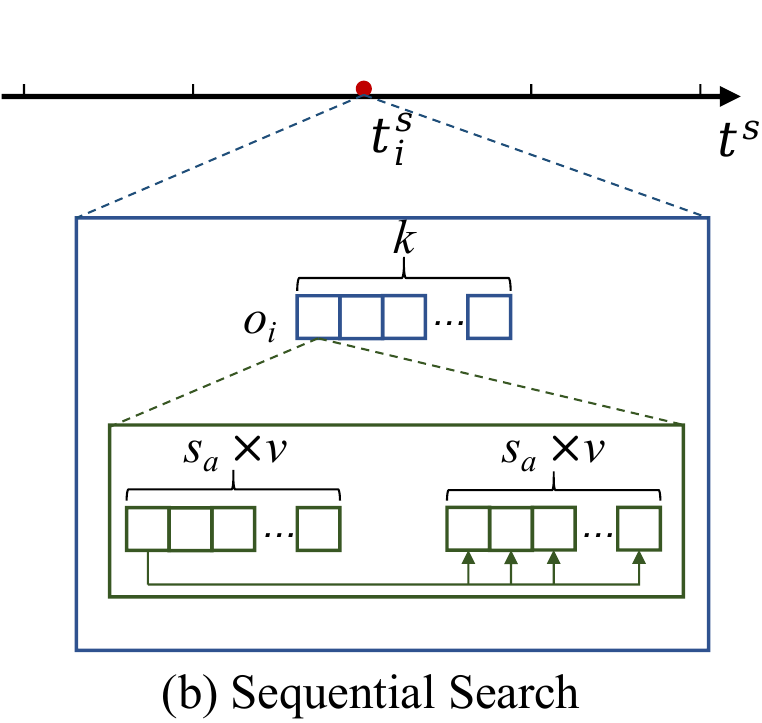}
        \label{fig:sequential}
    \end{minipage}

    \caption{Complexity analysis.}
    \label{fig:difference}
\end{figure}

\smallskip
\noindent\textbf{Time Complexity}.
Fig.~\ref{fig:difference} illustrates the complexity analysis of (a) the constrained search, and  for comparison, (b) the naive sequential search method without using the accelerate strategies.

The proposed constrained search (see Fig.~\ref{fig:difference}(a)) uses a $k$-sized buffer to maintain the previous contact probabilities computed for each candidate object.
Therefore, at each sampling time, the constrained search only processes the current sampling time. 
%
Let $\mathtt{o}_{p}$ be the number of processed candidate objects, where $\mathtt{o}_{p}\ll \mathtt{o}$ as benefited by  Strategy~\ref{strategy:skipping} 
that prunes the candidate objects maintained in the buffer if their probabilities at one time point are less than the threshold $\eta$. 
Note that a smaller threshold $\eta$ and a smaller $k$ will both lower $\mathtt{o}_{p}$. 
Given a query object and a candidate object, 
only $\mathcal{O}(\mathtt{v} \cdot \mathtt{s}_a^2)$ sample pairs are involved in the distances and probabilities calculation as enhanced by  Strategy~\ref{strategy:sample_pruning} that processes pairs in a partition-by-partition manner.
Furthermore, the actual number of distance computations between the sample pairs, denoted by $\beta$, is much smaller than $\mathcal{O}(\mathtt{v} \cdot \mathtt{s}_a^2)$ as reduced by Strategies~\ref{strategy:reverse_early_termination} and~\ref{strategy:early_termination} that enable early termination.
 As there are in total $(|T|/ \Delta t)$ sampling times to process, where $|T|$ is the query interval length and $\Delta t$ is the unit sampling time interval, the overall time complexity of constrained search is $\mathcal{O} \big( (|T|/ \Delta t) \cdot \mathtt{o}_{p} \cdot \beta \big)$.

For comparison, we also include the time complexity analysis of sequential search method as follows. The sequential search (see Fig.~\ref{fig:difference}(b)) calculates the contact probability for each {candidate object} whose samples appear in the same partition as the query object at each sampling time.
For each such object, $k$ consecutive timestamps starting from the current sampling time are processed, each of which involves the distance computation for $(\mathtt{v} \cdot \mathtt{s}_a)^2$ sample pairs.
As there are in total $|T|/ \Delta t$ sampling times to process,  the overall time complexity of the sequential search method is $\mathcal{O}\big( (|T|/ \Delta t) \cdot k \cdot \mathtt{o} \cdot \mathtt{v}^2 \cdot \mathtt{s}_a^2 \big)$. Therefore, the constrained search is at least $\mathcal{O}((\mathtt{o}/\mathtt{o}_p)\cdot k\cdot \mathtt{v})$ times faster than the sequential search, where $\mathtt{o} \gg \mathtt{o}_p$.

In the worst case that all samples of an object are located in one partition, $(\mathtt{v}\cdot \mathtt{s}_a)$ degenerates to $\mathtt{s}_m$, $\mathtt{o}$ and $\mathtt{o}_p$ degenerate to $|O|$, and $\beta$ degenerates to $\mathtt{s}_m^2$.
In this case, the constrained search has the time complexity of $\mathcal{O}\big( (|T|/ \Delta t) \cdot |O| \cdot \mathtt{s}_m^2 \big)$, while the sequential search method is in $\mathcal{O}\big( (|T|/ \Delta t) \cdot k \cdot |O| \cdot \mathtt{s}_m^2 \big)$.

\smallskip
\noindent\textbf{Space complexity}.
Overall, the space complexity analysis of the constrained search follows the same spirit of its time complexity, but some extra space overheads are needed.
In particular, it uses a set to maintain all $|O|$ objects, and also maintains a $k$-sized buffer for the $\mathtt{o}'$ candidate objects at each sampling time, where $\mathtt{o} \leq \mathtt{o}' \leq |O|$.
Besides, at each sampling time, the allocated space will be collected. Therefore, the space complexity is invariant to the number of sampling times, i.e., $(|T|/ \Delta t)$.
To sum up, the space complexity of the constrained search is 
$\mathcal{O}\big( \mathtt{o}_p  \cdot \beta + |O| + k \cdot \mathtt{o}' \big)$ on average and $\mathcal{O} \big( |O| \cdot (\mathtt{s}_m^2 + k + 1) \big)$ in the worst case.

\section{Experiments}
\label{sec:experiment}

\subsection{Overall Experimental Settings}
\label{ssec:settings}

All algorithms are implemented in Java and run on a PC with an Intel Core i5 3.10 GHz CPU and 8 GB memory.

\noindent\textbf{Baseline Methods}. 
As no method exists to solve indoor contact query, we design three baseline methods to compare with our proposed search method \CICQ{}.
First, an \SICQ{} method sequentially processes each sampling time and computes the concrete contact probabilities to decide if a candidate object has close contact with the query object\footnote{\SICQ{} is implemented using the overall framework in Algorithm~\ref{alg:ICQ} whose line~13 instead calls Algorithm~\ref{alg:sequential_search} in Appendix~\ref{sec:appendixC}.}.
Second, an \EICQ{} method finds the uncertainty region at each unseen sampling timestamp based on Euclidean distance rather than indoor distance.
Third, an \RICQ{} method searches for close contact objects over raw trajectories without uncertainty analyses at unseen timestamps.
An object is considered as a close contact if its consecutively observed instant contacts with the query object cover the required contact time.
All baselines employ the indoor distance-aware model~\cite{lu2012foundation} and a trajectory table for data organization.

\noindent\textbf{Parameters}. 
The settings of query inputs $|T|$, $\delta$, $\eta$, and $k$ are listed in Table~\ref{tab:parameter}, where the default values are in bold.
In practice, they can often be set to default values and users may only change them when necessary.
We also vary the object number $|O|$ and the side length $\mathit{ll}$ of lattices (see Section~\ref{ssec:sampling}) which is determined by the system developers and is usually stable values in real-world applications.
The unified sampling time interval $\Delta t$ is set to 10 seconds.
Therefore, we vary the instant contact number $k$ as integer multiples of 6 (i.e., 6, 12 and so on) such that the required contact period will be 1, 2 or more minutes.

\begin{table}[!htbp]
    \caption{Parameter Settings}\label{tab:parameter}
    \footnotesize
    \centering
    \begin{tabular}{|l|l|l|}
    \hline
    \textbf{Parameter} & \textbf{Meaning} & \textbf{Setting}                                          \\ \hline \hline
    $|T|$ (hour)      & query interval length & 6, 12, 18, \textbf{24}                                          \\ \hline
    $\delta$ (m)      & distance constraint      & 1, \textbf{2}, 3, 4, 5                                         \\ \hline
    $\eta$            & probability threshold  & 0.3, 0.4, \textbf{0.5}, 0.6, 0.7                                    \\ \hline
    $k$       & instant contact number  & 6, 12, \textbf{18}, 24, 30, 36, 42 \\ \hline
    $\mathit{ll}$ (m)     & lattice side length         & 0.2, \textbf{0.4}, 0.6, 0.8, 1.0                        \\
    \hline
    $|O|$      & object number         & 2k, \textbf{4k}, 6k, 8k                                    \\ \hline
    \end{tabular}
\end{table}

\noindent\textbf{Performance Metrics}.
To evaluate the efficiency of query processing, we run each query instance 30 times and measure the \textit{average running time} and \textit{average memory cost}.

As there are unseen timestamps in raw trajectories and we have to analyze the uncertain region of an object at an unseen timestamp, it will introduce inaccuracies. Therefore, we evaluate the effectiveness of the query result as follows.
We consider the metric \textit{recall}, the fraction of ground truth close contact objects that are included in a query result.
Moreover, we use the comprehensive metric \textit{F1 score} that is computed as $2 * (\mathit{recall} \times \mathit{precision}) / (\mathit{recall} + \mathit{precision})$, where $\mathit{precision}$ measures the fraction of returned close contact objects that are in the ground truth\footnote{We have also evaluated $\mathit{precision}$. However, the $\mathit{precision}$ measures in most tests are close to 1. Therefore, we omit them in the evaluation results.}.
The closer the F1 score is to 1, the better is the overall effectiveness of the query result.

\subsection{Experiments on Synthetic Data}

\subsubsection{Settings}

\textbf{Indoor Space}.
We generate a 5-floor indoor space based on a real-world floorplan\footnote{\url{https://longaspire.github.io/s/fp.html}} that is of 1,368 $\times$ 1,368 square metres (m$^2$) with 141 partitions and 216 doors on each floor. Four staircases, each set to $20$ m long, connect each two adjacent floors.

\noindent\textbf{Trajectory Data}. We simulate raw trajectories of moving objects in the indoor space for 24 hours (i.e., 86,400 seconds).
Each trajectory's lifespan is at least one hour (3,600 seconds) and its start time is random within the first 82,800 seconds.
For each object, we randomly select a list of destinations.
Between each two consecutive destinations, the object moves along a planned path with fluctuations in direction and moving speed.
The maximum speed is set to $5$ km/h (approximately 1.4~m/s). At each destination, the object stays for a random period of time from 0 to 480 seconds.

Raw positioning records are generated for each object every 10 seconds. The expiration $et - t$ in each record is fixed to 5 seconds.
To simulate the uncertainty in raw trajectories, we randomly remove 10\% of the generated records from each original trajectory.
After that, we use the preprocessing method described in Section~\ref{ssec:data_organization} to split the resultant trajectories.

\noindent\textbf{Query Instances}. For each specific parameter combination, 20 query instances are generated with random query objects and their trajectories.
Due to the randomness of trajectory data, it is hard to find close contacts that last for a long period.
To increase the likelihood of non-empty query results, we intentionally simulate 10 ``close contact trajectories'' for each query instance.
Each such trajectory is synthesized as follows.
From the current query instance's trajectory, we extract a segment that lasts for a random period of 500 to 1800 seconds. The extracted segment is used as the basis of a close contact trajectory.
To reduce the trajectory's data quality (and to increase the difficulty for contact tracing), we randomly delete 10\% of its positioning records and add some location offsets to some of its positioning records.

\subsubsection{Results}

\begin{figure}[htbp]
    \centering
    \subfigure[Time vs. $T$]{
        \begin{minipage}[t]{0.45\columnwidth}
            \centering
            \includegraphics[width=\columnwidth]{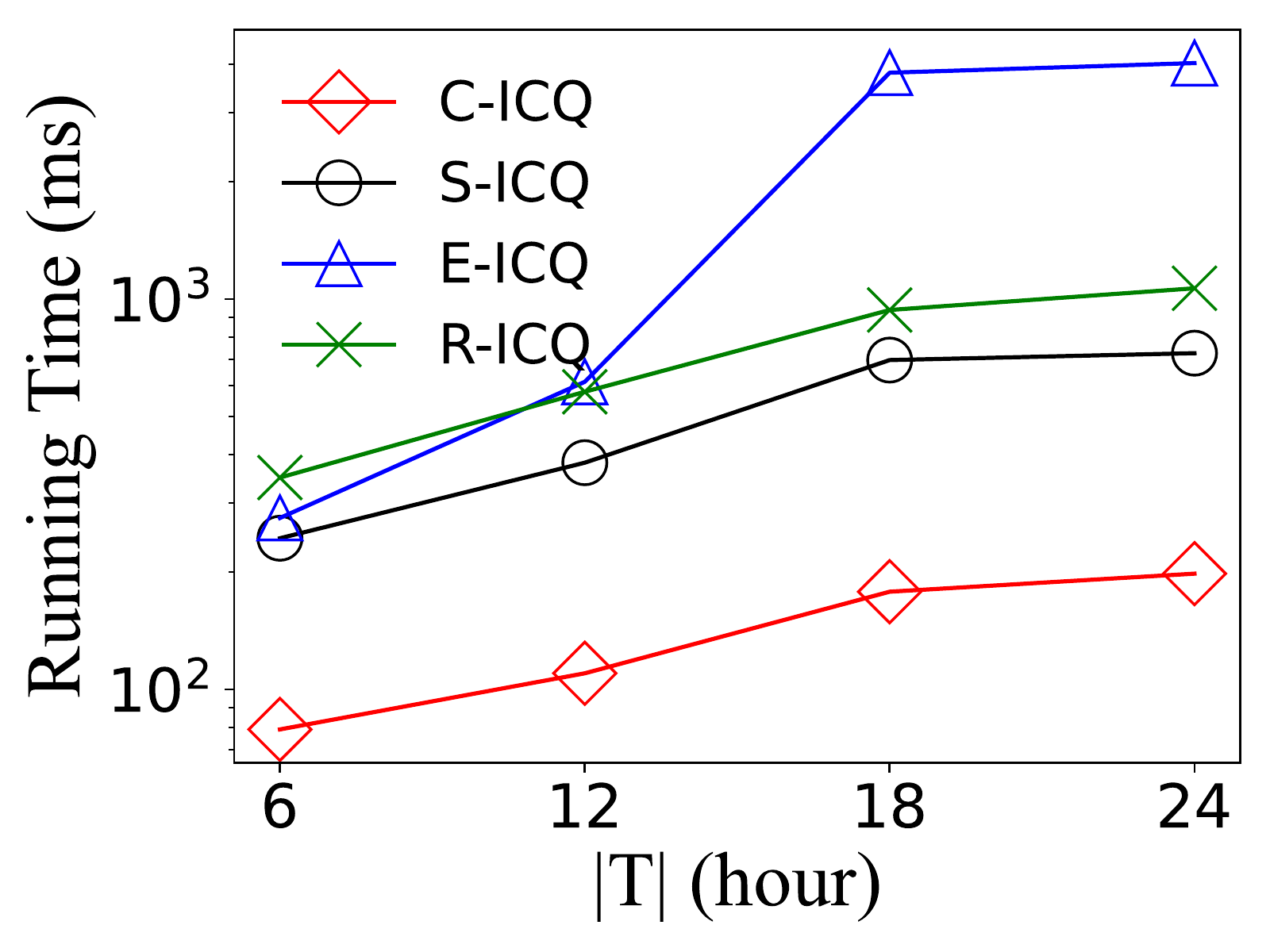}
            \label{fig:SYN-queryInterval-time}
        \end{minipage}
    }
    \subfigure[Mem. vs $|T|$]{
        \begin{minipage}[t]{0.45\columnwidth}
            \centering
            \includegraphics[width=\columnwidth]{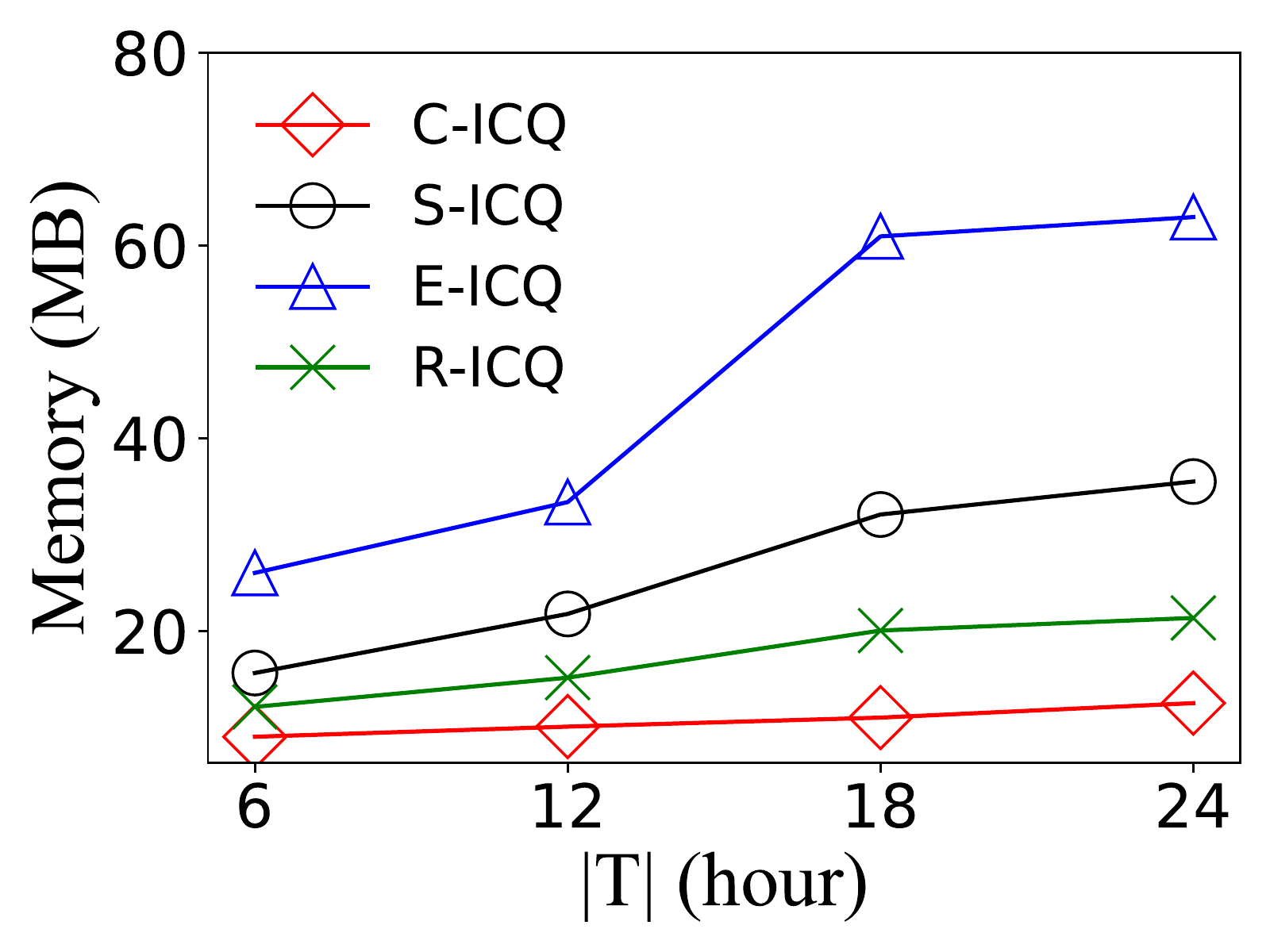}
            \label{fig:SYN-queryInterval-memory}
        \end{minipage}
    }
    
    \subfigure[Recall vs $|T|$]{
        \begin{minipage}[t]{0.45\columnwidth}
            \centering
            \includegraphics[width=\columnwidth]{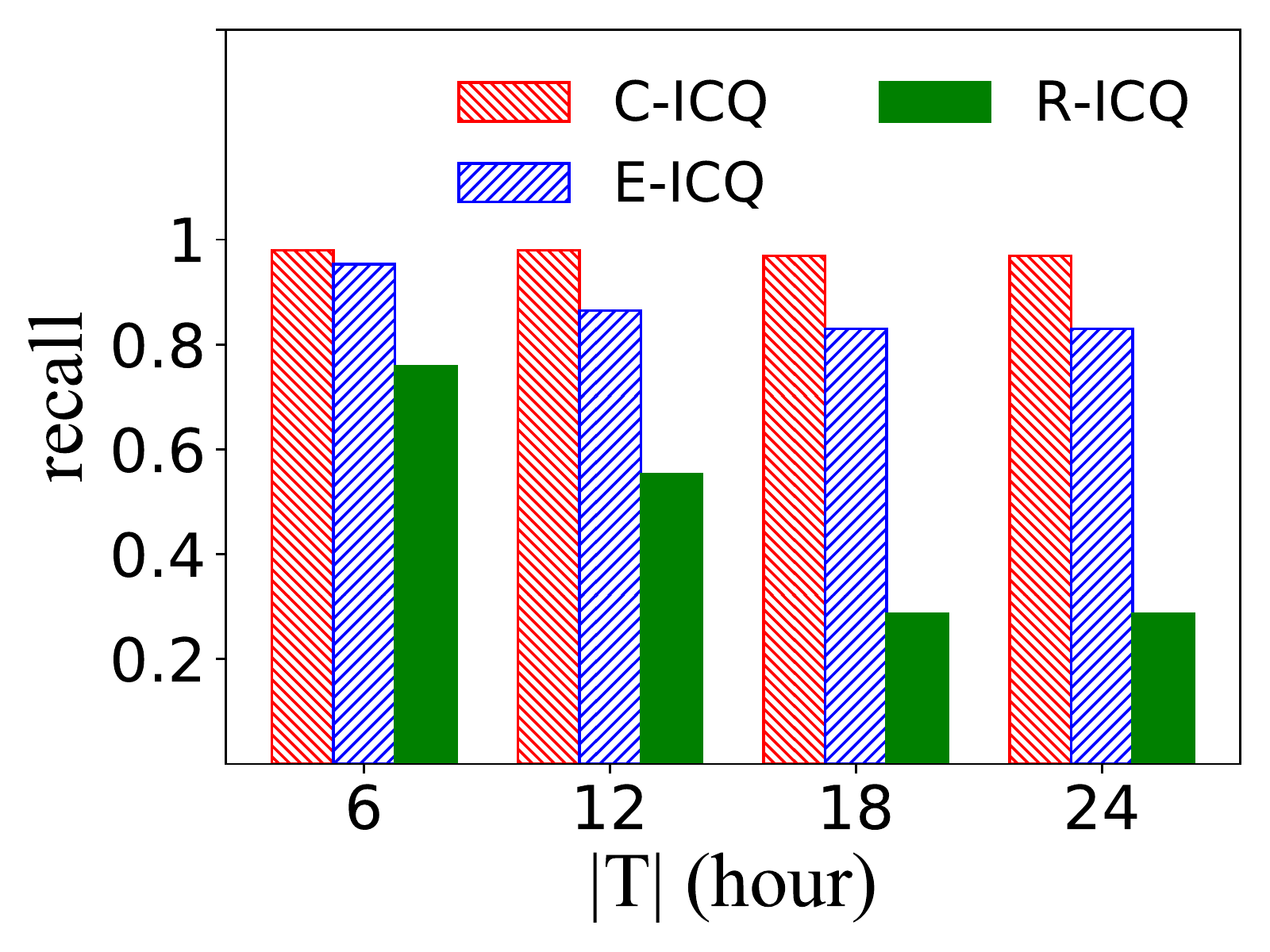}
            \label{fig:SYN-queryInterval-recall}
        \end{minipage}
    }
    \subfigure[F1 score vs $|T|$]{
        \begin{minipage}[t]{0.45\columnwidth}
            \centering
            \includegraphics[width=\columnwidth]{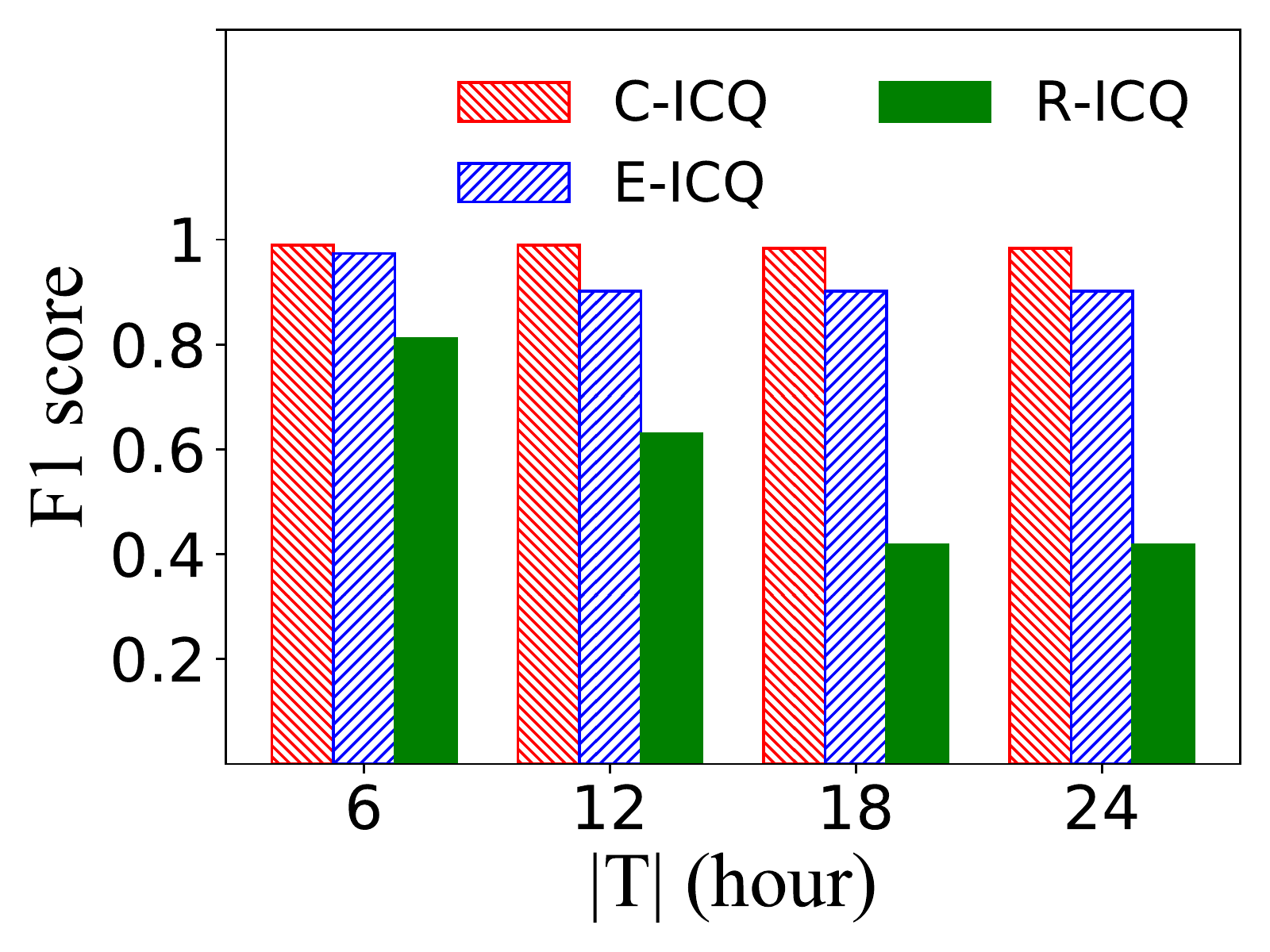}
            \label{fig:SYN-queryInterval-F}
        \end{minipage}
    }
    \centering
    \caption{Effect of $|T|$}
\end{figure}

\textbf{Effect of $|T|$.} We vary the query interval length $|T|$ from $6$ to $24$ hours.
The running time and memory cost of the four methods are reported in Figs.~\ref{fig:SYN-queryInterval-time} and~\ref{fig:SYN-queryInterval-memory}, respectively.
A longer $|T|$ leads to more time and memory costs as more candidate objects and sampling times are involved. 
For both efficiency measures, \CICQ{} performs the best. This is mainly due to its use of time skipping (Strategy~\ref{strategy:skipping}) for reducing the calls of instant contact determination. 
Compared to \SICQ{}, \CICQ{} additionally introduces early termination (Strategies~\ref{strategy:early_termination} and~\ref{strategy:reverse_early_termination}) and object sample pruning (Strategy~\ref{strategy:sample_pruning}) to instant contact determination, thereby incurring less time and memory overhead.
\RICQ{} and \EICQ{} cost more time and memory compared to our proposed methods. 
First, their used data structures (indoor distance-aware model plus a trajectory table) are less efficient than our enhanced graph model that is specialized to the retrieval of samples for relevant objects.
Second, for \EICQ{}, Euclidean distance leads to a larger uncertainty region with more location samples, which costs more time and memory in contact distance computations.
Third, unlike other methods, \RICQ{} needs to find and verify all candidate objects that have instant contact with the query object, thus incurring a longer running time.
Interestingly, \RICQ{} does not involve location samples and thus consumes relatively low memory.
Still, \CICQ{} consumes even less memory because of its efficient time skipping strategy.

The recall and F1 scores are reported in Figs.~\ref{fig:SYN-queryInterval-recall} and~\ref{fig:SYN-queryInterval-F}, respectively.
As \SICQ{} and \CICQ{} return the same set of close contact objects, we omit \SICQ{} in all effectiveness studies.
When increasing $|T|$, more candidate objects have to be analyzed.
As a result, all methods' two effectiveness measures degrade.
Nevertheless, \CICQ{} outperforms \EICQ{} and \RICQ{} at all $|T|$ values.
\RICQ{} does not analyze the uncertainty region, and its two effectiveness measures are clearly lower that the other methods using a sampling based analysis.
This indicates that our sample-based contact distance is effective at finding real close contacts.
Although \EICQ{} also uses the sampling based analysis, its effectiveness measures are inferior to those of \CICQ{}.
In \EICQ{}, samples are drawn from a coarser-grained uncertainty region derived based Euclidean distance.
This leads to a less accurate contact distance and thus a lower probability of finding true close contact objects.
The results show that our indoor uncertainty regions are more accurate for sampling locations.

Combining efficiency and effectiveness, \CICQ{} works better than the others in close contact tracing.

\begin{figure}[htbp]
    \centering
    \subfigure[Time vs $\delta$]{
        \begin{minipage}[t]{0.45\columnwidth}
            \centering
            \includegraphics[width=\columnwidth]{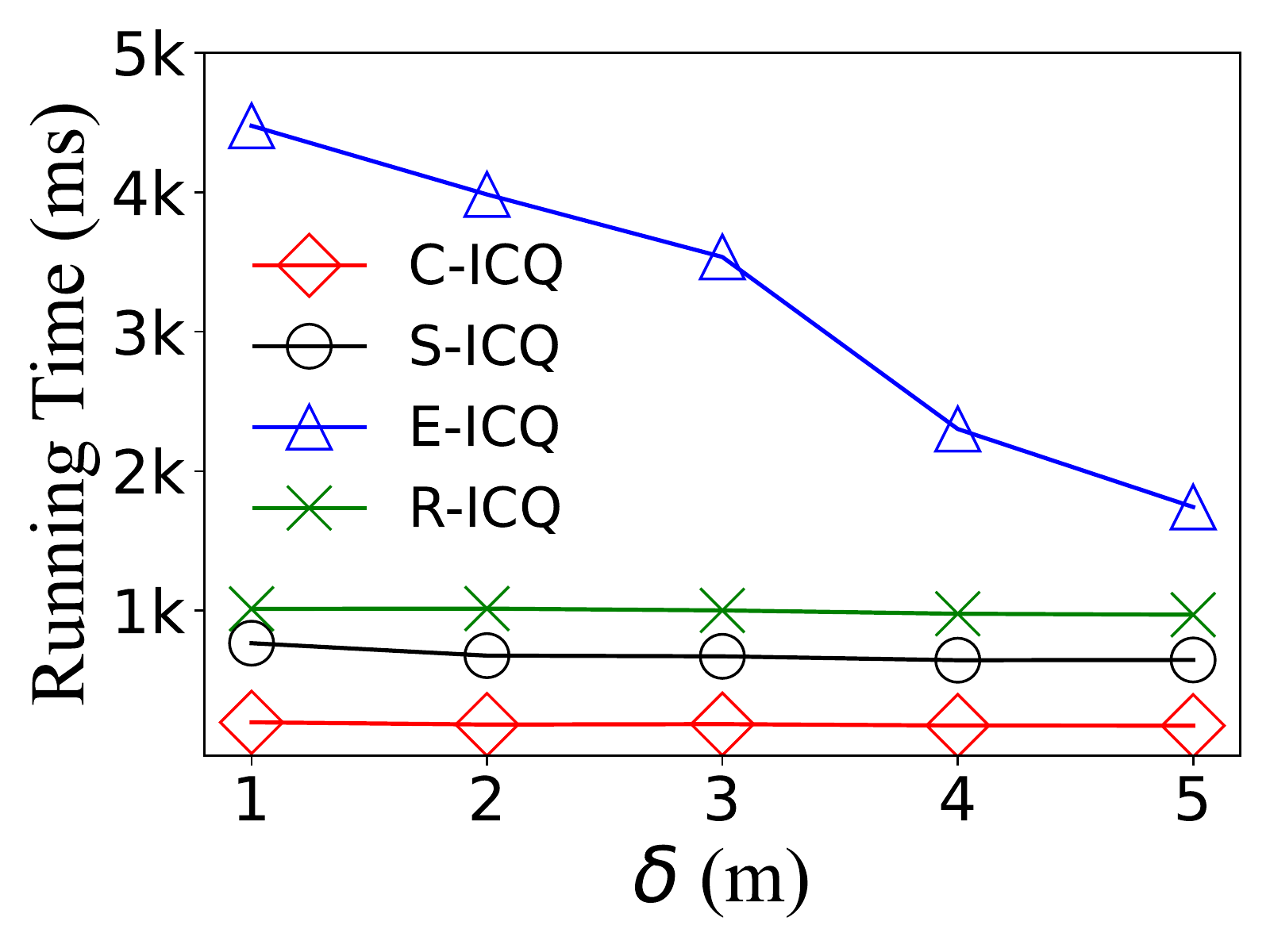}
            \label{fig:SYN-distance-time}
        \end{minipage}
    }
    \subfigure[Mem. vs $\delta$]{
        \begin{minipage}[t]{0.45\columnwidth}
            \centering
            \includegraphics[width=\columnwidth]{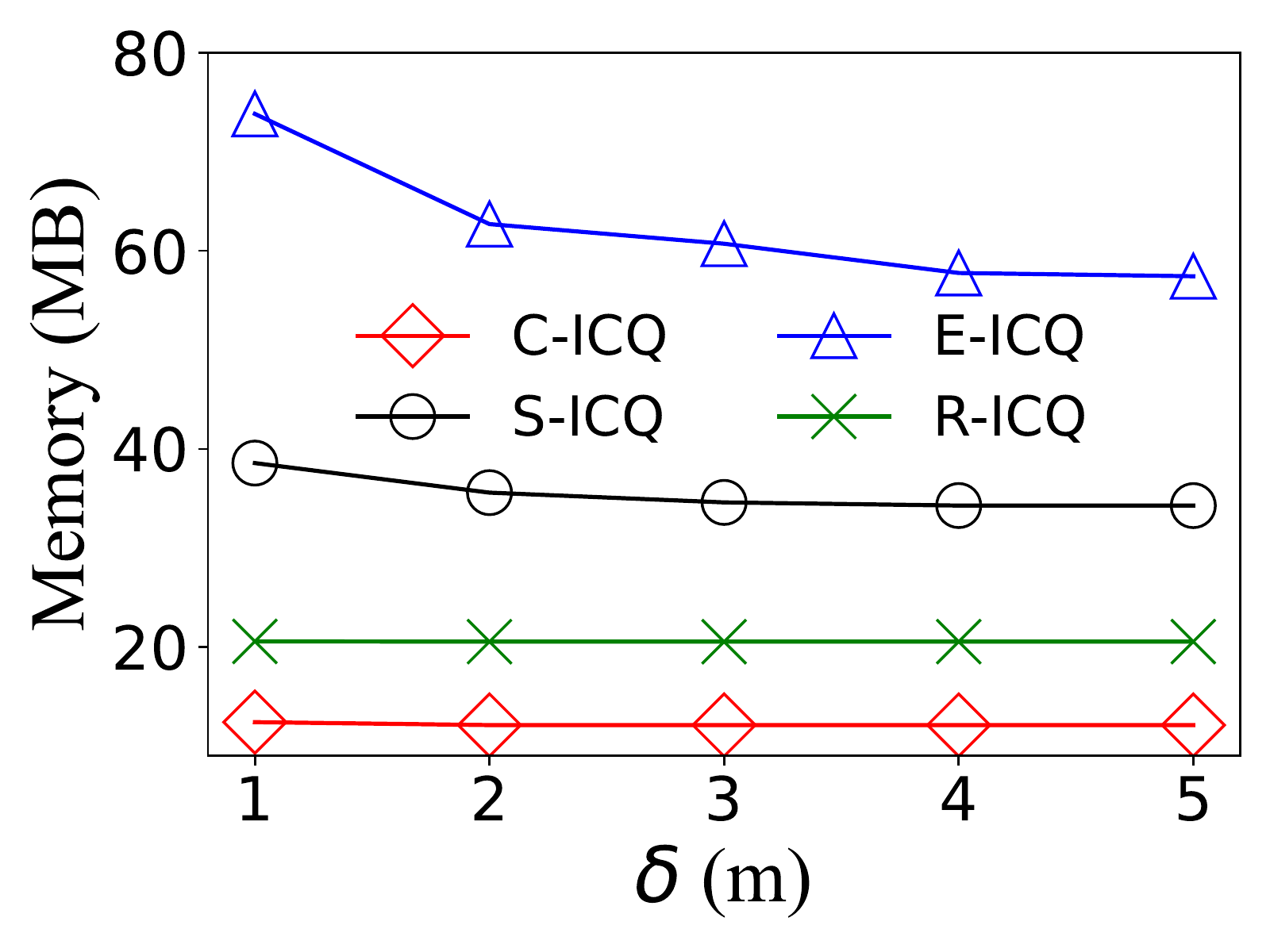}
            \label{fig:SYN-distance-memory}
        \end{minipage}
    }
    
    \subfigure[Recall vs $\delta$]{
        \begin{minipage}[t]{0.45\columnwidth}
            \centering
            \includegraphics[width=\columnwidth]{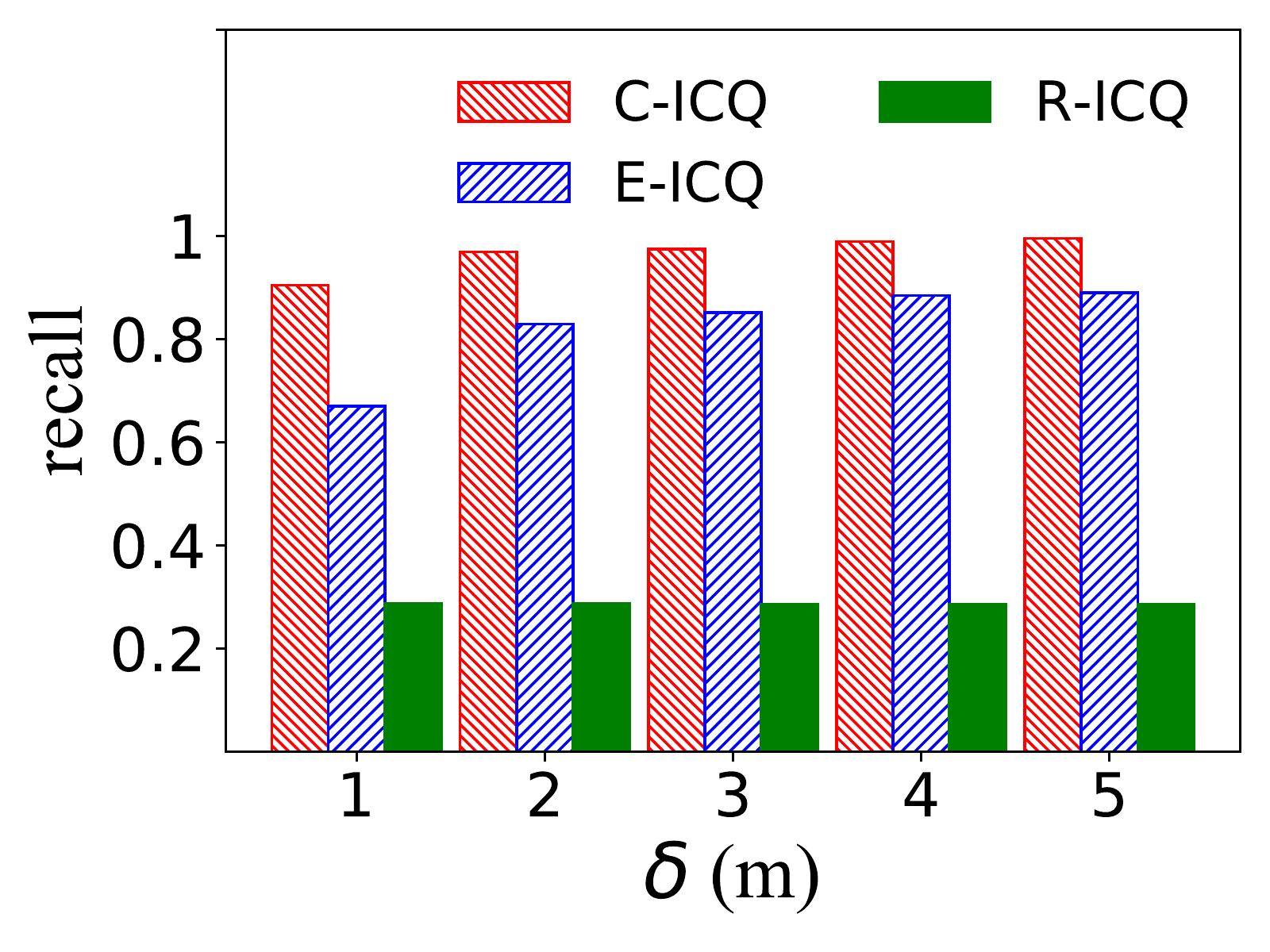}
            \label{fig:SYN-distance-recall}
        \end{minipage}
    }
    \subfigure[F1 score vs $\delta$]{
        \begin{minipage}[t]{0.45\columnwidth}
            \centering
            \includegraphics[width=\columnwidth]{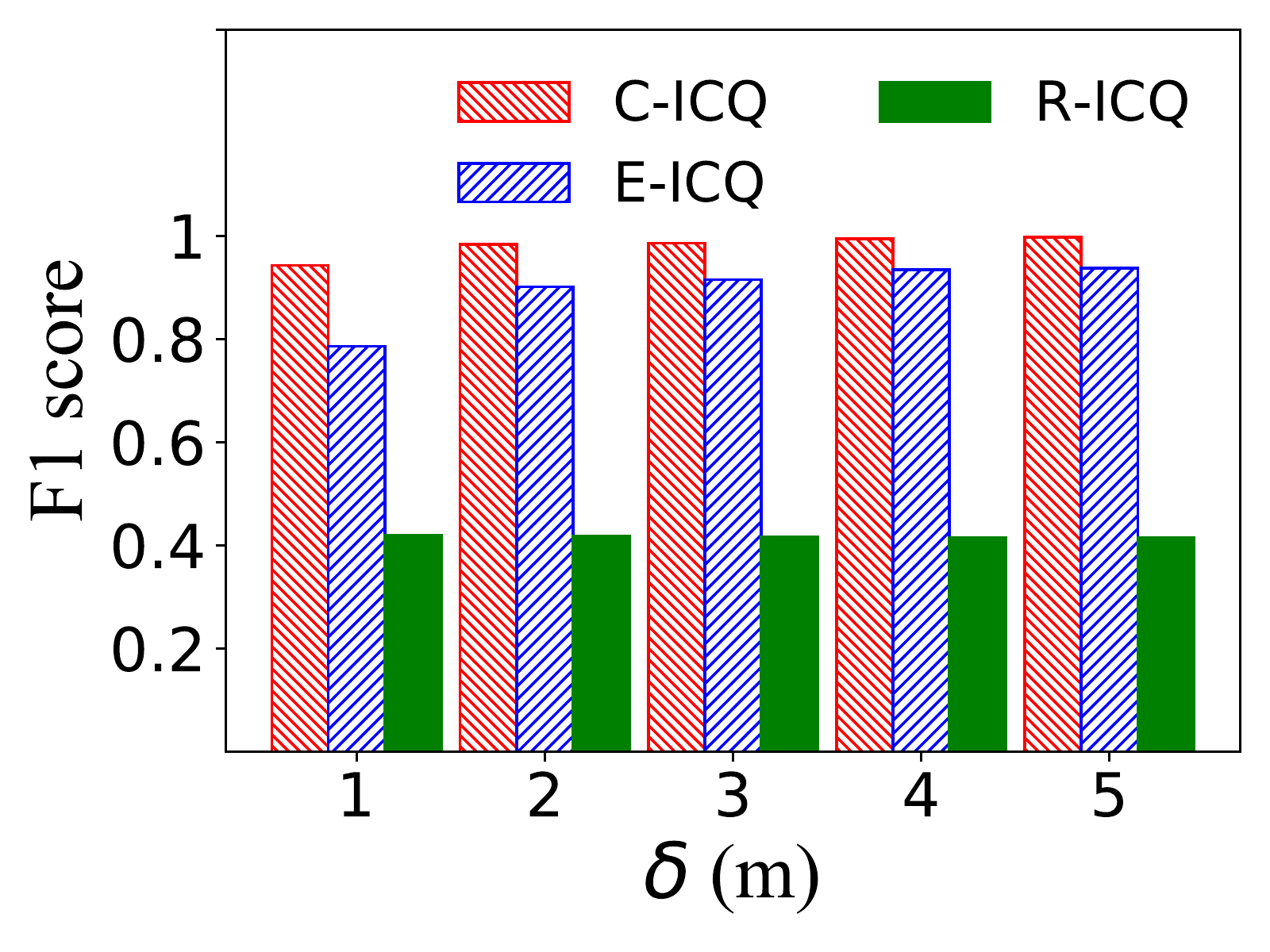}
            \label{fig:SYN-distance-F}
        \end{minipage}
    }
    \centering
    \caption{Effect of $\delta$}
\end{figure}

\noindent\textbf{Effect of $\delta$}. Figs.~\ref{fig:SYN-distance-time} and~\ref{fig:SYN-distance-memory} report the efficiency of four methods, \CICQ{} still costs least running time and memory because of our efficient model and strategies.
The running time and memory cost of \EICQ{} decrease when the distance constraint $\delta$ increases. 
This is because a higher $\delta$ makes it easier for the accumulating contact probability of object samples to exceed the fixed threshold $\eta$, causing a candidate object to be returned earlier and avoiding further processing in subsequent iterations.
In contrast, \CICQ{} and \SICQ{} use indoor distances to find the uncertainty region, which includes fewer objects, resulting in relatively stable efficiency with increased $\delta$.
The time and memory costs of \RICQ{} are insensitive to $\delta$ as there is no uncertainty analysis at unseen timestamps.

Figs.~\ref{fig:SYN-distance-recall} and~\ref{fig:SYN-distance-F} report on the recall and F1 score.
With a larger distance constraint, more objects tend to be counted as close contacts.
As a result, the recall and F1 score measures of both methods increase.
Compared to the two baselines, \CICQ{} still performs best because of the precise analysis of the uncertainty region.
\RICQ{} ignores unseen sampling timestamps, so the variation of the distance constraint has no impact on its effectiveness.

\begin{figure}[htbp]
    \centering
    \subfigure[Time vs $\eta$]{
        \begin{minipage}[t]{0.45\columnwidth}
            \centering
            \includegraphics[width=\columnwidth]{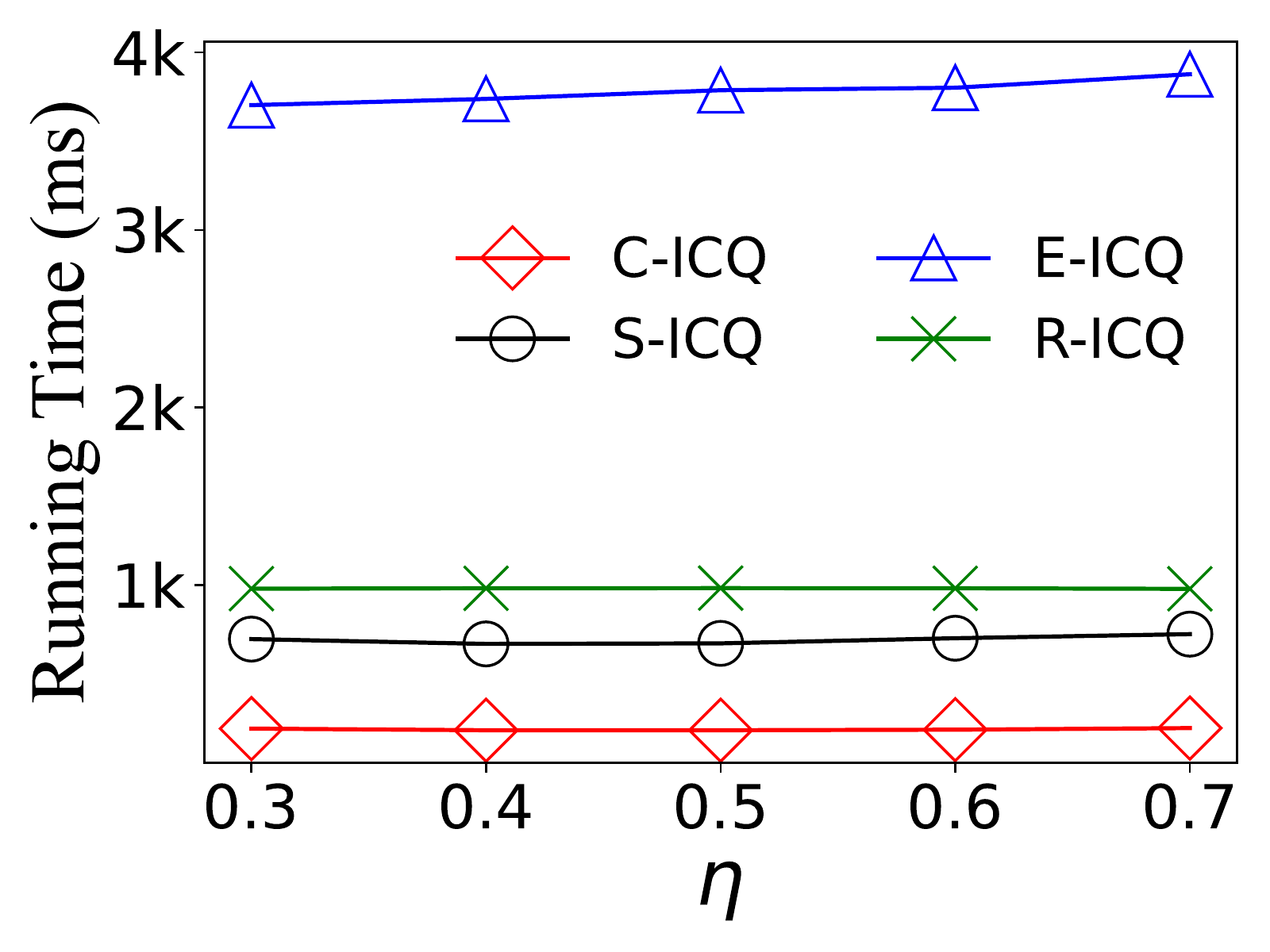}
            \label{fig:SYN-probability-time}
        \end{minipage}
    }
    \subfigure[Mem. vs $\eta$]{
        \begin{minipage}[t]{0.45\columnwidth}
            \centering
            \includegraphics[width=\columnwidth]{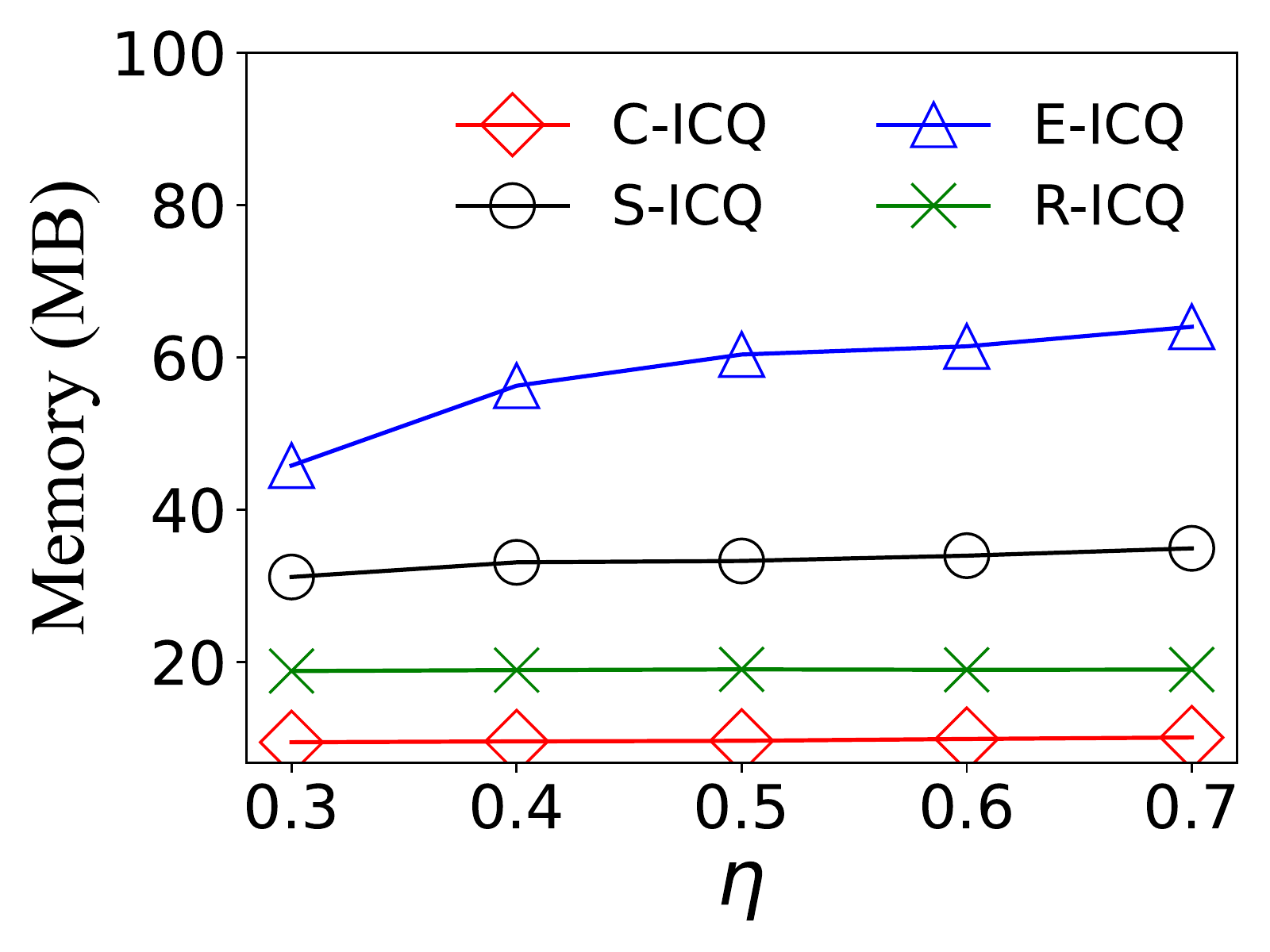}
            \label{fig:SYN-probability-memory}
        \end{minipage}
    }
    
    \subfigure[Recall vs $\eta$]{
        \begin{minipage}[t]{0.45\columnwidth}
            \centering
            \includegraphics[width=\columnwidth]{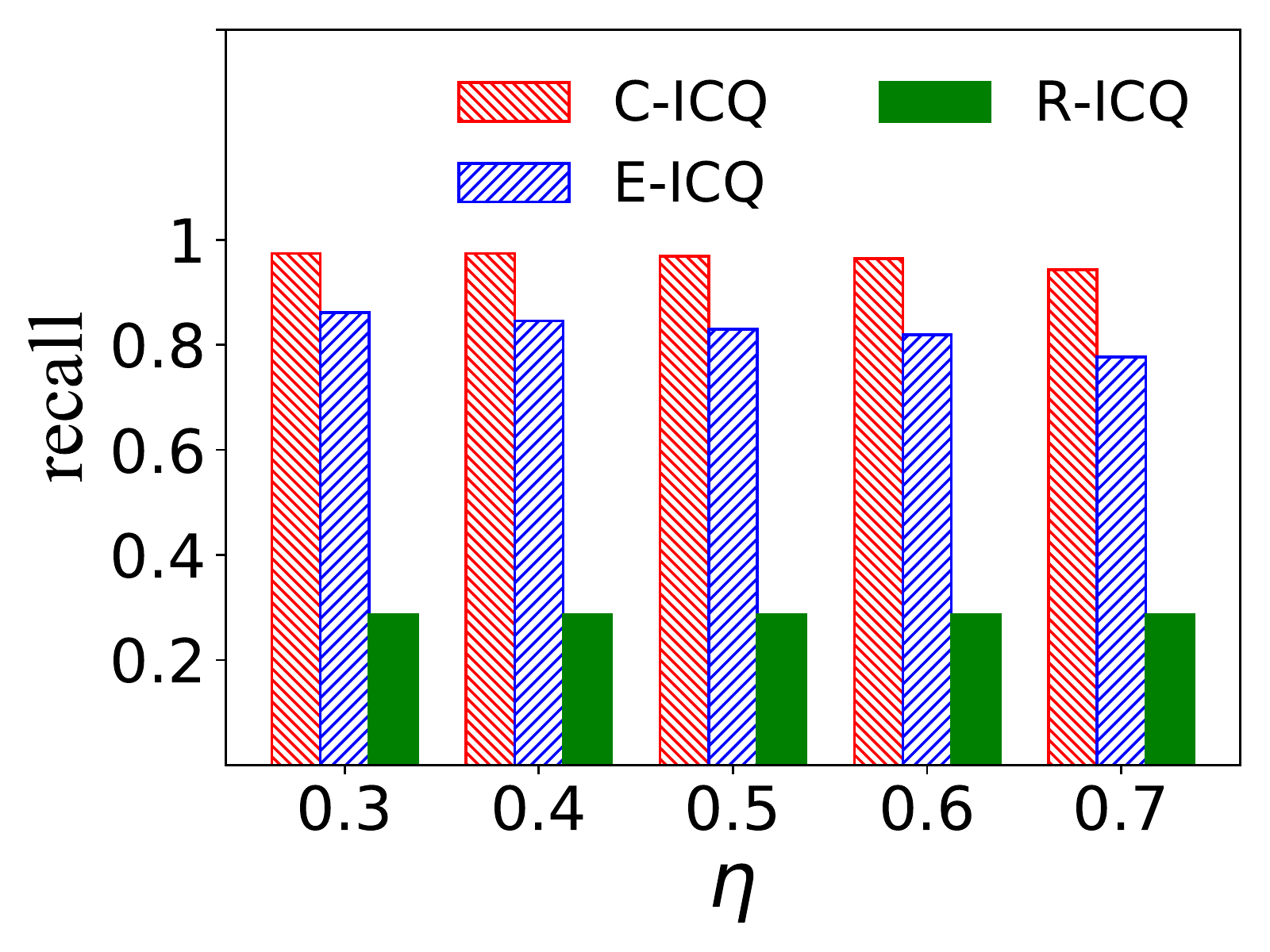}
            \label{fig:SYN-probability-recall}
        \end{minipage}
    }
    \subfigure[F1 score vs $\eta$]{
        \begin{minipage}[t]{0.45\columnwidth}
            \centering
            \includegraphics[width=\columnwidth]{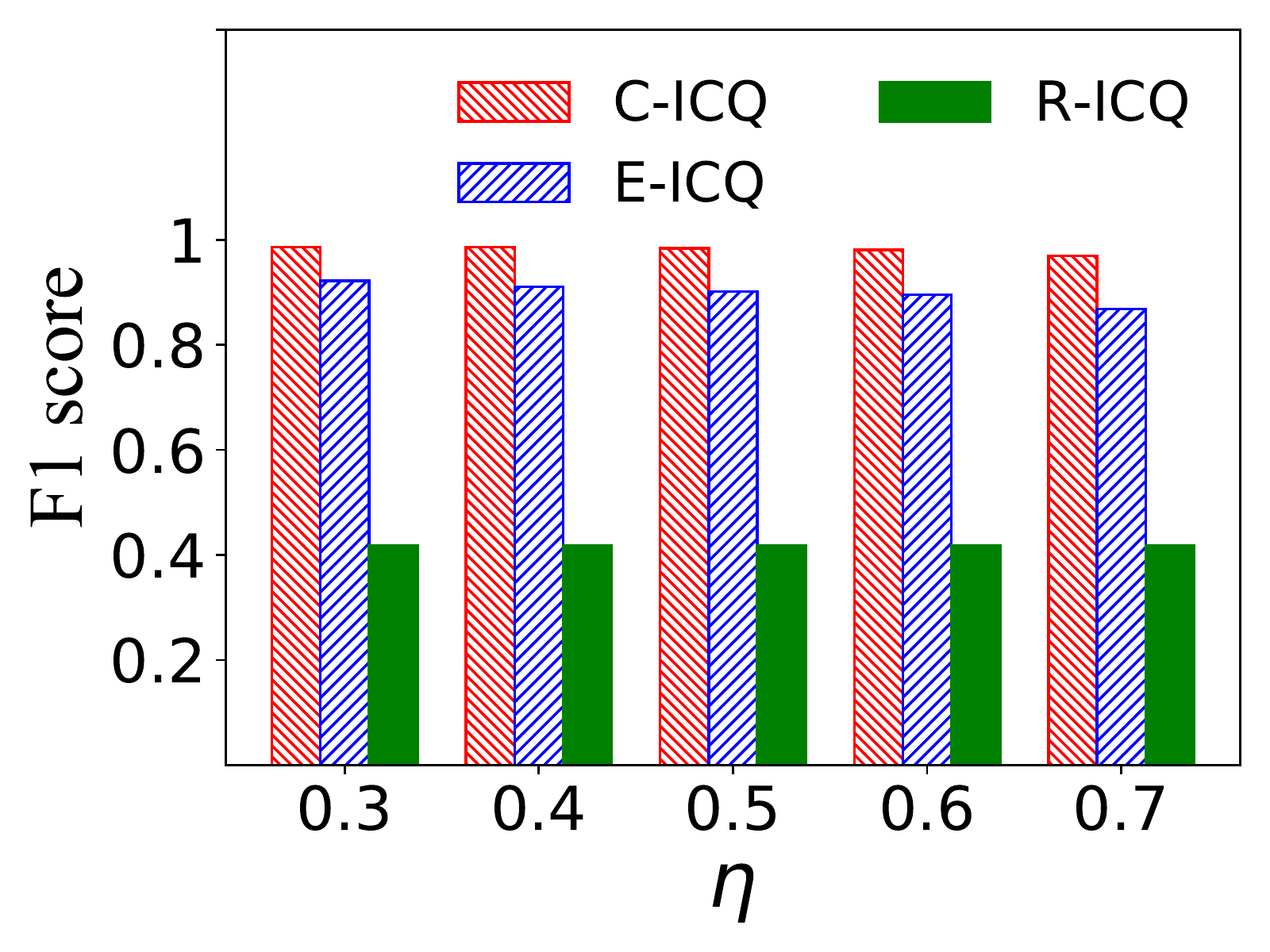}
            \label{fig:SYN-probability-F}
        \end{minipage}
    }
    \centering
    \caption{Effect of $\eta$}
\end{figure}

\noindent\textbf{Effect of $\eta$}. Referring to Figs.~\ref{fig:SYN-probability-time} and~\ref{fig:SYN-probability-memory}, increasing the contact probability threshold $\eta$ has little impact on the running time and memory consumption of all methods. Indeed, varying $\eta$ does not change the number of calls of instant contact determinations.

Figs.~\ref{fig:SYN-probability-recall} and~\ref{fig:SYN-probability-F} report on the two effectiveness measures of the three methods when varying $\eta$. 
As mentioned above, \RICQ{} omits unseen sampling timestamps.
Therefore, both recall and F1 score of \RICQ{} stay no change.
In contrast, a higher $\eta$ imposes stricter requirements on the establishment of instant contact, thereby making it less accurate the instant contact determination based on the derived samples.
As a result, both recall and F1 score of \CICQ{} and \EICQ{} decrease with $\eta$ increased.
Nevertheless, \CICQ{} still performs the best.

\begin{figure}[htbp]
    \centering
    \subfigure[Time vs $k$]{
        \begin{minipage}[t]{0.45\columnwidth}
            \centering
            \includegraphics[width=\columnwidth]{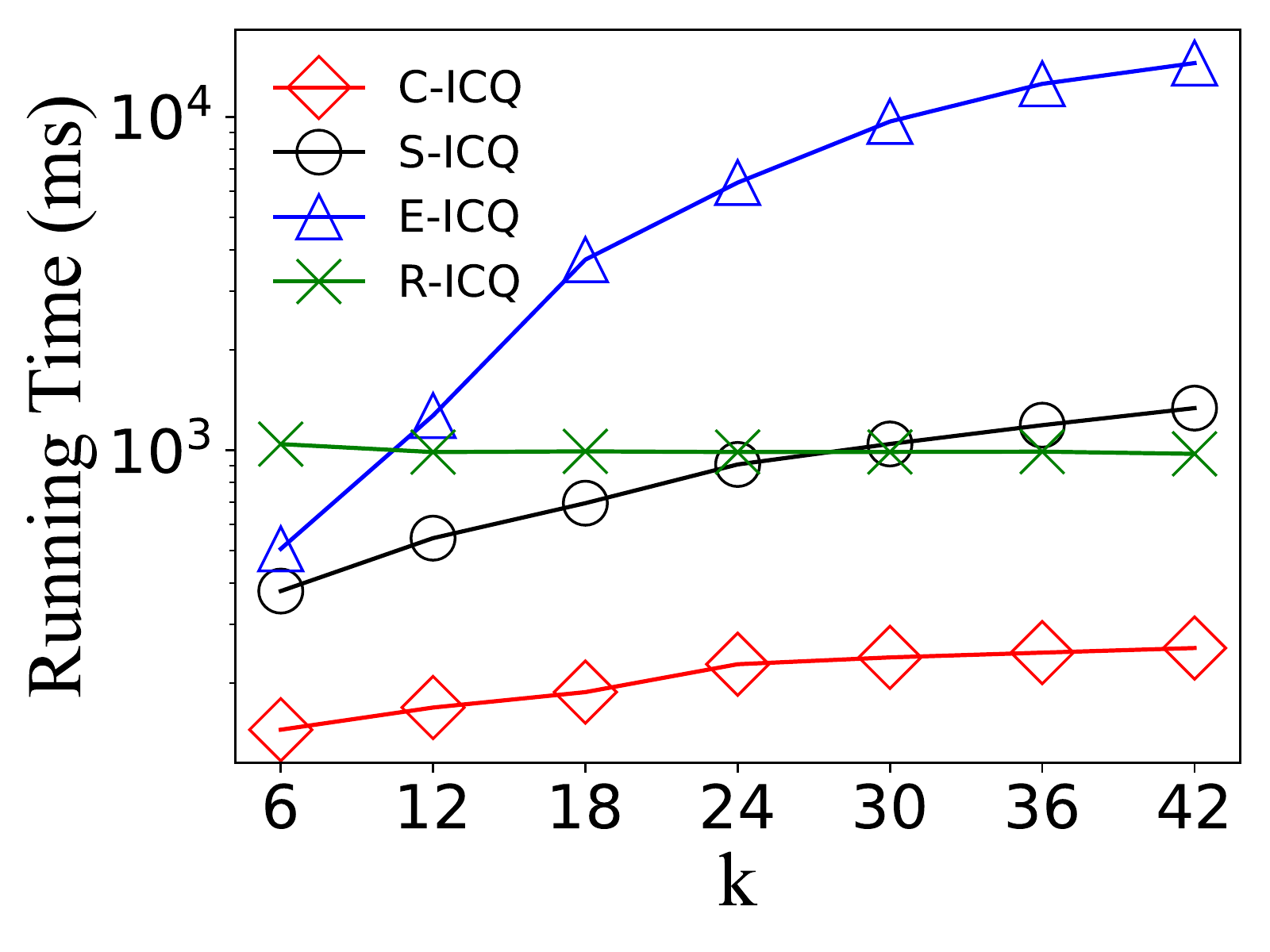}
            \label{fig:SYN-duration-time}
        \end{minipage}
    }
    \subfigure[Mem. vs $k$]{
        \begin{minipage}[t]{0.45\columnwidth}
            \centering
            \includegraphics[width=\columnwidth]{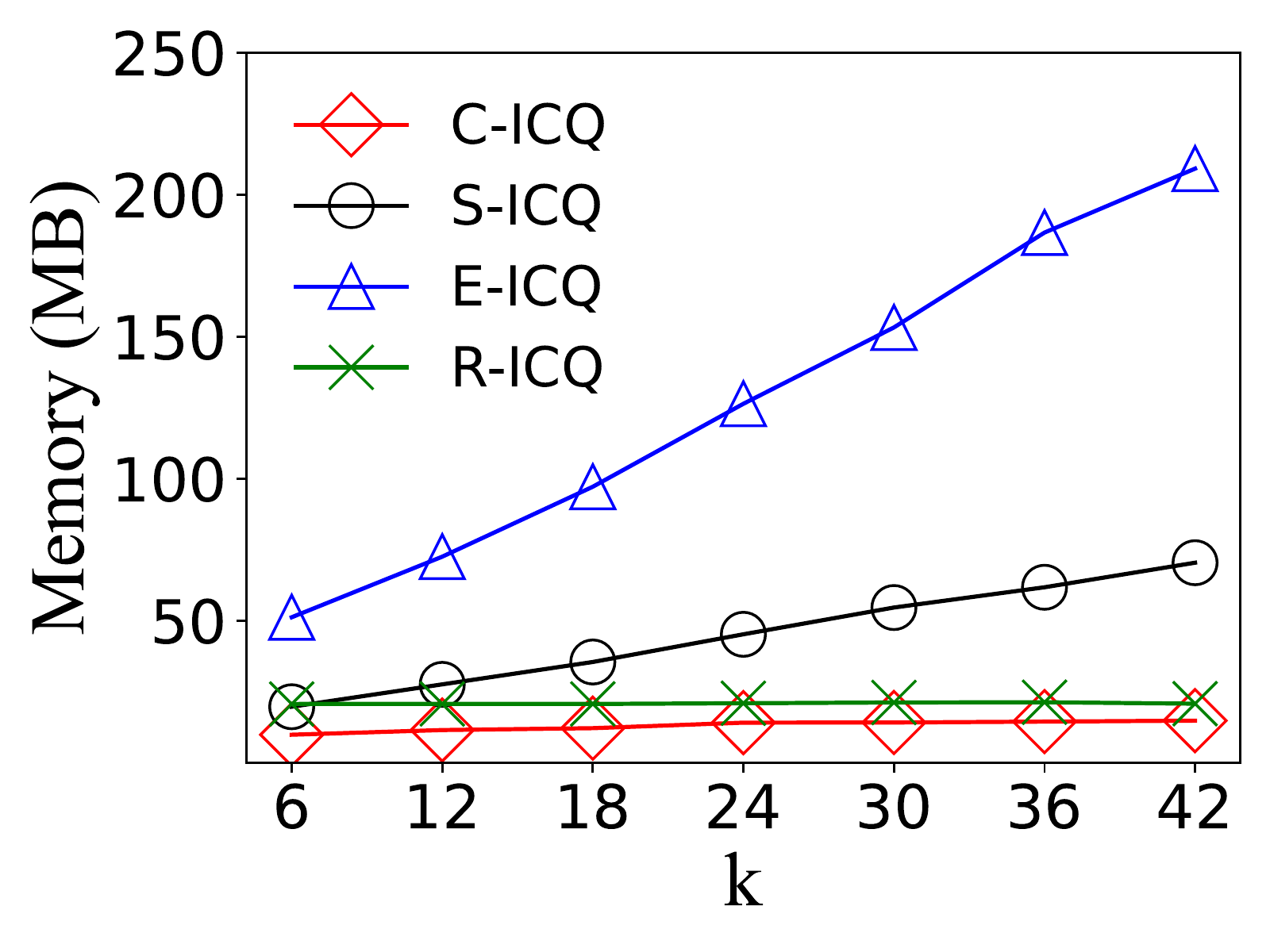}
            \label{fig:SYN-duration-memory}
        \end{minipage}
    }
    
    \subfigure[Recall vs $k$]{
        \begin{minipage}[t]{0.45\columnwidth}
            \centering
            \includegraphics[width=\columnwidth]{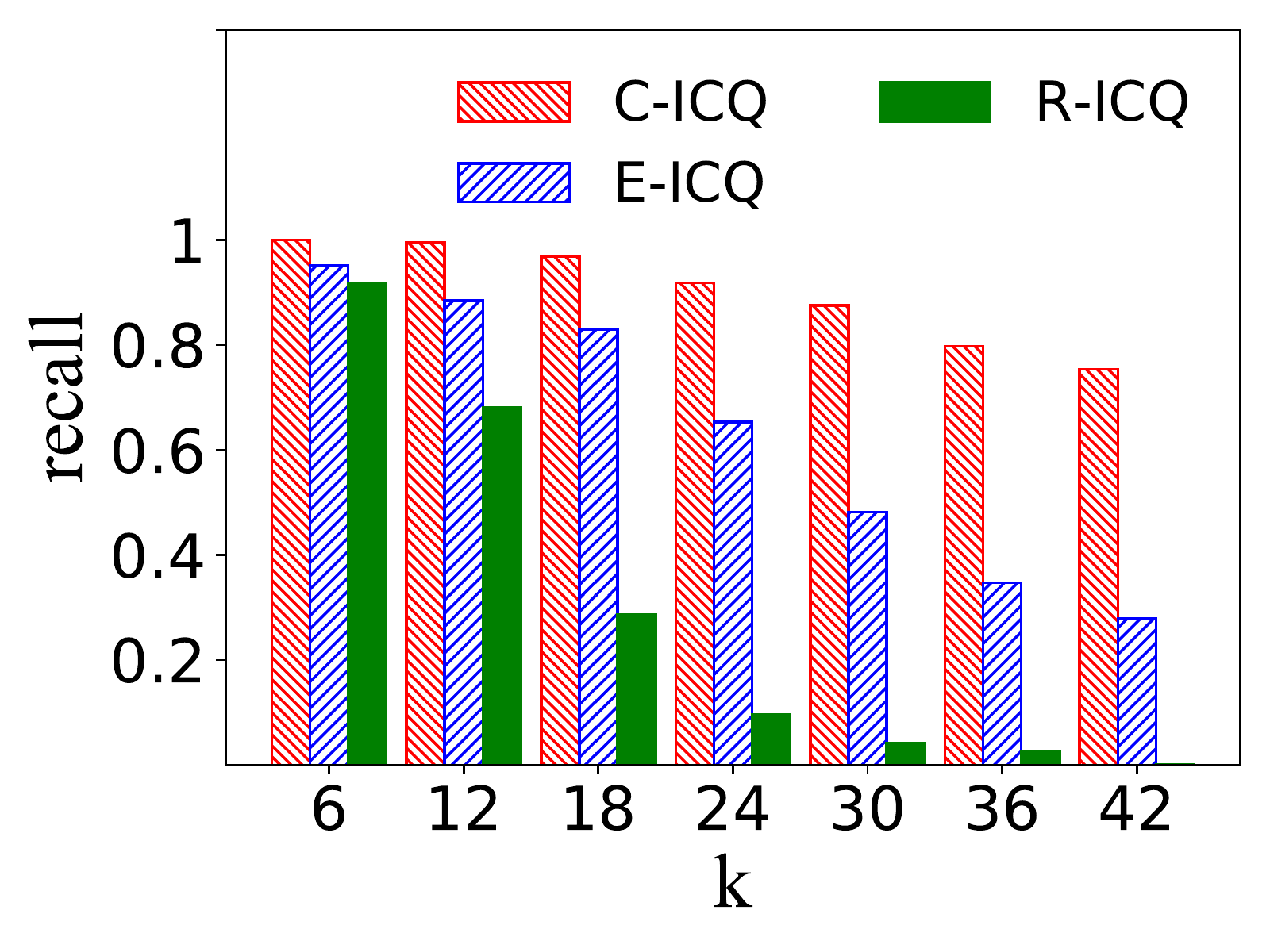}
            \label{fig:SYN-duration-recall}
        \end{minipage}
    }
    \subfigure[F1 score vs $k$]{
        \begin{minipage}[t]{0.45\columnwidth}
            \centering
            \includegraphics[width=\columnwidth]{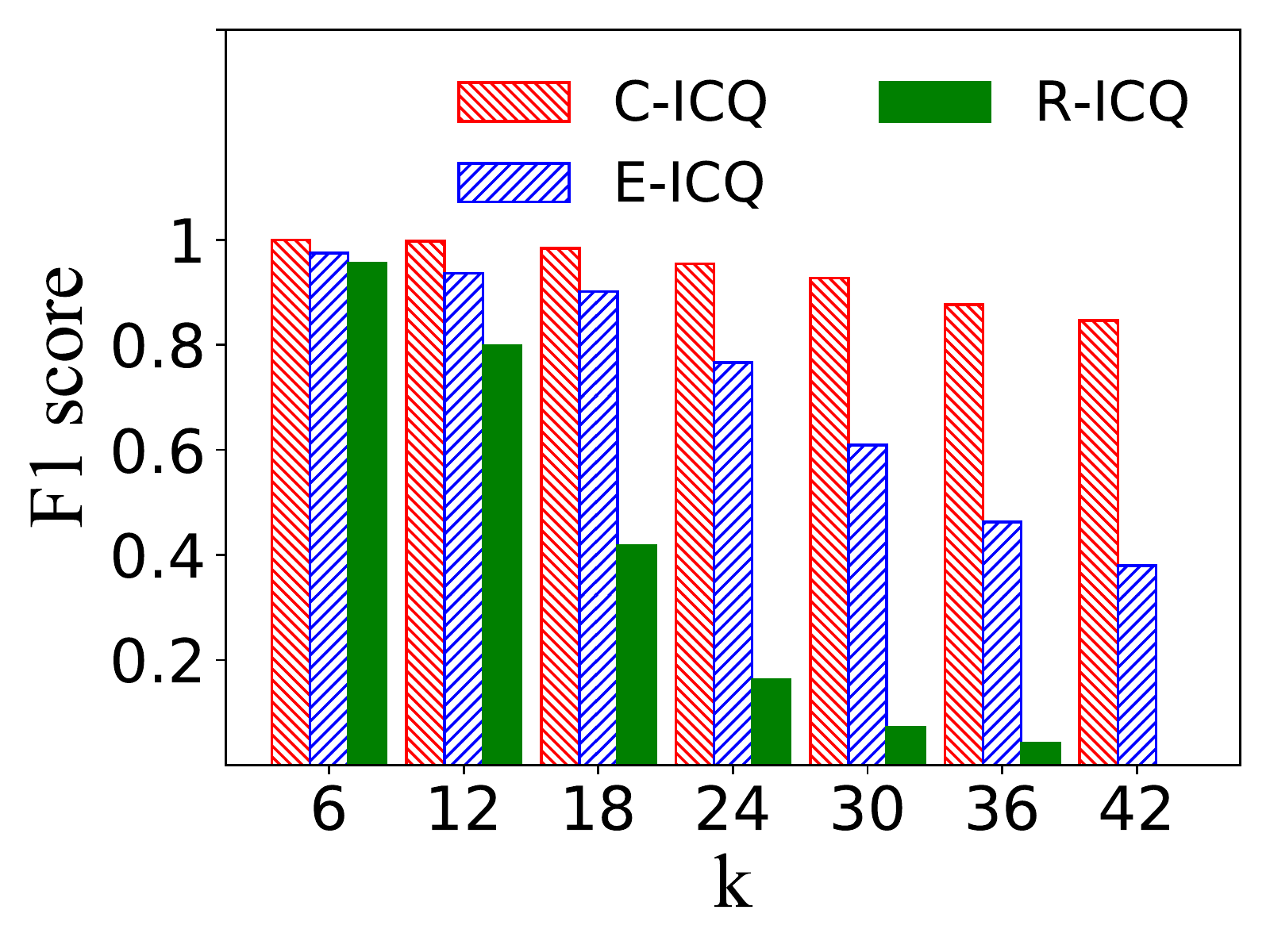}
            \label{fig:SYN-duration-F}
        \end{minipage}
    }
    \centering
    \caption{Effect of $k$}
\end{figure}

\noindent\textbf{Effect of $k$}. We vary $k$ from 6 to 42 and report the running time and memory cost in Figs.~\ref{fig:SYN-duration-time} and~\ref{fig:SYN-duration-memory}, respectively.
As no probabilistic samples are derived and processed, the running time and memory cost of \RICQ{} are almost insensitive to $k$.
In contrast, both running time and memory cost of other three methods increase with an increasing $k$.
For \EICQ{}, a larger $k$ involves more sampling times to be checked consecutively in determining the close contact between two objects. 
This requires a longer running time and a larger memory cost.
Compared to \SICQ{}, \CICQ{}'s both measures increase much more slowly.
This is due to its efficient time skipping strategy, which reduces the unnecessary heavy computations on consecutive sampling times. 
Moreover, the time and memory costs of \CICQ{} tend to be stable when $k$ is sufficiently large. 
This demonstrates that our proposed \CICQ{} performs very well when the close contact tracing is conducted with a relatively large $k$.

As shown in Figs.~\ref{fig:SYN-duration-recall} and~\ref{fig:SYN-duration-F}, the recall and F1 score of all methods deteriorate when $k$ becomes larger.
A larger $k$ means a stricter requirement on close contact (see our research problem defined in Section~\ref{ssec:problem_formulation}), which renders the close contact tracing on uncertain positioning data less effective.
Nevertheless, \CICQ{} using probabilistic samples clearly outperforms \EICQ{} analyzing uncertainty region with Euclidean distance and \RICQ{}.
Moreover, the performance gap enlarges when  $k$ increases to 42.
In general, our proposed \CICQ{} is more efficient and effective than the two baselines when a larger $k$ is used.

\begin{figure}[htbp]
    \centering
    \subfigure[Time vs $\mathit{ll}$]{
        \begin{minipage}[t]{0.45\columnwidth}
            \centering
            \includegraphics[width=\columnwidth]{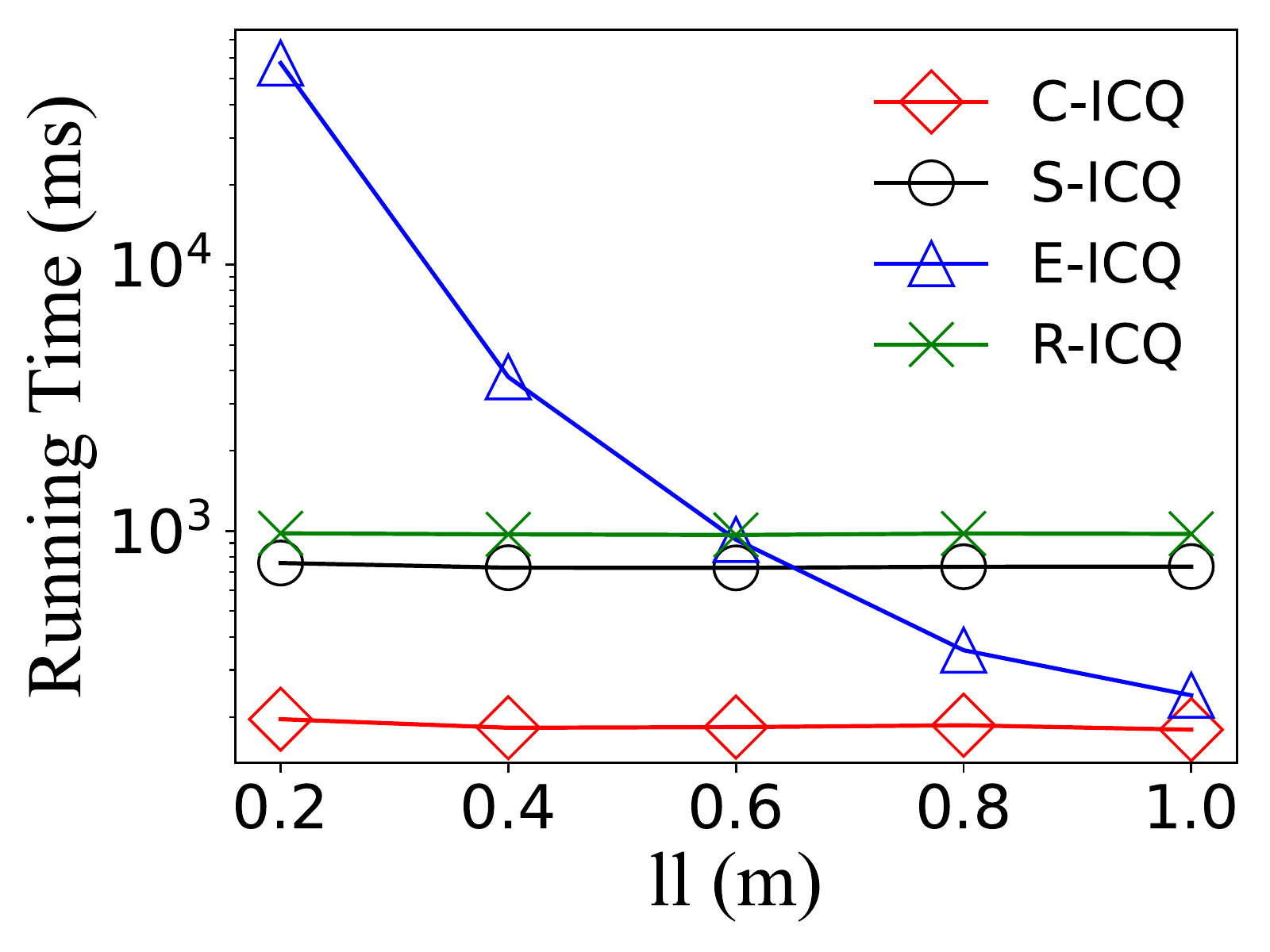}
            \label{fig:SYN-granularity-time}
        \end{minipage}
    }
    \subfigure[Mem. vs $\mathit{ll}$]{
        \begin{minipage}[t]{0.45\columnwidth}
            \centering
            \includegraphics[width=\columnwidth]{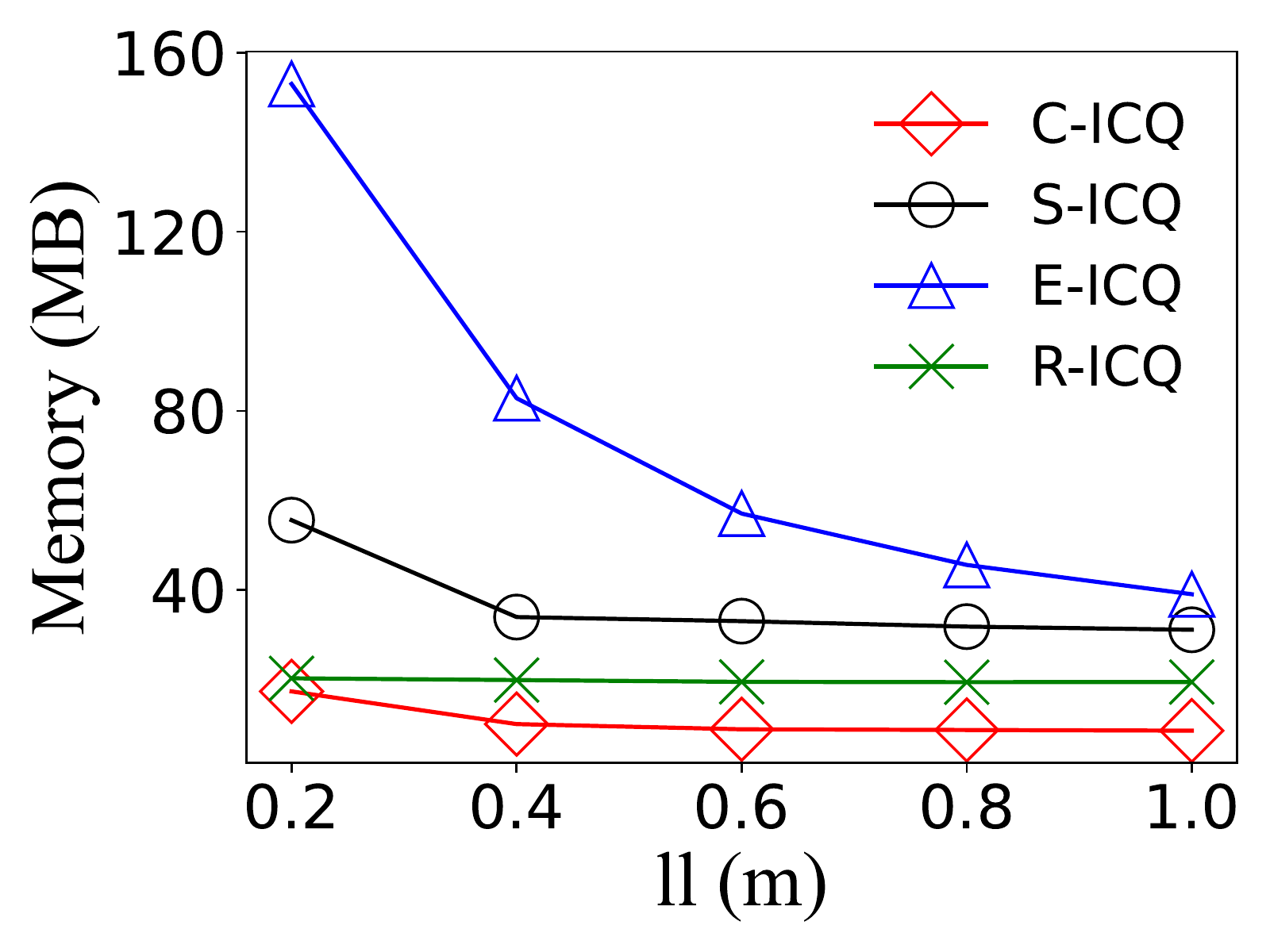}
            \label{fig:SYN-granularity-memory}
        \end{minipage}
    }
    
    \subfigure[Recall vs $\mathit{ll}$]{
        \begin{minipage}[t]{0.45\columnwidth}
            \centering
            \includegraphics[width=\columnwidth]{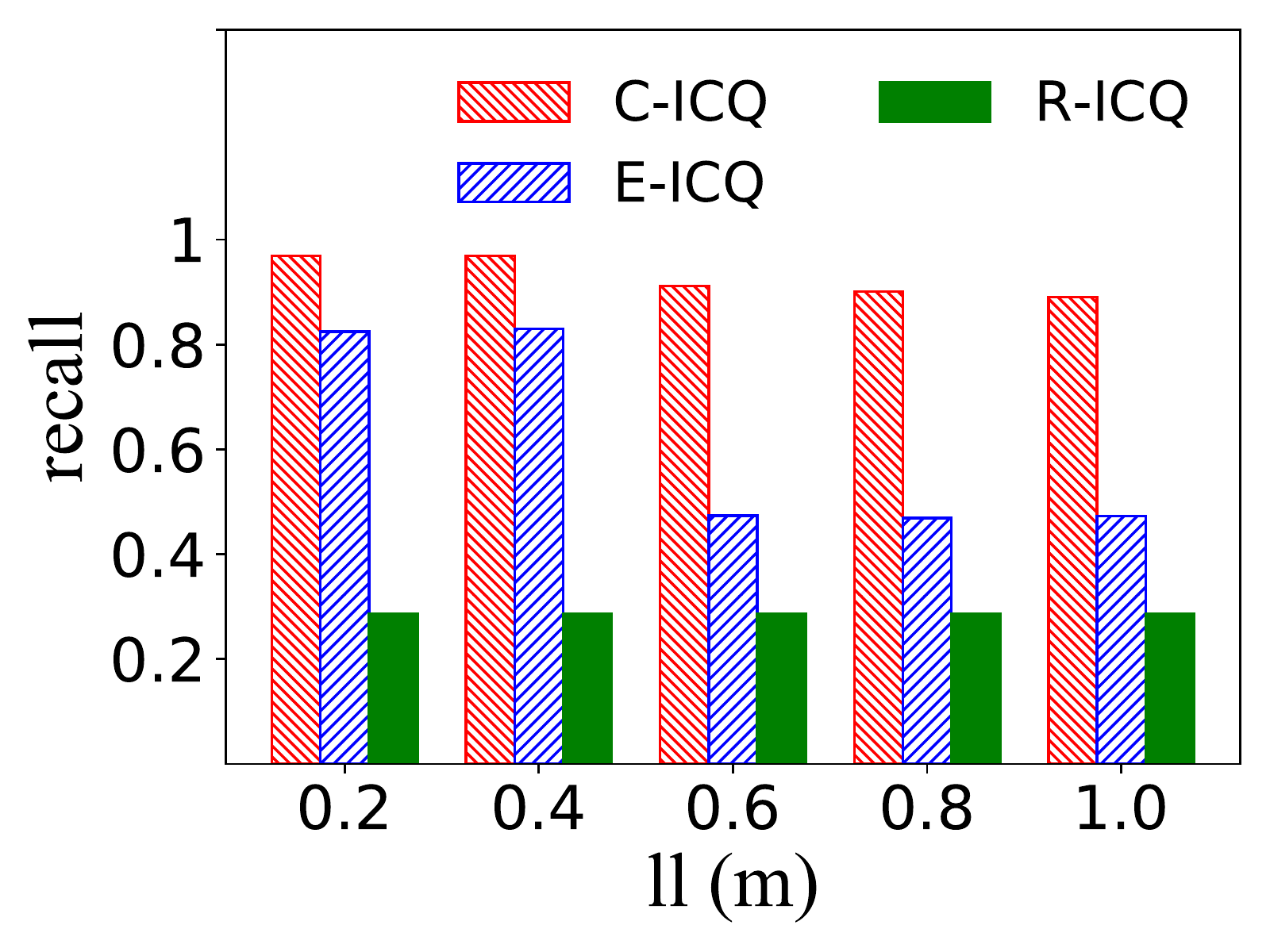}
            \label{fig:SYN-granularity-recall}
        \end{minipage}
    }
    \subfigure[F1 score vs $\mathit{ll}$]{
        \begin{minipage}[t]{0.45\columnwidth}
            \centering
            \includegraphics[width=\columnwidth]{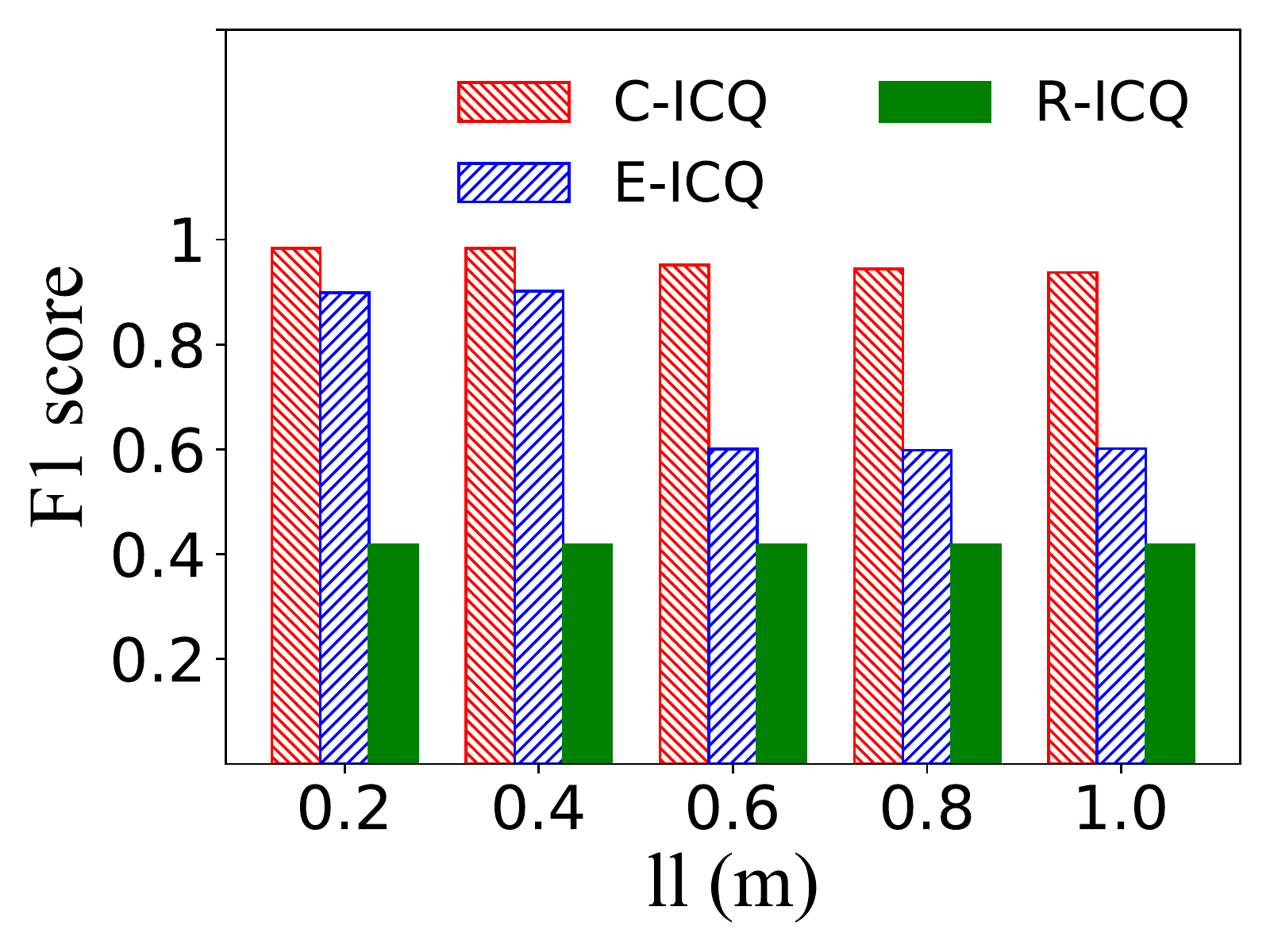}
            \label{fig:SYN-granularity-F}
        \end{minipage}
    }
    \centering
    \caption{Effect of $\mathit{ll}$}
\end{figure}

\noindent\textbf{Effect of $\mathit{ll}$}. We vary the lattice side length $\mathit{ll}$ from $0.2$m to $1$m.
Referring to Figs.~\ref{fig:SYN-granularity-time} and~\ref{fig:SYN-granularity-memory}, in terms of the running time and memory cost, the efficiency of \EICQ{} increases with a larger $\mathit{ll}$ because fewer samples are introduced in computations.
\RICQ{}'s time and memory costs remain stable as no analysis is involved.
Still, \CICQ{} runs much faster and cost less memory than other methods.

Referring to Figs.~\ref{fig:SYN-granularity-recall} and~\ref{fig:SYN-granularity-F}, the recall and F1 score of \CICQ{} degrade when a larger $\mathit{ll}$ is used. 
As fewer samples are used, the accuracy of instant contact determination decreases, and so do the recall and precision of the query result.
In our experiments, \CICQ{} achieves better recall and F1 score compared to the two baselines.

\begin{figure}[htbp]
    \centering
    \subfigure[Time vs $|O|$]{
        \begin{minipage}[t]{0.45\columnwidth}
            \centering
            \includegraphics[width=\columnwidth]{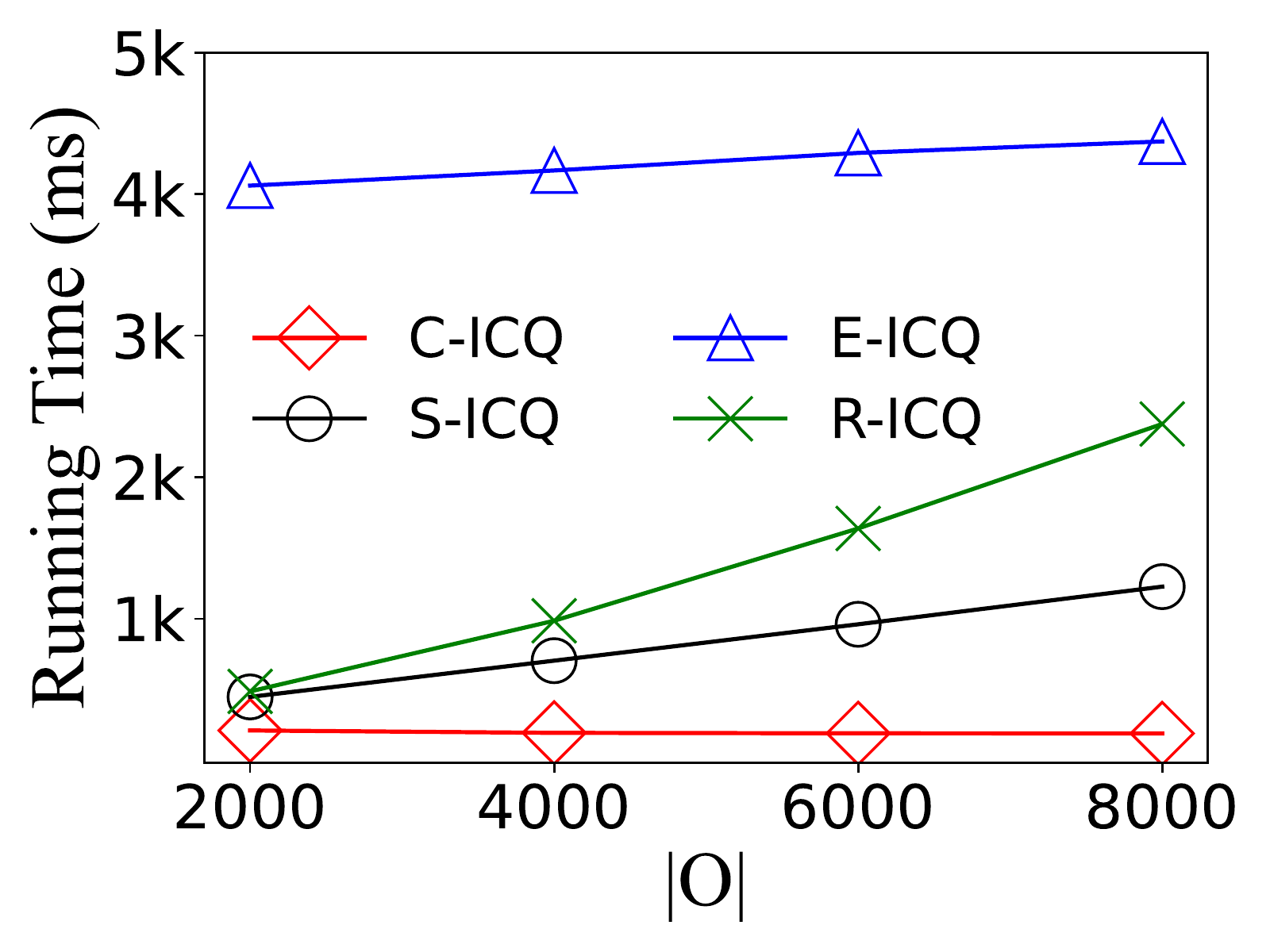}
            \label{fig:SYN-objectNum-time}
        \end{minipage}
    }
    \subfigure[Mem. vs $|O|$]{
        \begin{minipage}[t]{0.45\columnwidth}
            \centering
            \includegraphics[width=\columnwidth]{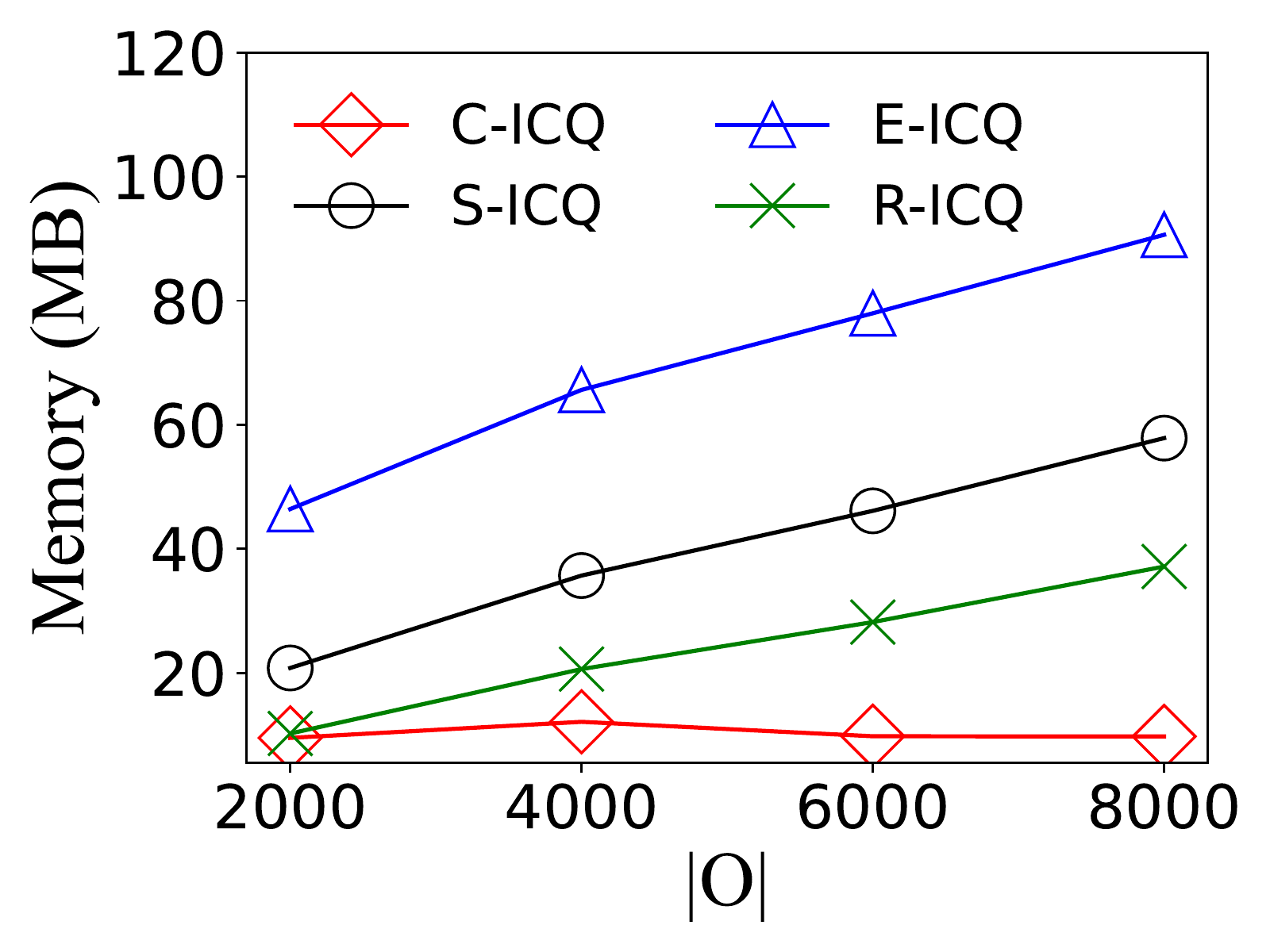}
            \label{fig:SYN-objectNum-memory}
        \end{minipage}
    }
    
    \subfigure[Recall vs $|O|$]{
        \begin{minipage}[t]{0.45\columnwidth}
            \centering
            \includegraphics[width=\columnwidth]{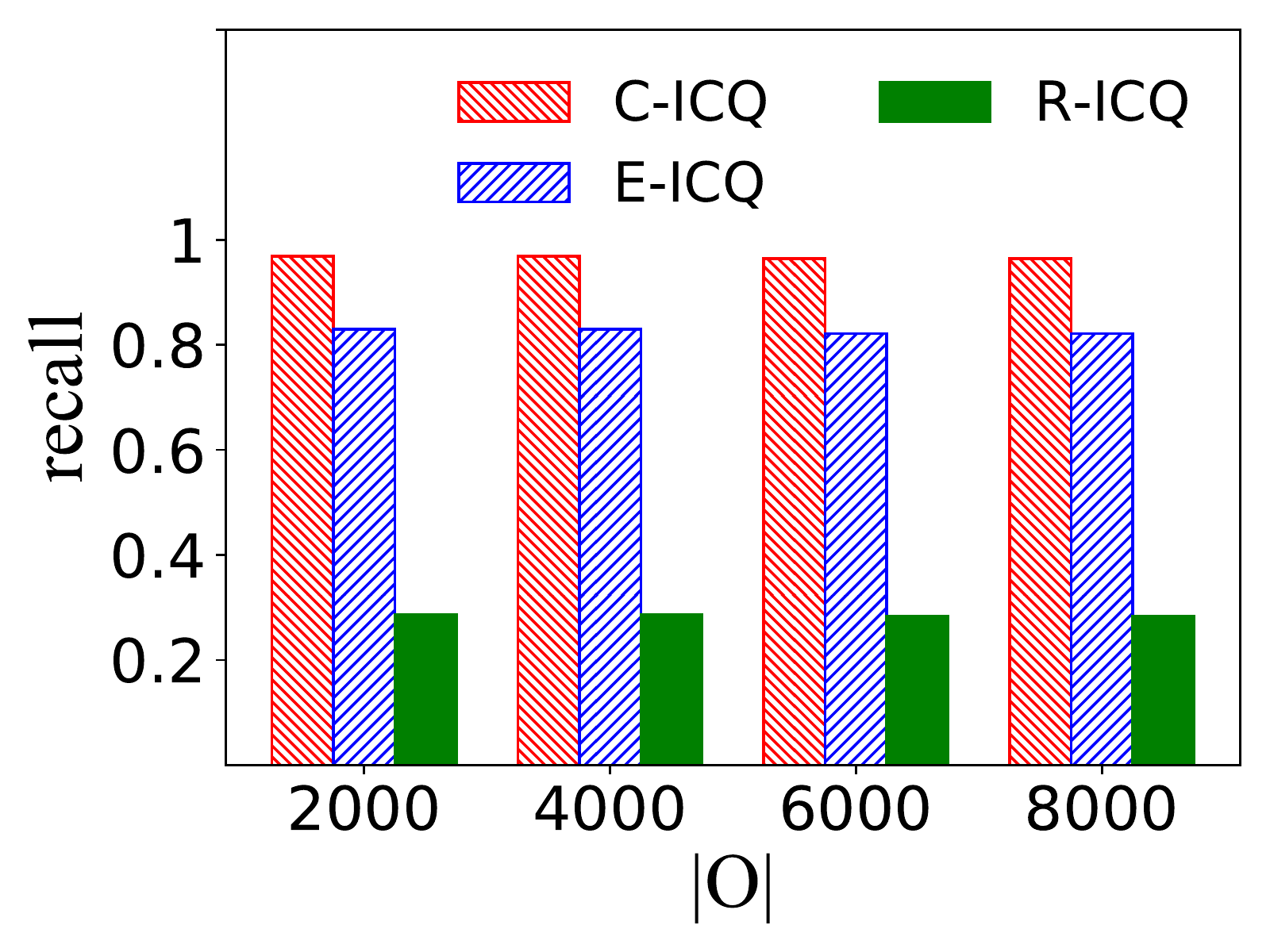}
            \label{fig:SYN-objectNum-recall}
        \end{minipage}
    }
    \subfigure[F1 score vs $|O|$]{
        \begin{minipage}[t]{0.45\columnwidth}
            \centering
            \includegraphics[width=\columnwidth]{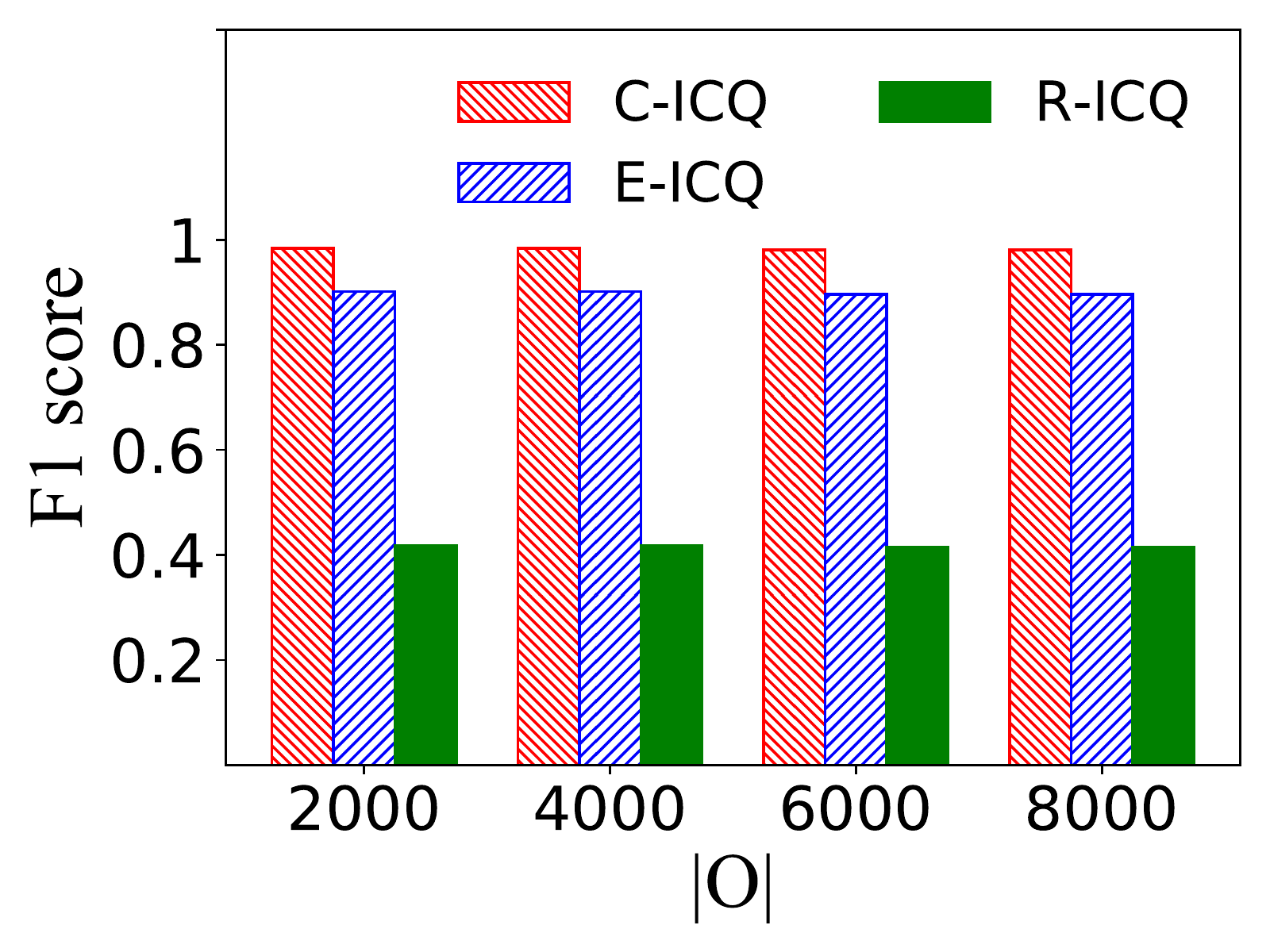}
            \label{fig:SYN-objectNum-F}
        \end{minipage}
    }
    \centering
    \caption{Effect of $|O|$}
\end{figure}

\noindent\textbf{Effect of $|O|$}. To test the scalability of all methods, we vary the object number $|O|$ from 2k to 8k. Referring Figs.~\ref{fig:SYN-objectNum-time} and \ref{fig:SYN-objectNum-memory}, the time and memory costs of \EICQ{}, \RICQ{} and \SICQ{} increases as $|O|$ increases, whereas \CICQ{} stays stable. 
When more candidate objects are involved, \EICQ{}, \RICQ{} and \SICQ{} need to check a larger fraction of potential close contact objects at each sampling time.
In contrast, \CICQ{} can rule out those unpromising objects and process only those necessary sampling times due to the time skipping strategy that it employs.

Referring to Figs.~\ref{fig:SYN-objectNum-recall} and \ref{fig:SYN-objectNum-F}, both effectiveness measures of the three methods stay nearly stable when $|O|$ increases. Still, \CICQ{} outperforms \EICQ{} and \RICQ{} due to its higher effectiveness of sample-based close contact tracing.
Therefore, our proposed \CICQ{} method works effectively in large-scale data scenarios.

\subsection{Experiments on Real Data}

\subsubsection{Settings}

We also evaluate our two proposals on a real dataset collected from a shopping mall in Hangzhou, China. 
The shopping mall occupies $108$m $\times$ $80$m and 7 floors with 10 staircases. 
Each staircase is roughly $20$m long. 
We acquire indoor positioning records from the building on 2018/01/01.
In total, 4,616 distinct moving objects (MAC addresses) and 727,263 positioning records are obtained. 

The query instance generation and the parameter settings follow the same as the experiments on the synthetic dataset (see Section~\ref{ssec:settings}).
We do not vary $|O|$ as it is fixed in the real dataset.
As we have no information about the true object movements to compute the recall and F1-score in the real data setting, we only compare the query processing methods in terms of efficiency metrics.

\subsubsection{Results}

\begin{figure}[htbp]
    \centering
    \subfigure[Time vs $|T|$]{
        \begin{minipage}[t]{0.45\columnwidth}
            \centering
            \includegraphics[width=\columnwidth]{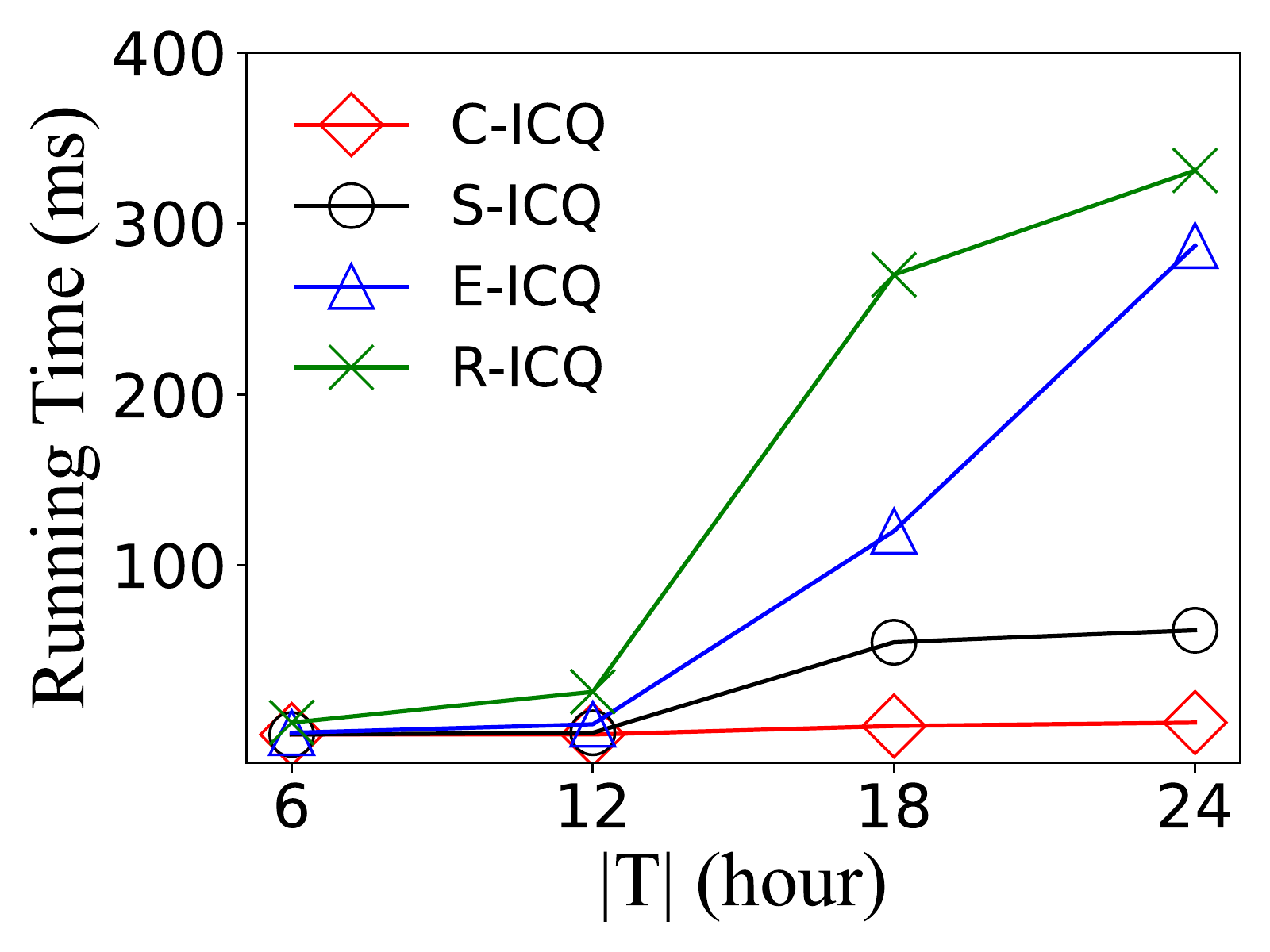}
            \label{fig:HSM-queryInterval-time}
        \end{minipage}
    }
    \subfigure[Mem. vs $|T|$]{
        \begin{minipage}[t]{0.45\columnwidth}
            \centering
            \includegraphics[width=\columnwidth]{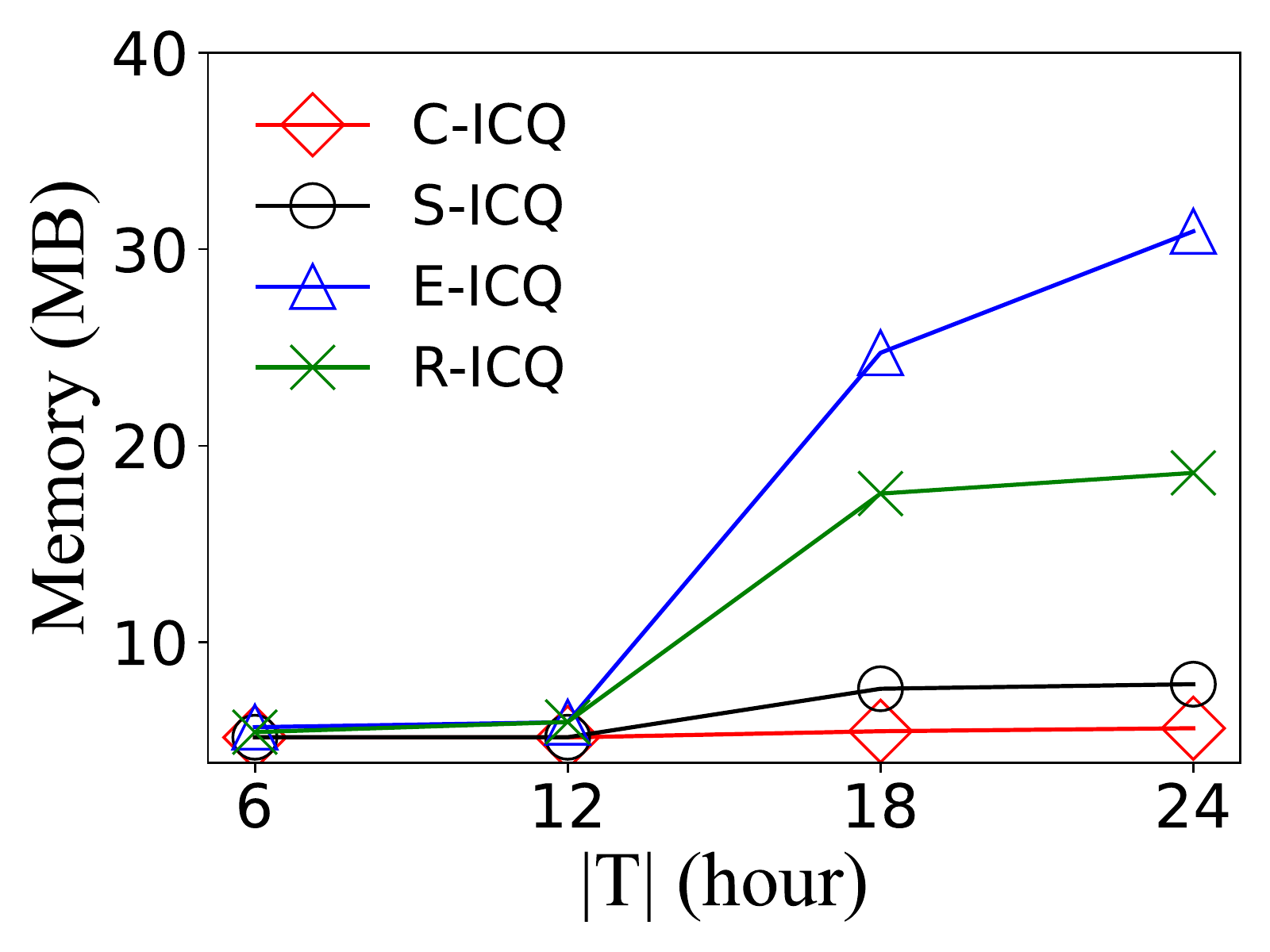}
            \label{fig:HSM-queryInterval-memory}
        \end{minipage}
    }
    \centering
    \caption{Effect of $|T|$}
\end{figure}

\noindent\textbf{Effect of $|T|$}. We set the query time interval $T$ to $(0, 6]$, $(0, 12]$, $(0, 18]$, and $(0, 24]$.
The running time and memory costs are reported in Figs.~\ref{fig:HSM-queryInterval-time} and~\ref{fig:HSM-queryInterval-memory}, respectively.
As the query interval length $|T|$ becomes larger, the running time and memory cost increase because more candidate objects and more sampling times are involved.
Since fewer moving objects appeared in the morning, i.e., within (0, 12], but much more in the afternoon, i.e., within (12, 18], both costs of all methods increase significantly when $T$ is changed from $(0,12]$ to $(0,18]$.
In all tested cases, \CICQ{} outperforms the three baselines in terms of efficiency.

\begin{figure}[htbp]
    \centering
    \subfigure[Time vs $\delta$]{
        \begin{minipage}[t]{0.45\columnwidth}
            \centering
            \includegraphics[width=\columnwidth]{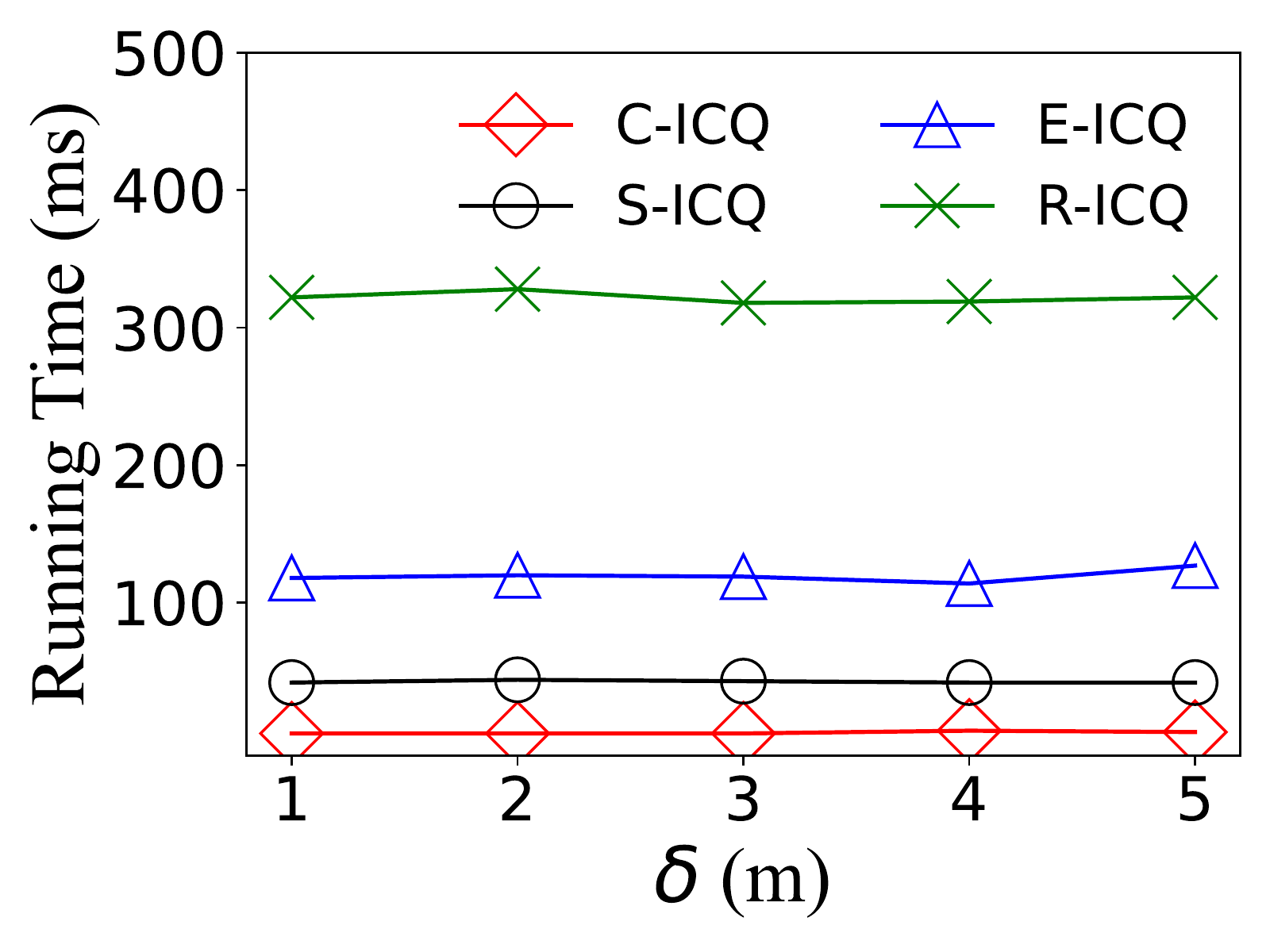}
            \label{fig:HSM-distance-time}
        \end{minipage}
    }
    \subfigure[Mem. vs $\delta$]{
        \begin{minipage}[t]{0.45\columnwidth}
            \centering
            \includegraphics[width=\columnwidth]{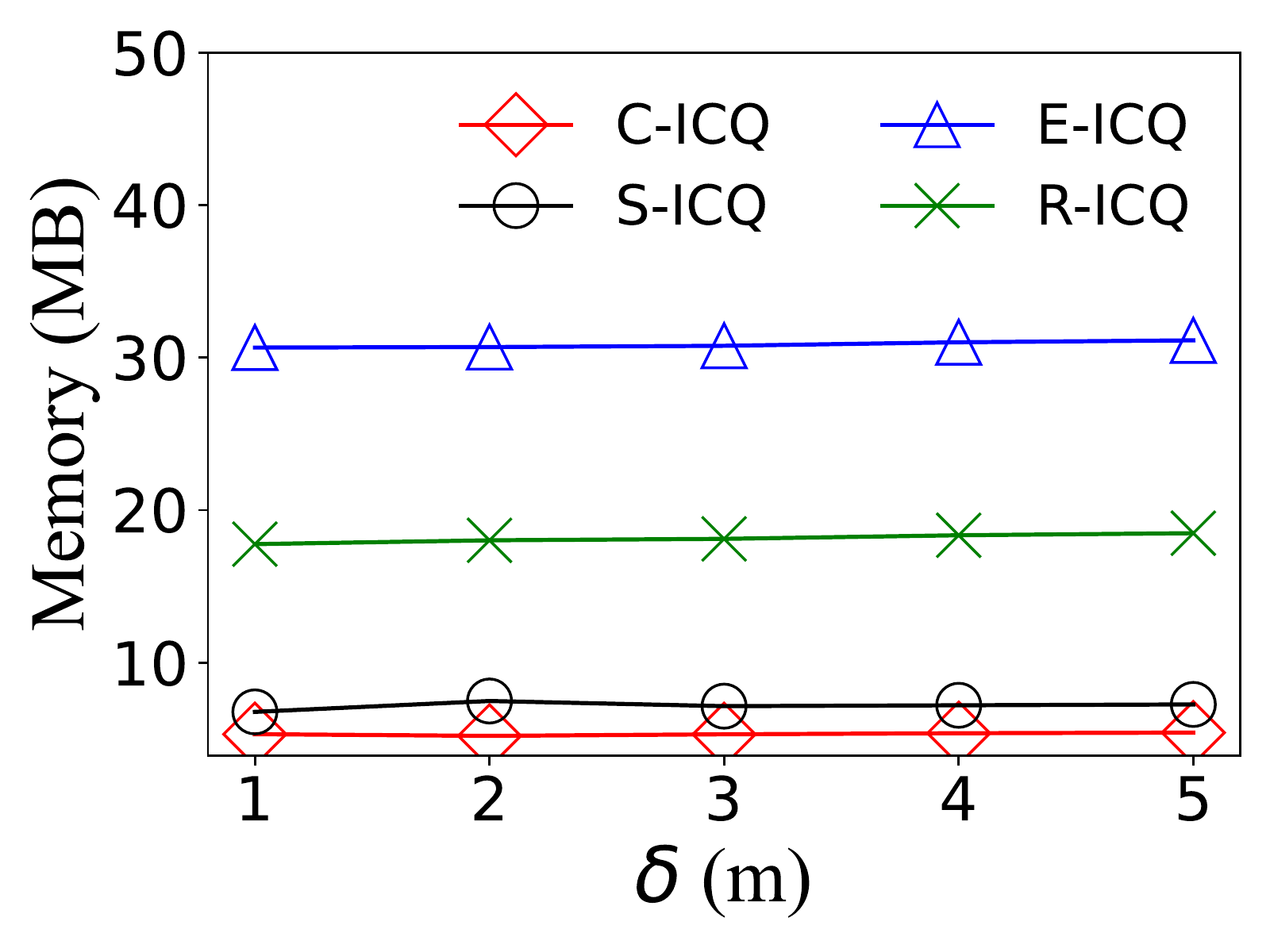}
            \label{fig:HSM-distance-memory}
        \end{minipage}
    }
    \centering
    \caption{Effect of $\delta$}
\end{figure}

\noindent\textbf{Effect of $\delta$}. According to the results reported in  Figs.~\ref{fig:HSM-distance-time} and~\ref{fig:HSM-distance-memory}, both the running time and memory consumption stay stable for all methods as $\delta$ increases. Nevertheless, \CICQ{} still performs best in terms of efficiency.
By analyzing the distribution of the contact distances of the object pairs in the same partition, we find that there are just a few within a distance less than 5m.

\begin{figure}[htbp]
    \centering
    \subfigure[Time vs $\eta$]{
        \begin{minipage}[t]{0.45\columnwidth}
            \centering
            \includegraphics[width=\columnwidth]{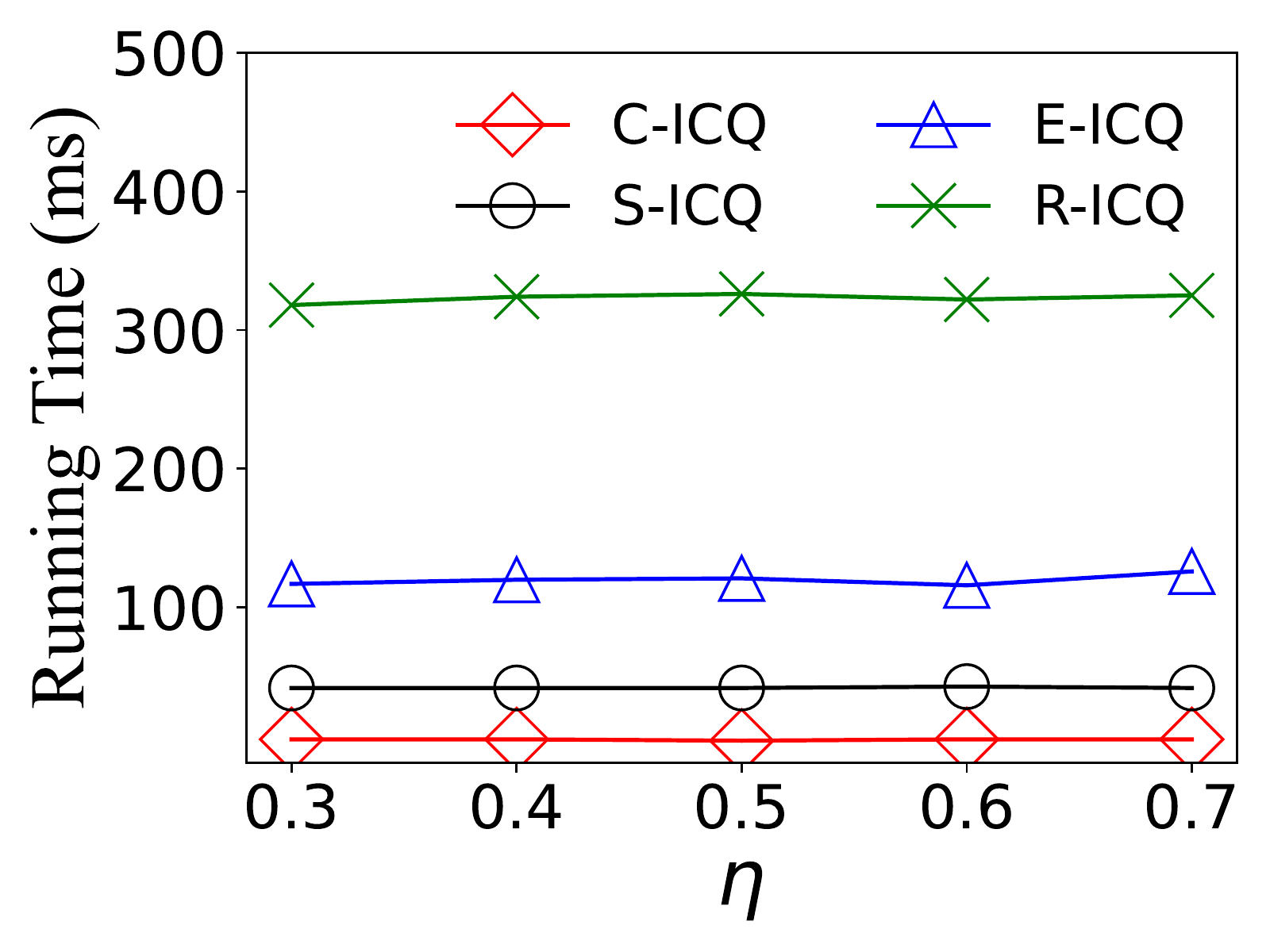}
            \label{fig:HSM-probability-time}
        \end{minipage}
    }
    \subfigure[Mem. vs $\eta$]{
        \begin{minipage}[t]{0.45\columnwidth}
            \centering
            \includegraphics[width=\columnwidth]{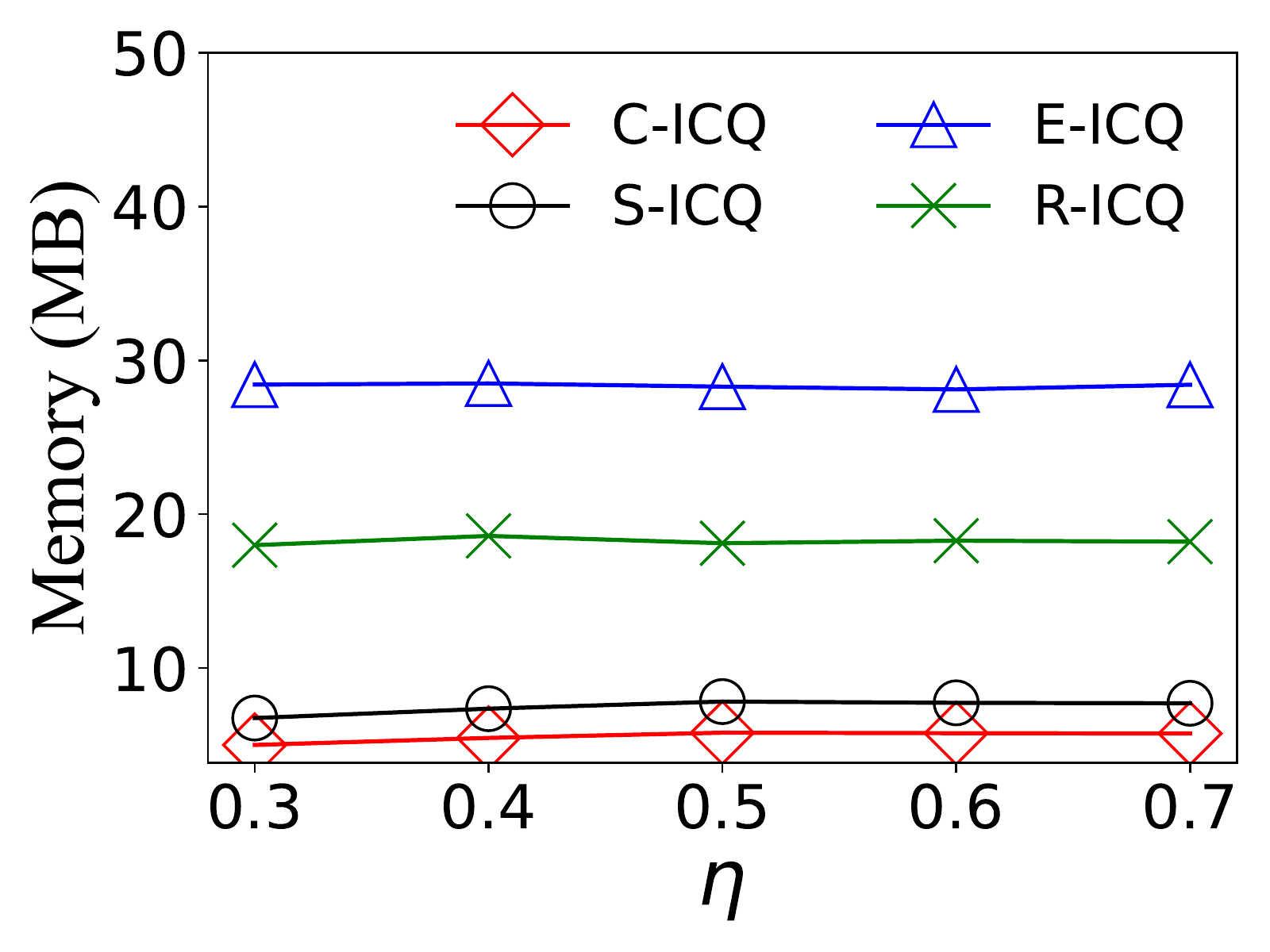}
            \label{fig:HSM-probability-memory}
        \end{minipage}
    }
    \centering
    \caption{Effect of $\eta$}
\end{figure}

\noindent\textbf{Effect of $\eta$}. As shown in Figs.~\ref{fig:HSM-probability-time} and~\ref{fig:HSM-probability-memory}, the contact probability threshold $\eta$ has no impact on the efficiency of \RICQ{} because it omits the unseen sample timestamps.
A larger $\eta$ has a slight impact on \CICQ{} since it has employed strategies for early termination of instant contact determination.

\begin{figure}[htbp]
    \centering
    \subfigure[Time vs $k$]{
        \begin{minipage}[t]{0.45\columnwidth}
            \centering
            \includegraphics[width=\columnwidth]{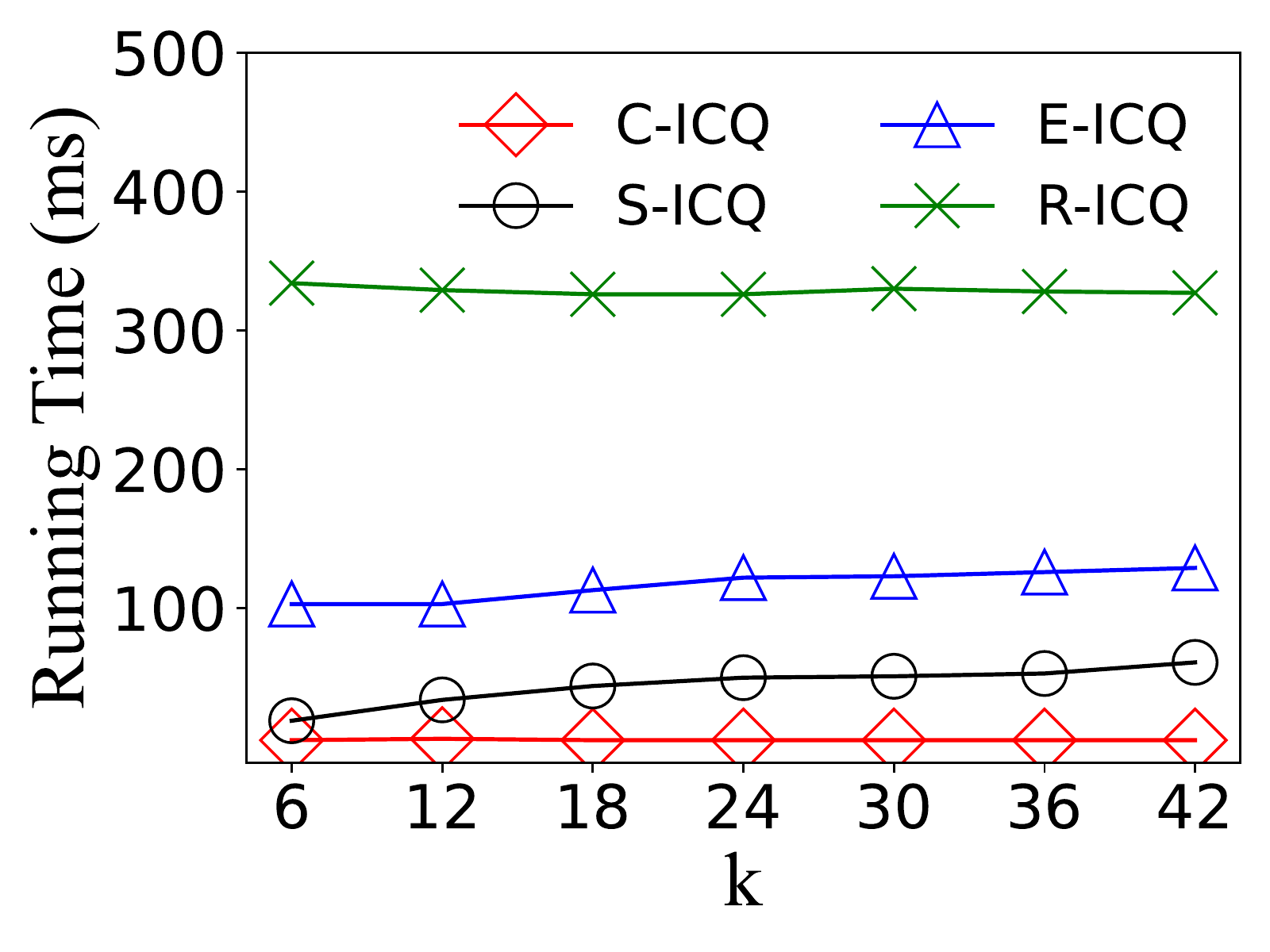}
            \label{fig:HSM-duration-time}
        \end{minipage}
    }
    \subfigure[Mem. vs $k$]{
        \begin{minipage}[t]{0.45\columnwidth}
            \centering
            \includegraphics[width=\columnwidth]{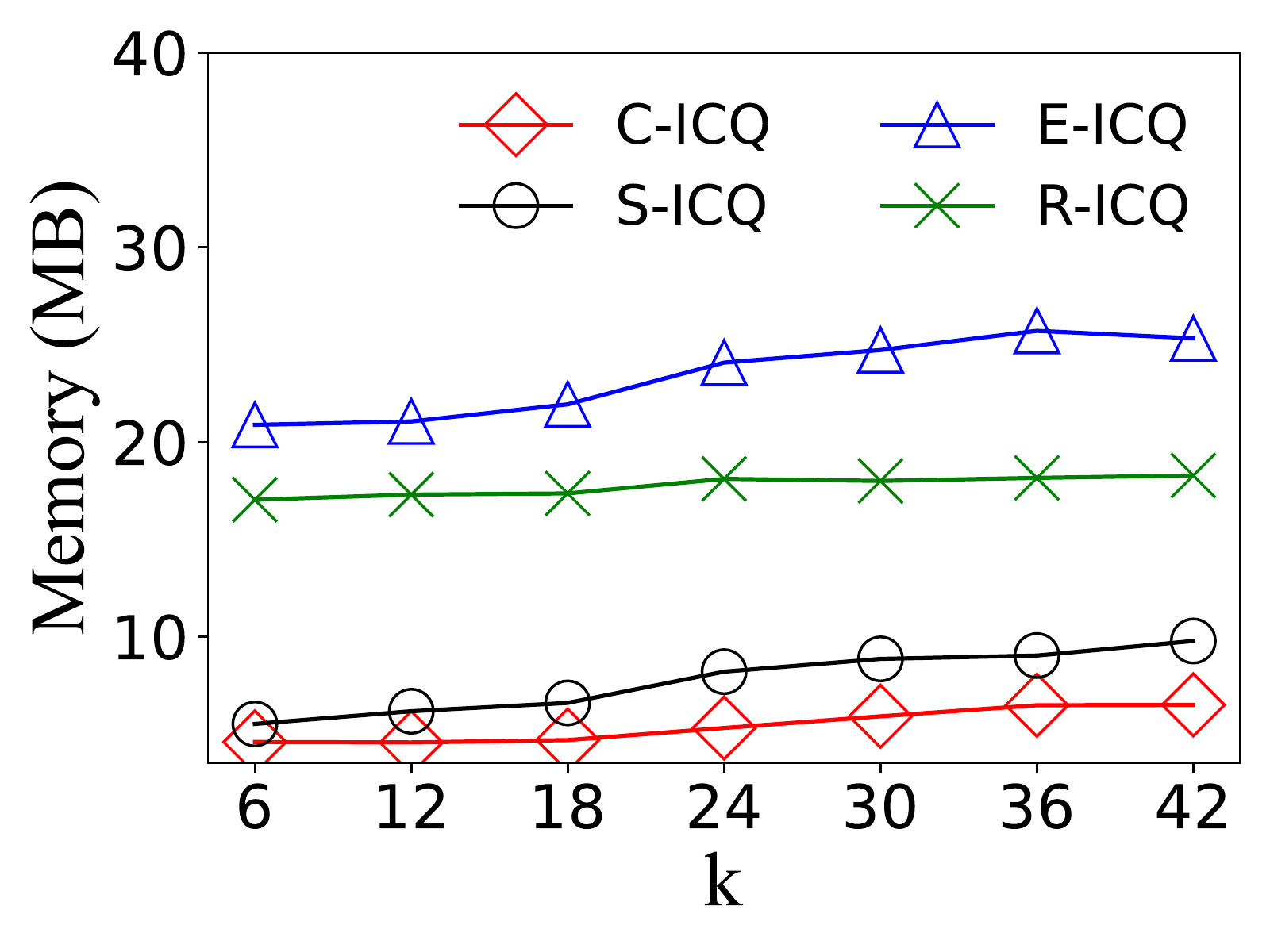}
            \label{fig:HSM-duration-memory}
        \end{minipage}
    }
    \centering
    \caption{Effect of $k$}
\end{figure}

\noindent\textbf{Effect of $k$}. Figs.~\ref{fig:HSM-duration-time} and~\ref{fig:HSM-duration-memory} report the running time and memory usage of all methods. 
Compared to \CICQ{}, the baseline methods \EICQ{} and \RICQ{} cost more time and memory with an increasing $k$ because our methods use the enhanced graph model that efficiently organizes the spatial data and trajectory and well supports the indoor contact tracing.
The efficiency of both \SICQ{} and \CICQ{} decreases slightly when $k$ increases.
Increasing $k$ involves more sampling times and thus increases both time and memory costs. 

\begin{figure}[htbp]
    \centering
    \subfigure[Time vs $\mathit{ll}$]{
        \begin{minipage}[t]{0.45\columnwidth}
            \centering
            \includegraphics[width=\columnwidth]{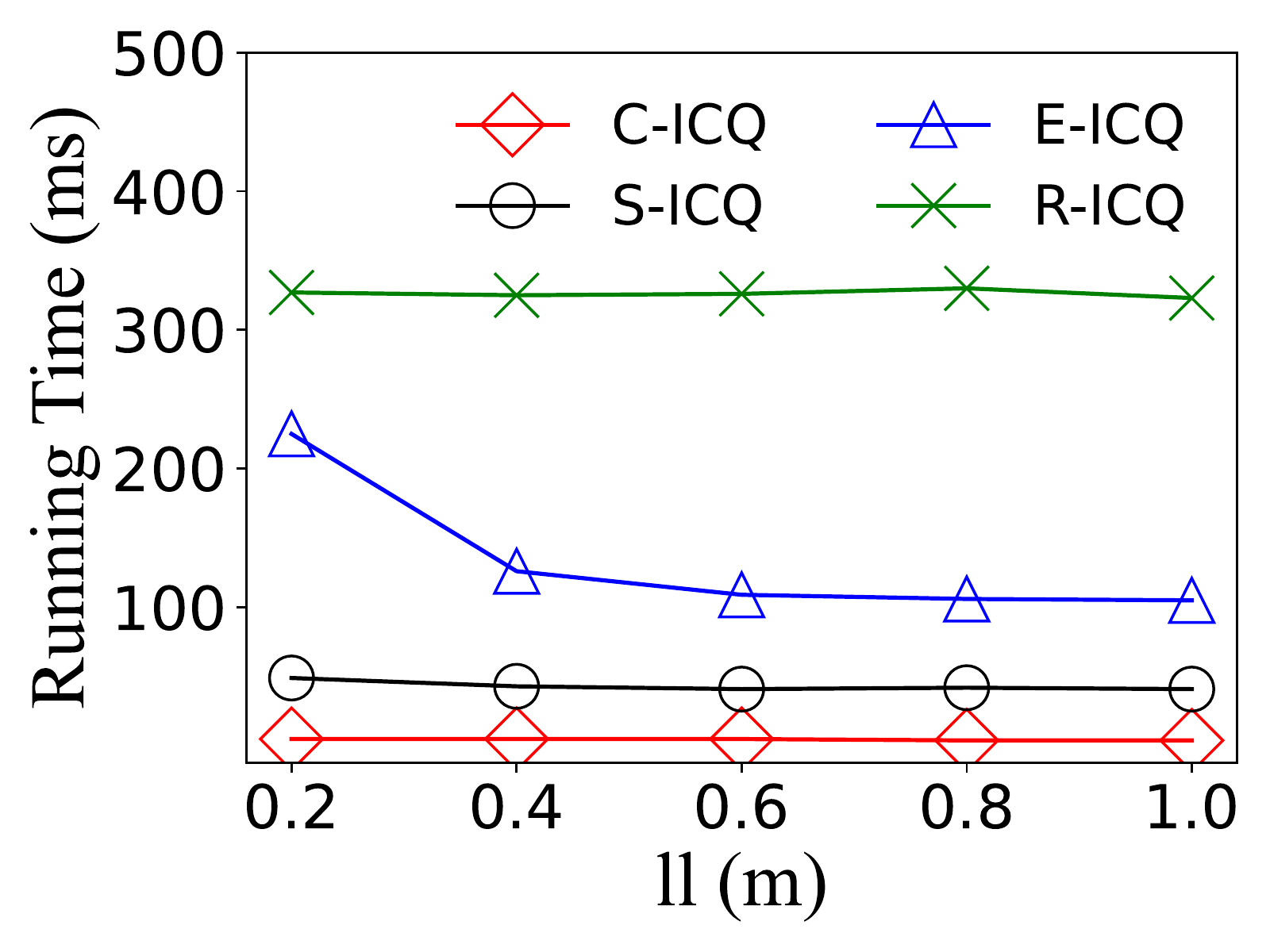}
            \label{fig:HSM-granularity-time}
        \end{minipage}
    }
    \subfigure[Mem. vs $\mathit{ll}$]{
        \begin{minipage}[t]{0.45\columnwidth}
            \centering
            \includegraphics[width=\columnwidth]{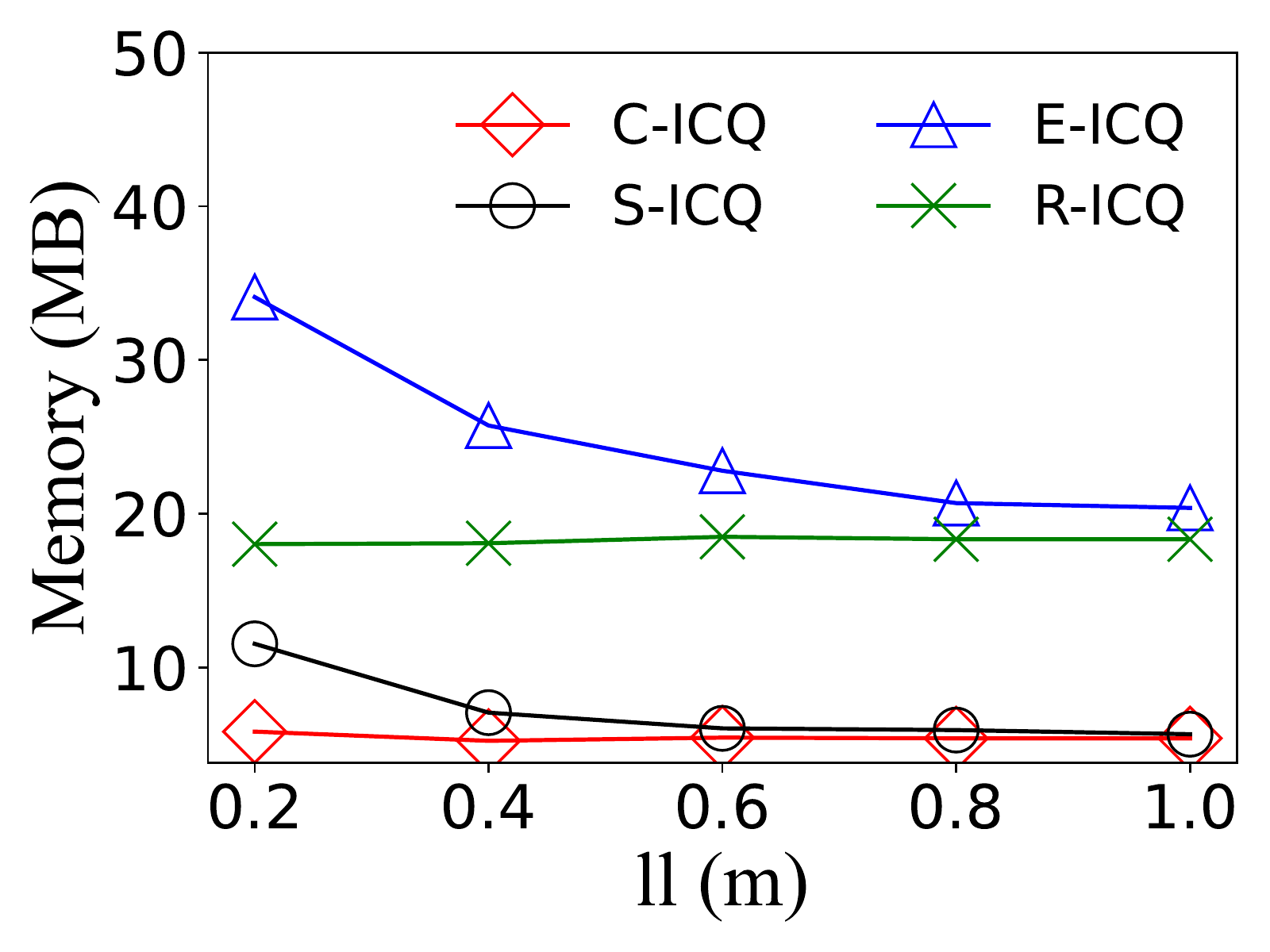}
            \label{fig:HSM-granularity-memory}
        \end{minipage}
    }
    \centering
    \caption{Effect of $\mathit{ll}$}
\end{figure}

\noindent\textbf{Effect of $\mathit{ll}$}. According to the results reported in Figs.~\ref{fig:HSM-granularity-time} and~\ref{fig:HSM-granularity-memory}, the running time and memory consumption decrease as $ll$ increases.
In general, increasing $\mathit{ll}$ has no impact on \RICQ{} and only slight impact on \CICQ{}, but improves significantly the efficiency of \EICQ{}. These observations are consistent with what we have seen on the synthetic data. 

\subsection{Summary of Experimental Results}
We summarize our findings from the results as follows.

In terms of efficiency, \CICQ{} performs best because it avoids a lot of heavy computations with efficient acceleration strategies, while \SICQ{} needs more time and memory costs because it analyzes the close contacts for all relevant sampling times without any acceleration. 

On the other hand, \SICQ{} and \CICQ{} return the same results. Their query result effectiveness is always better than that of \EICQ{} and \RICQ{}. This demonstrates the necessity of our proposed sample-based close contact tracing, which captures indoor moving objects' spatiotemporal dynamics much better than the \EICQ{} using Euclidean distance and the \RICQ{} ignoring the unseen sampling timestamps.

All in all, considering both efficiency and effectiveness, \CICQ{} is the best among all methods in comparison.
\section{Related Work}
\label{sec:related}

\noindent\textbf{Contact Tracing}.
There are two mainstream contact tracing technologies, namely proximity-based solutions~\cite{tedeschi2021iotrace,rivest2020pact,ng2022epidemic, bay2020bluetrace, li2021vcontact, shrestha2021bluetooth, di2020bluetooth, nguyen2022digital,blasimme2021digital,reichert2021survey,ahmed2020survey,hatamian2021privacy} and location-based solutions~\cite{xiong2020react, eusuf2020web, kato2021pct, zhang2021efficient, chao2021efficient, ami2021computation, hasan2021covid, li2022towards, chan2022continuous}.

Relying on wireless technologies such as Bluetooth, the proximity-based contact tracing solutions regard two devices as close contacts of each other if one of them falls within the wireless detection range of the other for a certain time period. 
Some studies review and compare proximity-based contact tracing methods, from the perspectives of system design~\cite{reichert2021survey, ahmed2020survey}, privacy issues~\cite{hatamian2021privacy}, and future development~\cite{nguyen2022digital, blasimme2021digital}.
Some recent representative works are introduced below.
Tedeschi et al.~\cite{tedeschi2021iotrace} propose an IoT-based architecture for saving energy consumption and computational cost on end-user devices.
Ng et al.~\cite{ng2022epidemic} predict the exposure risk of a user using a machine learning classifier. 
Shrestha et al.~\cite{shrestha2021bluetooth} propose a service-oriented architecture to enable Bluetooth-based mobile contact tracing. 
Li et al.~\cite{li2021vcontact} study privacy-aware, indirected close contact using the hashed WiFi access point IDs in a distributed environment.
The proximity-based methods, however, suffer from several shortcomings. 
First, these methods require mobile app installation, which is less user-friendly.
Second, they fail to support adjusting criteria of close contacts, e.g., customizing the distance restriction. 
Third, they neglect spatial information, not to mention indoor topology.

Location-based contact tracing methods identify close contact pairs using historical location data instead of wireless proximity information. 
Different from our work on efficient contact tracing over historical data, many works focus on developing the contact tracer as an administrative web-based system~\cite{eusuf2020web}, a privacy-enhanced mobile application~\cite{xiong2020react,kato2021pct,ami2021computation}, or a decentralized Blockchain-based system~\cite{hasan2021covid}.
Several works~\cite{zhang2021efficient, chao2021efficient, li2022towards} consider the efficiency of close contact search.
In particular, Chao et al.~\cite{chao2021efficient} devise an iterative algorithm and a grid-based time interval tree to find contacted trajectories along the transmission chains.
Li et al.~\cite{li2022towards} develop group processing algorithms for online continuous monitoring of the contacts with the given objects.
Zhang et al.~\cite{zhang2021efficient} resolve contact similarity search on uncertain trajectories based on an uncertain trajectory M-tree applied to necklace-based trajectories.
However, all these aforementioned studies apply to GPS data and determine the contact based on free-space distances. As a result, they cannot resolve our problem in indoor settings characterized by unique indoor topology and distances.
Some works focus on indoor contact tracing.
Alarabi et al.~\cite{alarabi2021traceall} adapt a three-dimensional R-Tree to answer the contact tracing queries. However, they do not consider the indoor restrictions such as walls and doors.
Ho et al.~\cite{ho2022clustering} develop a system to analyze the movement of users, as a foundation to further help contact tracing, but it can not solve contact tracing directly. 
Moreover, neither of the works~\cite{alarabi2021traceall, ho2022clustering} considers the uncertainty of indoor data.

\noindent\textbf{Analysis and Queries over Indoor Uncertain Data}.
Lu et al.~\cite{lu2011spatio} propose and study probabilistic, spatio-temporal joins on historical {RFID}-like tracking data, which aim to find the object pairs in the same room at a particular timestamp.
However, such joins do not consider concrete indoor distances.
Xie et al.~\cite{xie2014distance} study indoor distance-aware join on moving objects in indoor space.
However, the join query returns indoor object pairs at a certain time and it does not consider the duration.
Lu et al.~\cite{lu2016finding} mine frequently visited POIs from symbolic indoor tracking data collected through Bluetooth technologies.
The data formats of these works~\cite{lu2011spatio, xie2014distance, lu2016finding} are different from that in \ICQ{}.
Li et al.~\cite{li2018search} propose a framework for finding the current top-$k$ indoor dense regions from a set of user-defined query regions.
Li et al.~\cite{li2018finding} tackle the problem of finding the top-$k$ popular indoor semantic locations with the highest flow values using uncertain historical indoor mobility data.
Nonetheless, these works~\cite{lu2016finding, li2018search, li2018finding} return dense or popular indoor regions, while our work concerns the relationship between two individual objects.
Liu et al.~\cite{liu2021towards} study crowd-aware indoor path planning queries by taking moving objects into account in an online setting. 
Unlike \ICQ{}, this work~\cite{liu2021towards} does not analyze the uncertain regions of individual objects.
A recent work~\cite{chan2022continuous} considers indoor topology and uncertainty modeling in continuously monitoring the distances between object pairs. 
However, the proposed solution aims for early warning of all approaching object pairs by continuously calculating and forecasting their indoor distances.
In contrast, our study focuses on retrieving from historical data past close contacts with a confirmed infected object.

\section{Conclusion and Future Work}
\label{sec:conclusion}

In this work, we study the novel Indoor Contact Query (\ICQ{}). It conducts contact tracing over uncertain indoor positioning data to find all moving objects that had close contact with a given object $o$. To process \ICQ{}, we propose a set of data management techniques: 1) an enhanced indoor graph model to organize a multitude of relevant data; 2) an uncertainty-aware method for determining instant close contact; 3) a unified query processing framework with a close contact determination method, a search algorithm, and acceleration strategies. Experiments on synthetic and real datasets demonstrate the efficiency and effectiveness of our proposals.

Several directions exist for future research.
First, it is interesting to support other criteria of close contact within our proposed framework.
Also, it is possible to generalize our solution to support contact tracing over different types of indoor positioning data. 
Moreover, it is relevant to consider contact tracing over a mix of indoor and outdoor mobility data when such data is available.
Last but not least,  in case that COVID-19 rebounds, it is interesting to construct and deploy a prototype of an \ICQ{} system in a real-world setting, such as a university campus, to help contain virus spread.

\ifCLASSOPTIONcaptionsoff
  \newpage
\fi

\section*{Acknowledgement}
This work was supported by Independent Research Fund Denmark (No. 8022-00366B) and Australian Research Council (DP230100081 and FT180100140).

\bibliographystyle{IEEEtran}
\bibliography{spatial.bib}

\begin{thebibliography}{10}
\providecommand{\url}[1]{#1}
\csname url@samestyle\endcsname
\providecommand{\newblock}{\relax}
\providecommand{\bibinfo}[2]{#2}
\providecommand{\BIBentrySTDinterwordspacing}{\spaceskip=0pt\relax}
\providecommand{\BIBentryALTinterwordstretchfactor}{4}
\providecommand{\BIBentryALTinterwordspacing}{\spaceskip=\fontdimen2\font plus
\BIBentryALTinterwordstretchfactor\fontdimen3\font minus
  \fontdimen4\font\relax}
\providecommand{\BIBforeignlanguage}[2]{{%
\expandafter\ifx\csname l@#1\endcsname\relax
\typeout{** WARNING: IEEEtran.bst: No hyphenation pattern has been}%
\typeout{** loaded for the language `#1'. Using the pattern for}%
\typeout{** the default language instead.}%
\else
\language=\csname l@#1\endcsname
\fi
#2}}
\providecommand{\BIBdecl}{\relax}
\BIBdecl

\bibitem{health-disease}
\url{https://knowablemagazine.org/article/health-disease/2020/
  pandemics-recent-history}.

\bibitem{baniukevic2011improving}
A.~Baniukevic, D.~Sabonis, C.~S. Jensen, and H.~Lu, ``Improving wi-fi based
  indoor positioning using bluetooth add-ons,'' in \emph{MDM}, vol.~1.\hskip
  1em plus 0.5em minus 0.4em\relax IEEE, 2011, pp. 246--255.

\bibitem{tedeschi2021iotrace}
P.~Tedeschi, S.~Bakiras, and R.~Di~Pietro, ``Iotrace: A flexible, efficient,
  and privacy-preserving iot-enabled architecture for contact tracing,''
  \emph{IEEE Communications Magazine}, vol.~59, no.~6, pp. 82--88, 2021.

\bibitem{rivest2020pact}
R.~L. Rivest, D.~Weitzner, L.~Ivers, I.~Soibelman, and M.~Zissman, ``Pact:
  Private automated contact tracing,'' \emph{Retrieved December}, vol.~2, p.
  2020, 2020.

\bibitem{ng2022epidemic}
P.~C. Ng, P.~Spachos, S.~Gregori, and K.~N. Plataniotis, ``Epidemic exposure
  tracking with wearables: A machine learning approach to contact tracing,''
  \emph{IEEE Access}, 2022.

\bibitem{bay2020bluetrace}
J.~Bay, J.~Kek, A.~Tan, C.~S. Hau, L.~Yongquan, J.~Tan, and T.~A. Quy,
  ``Bluetrace: A privacy-preserving protocol for community-driven contact
  tracing across borders,'' \emph{GovTech}, vol.~18, 2020.

\bibitem{li2021vcontact}
G.~Li, S.~Hu, S.~Zhong, W.~L. Tsui, and S.-H.~G. Chan, ``{vContact}: Private
  {WiFi}-based {IoT} contact tracing with virus lifespan,'' \emph{IEEE IoT-J},
  vol.~9, no.~5, pp. 3465--3480, 2021.

\bibitem{shrestha2021bluetooth}
M.~Shrestha, X.~Liu, and M.~Sreekanthan, ``A {Bluetooth}-based contact-tracing
  mobile app for airborne-based epidemic control,'' in \emph{ISNCC}.\hskip 1em
  plus 0.5em minus 0.4em\relax IEEE, 2021, pp. 1--5.

\bibitem{di2020bluetooth}
P.~Di~Marco, P.~Park, M.~Pratesi, and F.~Santucci, ``A bluetooth-based
  architecture for contact tracing in healthcare facilities,'' \emph{JSAN},
  vol.~10, no.~1, p.~2, 2020.

\bibitem{xiong2020react}
L.~Xiong, C.~Shahabi, Y.~Da, R.~Ahuja, V.~Hertzberg, L.~Waller, X.~Jiang, and
  A.~Franklin, ``{REACT}: Real-time contact tracing and risk monitoring using
  privacy-enhanced mobile tracking,'' \emph{SIGSPATIAL Special}, vol.~12,
  no.~2, pp. 3--14, 2020.

\bibitem{eusuf2020web}
S.~S. Eusuf, K.~A. Islam, M.~E. Ali, S.~M. Abdullah, and A.~S. Azad, ``A
  {Web}-based system for efficient contact tracing query in a large
  spatio-temporal database,'' in \emph{SIGSPATIAL}, 2020, pp. 473--476.

\bibitem{kato2021pct}
F.~Kato, Y.~Cao, and M.~Yoshikawa, ``{PCT-TEE}: Trajectory-based private
  contact tracing system with trusted execution environment,'' \emph{TSAS},
  vol.~8, no.~2, pp. 1--35, 2021.

\bibitem{zhang2021efficient}
X.~Zhang, S.~Ray, F.~Shoeleh, and R.~Lu, ``Efficient contact similarity query
  over uncertain trajectories,'' in \emph{EDBT}, 2021, pp. 403--408.

\bibitem{chao2021efficient}
P.~Chao, D.~He, L.~Li, M.~Zhang, and X.~Zhou, ``Efficient trajectory contact
  query processing,'' in \emph{DASFAA}.\hskip 1em plus 0.5em minus 0.4em\relax
  Springer, 2021, pp. 658--666.

\bibitem{ami2021computation}
J.~Ami, K.~Ishii, Y.~Sekimoto, H.~Masui, I.~Ohmukai, Y.~Yamamoto, and
  T.~Okumura, ``Computation of infection risk via confidential locational
  entries: A precedent approach for contact tracing with privacy protection,''
  \emph{IEEE Access}, 2021.

\bibitem{hasan2021covid}
H.~R. Hasan, K.~Salah, R.~Jayaraman, I.~Yaqoob, M.~Omar, and S.~Ellahham,
  ``{COVID-19} contact tracing using {Blockchain},'' \emph{IEEE Access},
  vol.~9, pp. 62\,956--62\,971, 2021.

\bibitem{li2022towards}
K.~Li, L.~Chen, S.~Shang, Y.~Liu, H.~Wang, P.~Kalnis, and B.~Yao, ``Towards
  controlling the transmission of diseases: Continuous exposure discovery over
  massive-scale moving objects,'' in \emph{IJCAI}.\hskip 1em plus 0.5em minus
  0.4em\relax ijcai. org, 2022.

\bibitem{lu2011spatio}
H.~Lu, B.~Yang, and C.~S. Jensen, ``Spatio-temporal joins on symbolic indoor
  tracking data,'' in \emph{ICDE}.\hskip 1em plus 0.5em minus 0.4em\relax IEEE,
  2011, pp. 816--827.

\bibitem{li2018search}
H.~Li, H.~Lu, L.~Shou, G.~Chen, and K.~Chen, ``In search of indoor dense
  regions: An approach using indoor positioning data,'' \emph{IEEE TKDE},
  vol.~30, no.~8, pp. 1481--1495, 2018.

\bibitem{li2018finding}
------, ``Finding most popular indoor semantic locations using uncertain
  mobility data,'' \emph{IEEE TKDE}, vol.~31, no.~11, pp. 2108--2123, 2018.

\bibitem{chan2022continuous}
H.~K.-H. Chan, H.~Li, X.~Li, and H.~Lu, ``Continuous social distance monitoring
  in indoor space,'' \emph{Proc. {VLDB} Endow.}, vol.~15, no.~7, pp.
  1390--1402, 2022.

\bibitem{lu2012foundation}
H.~Lu, X.~Cao, and C.~S. Jensen, ``A foundation for efficient indoor
  distance-aware query processing,'' in \emph{ICDE}.\hskip 1em plus 0.5em minus
  0.4em\relax IEEE, 2012, pp. 438--449.

\bibitem{boysen2014journey}
M.~Boysen, C.~de~Haas, H.~Lu, and X.~Xie, ``A journey from ifc files to indoor
  navigation,'' in \emph{W2GIS}.\hskip 1em plus 0.5em minus 0.4em\relax
  Springer, 2014, pp. 148--165.

\bibitem{boysen2014constructing}
M.~Boysen, C.~de~Haas, H.~Lu, X.~Xie, and A.~Pilvinyte, ``Constructing indoor
  navigation systems from digital building information,'' in \emph{ICDE}.\hskip
  1em plus 0.5em minus 0.4em\relax IEEE, 2014, pp. 1194--1197.

\bibitem{nguyen2022digital}
T.~D. Nguyen, M.~Miettinen, A.~Dmitrienko, A.-R. Sadeghi, and I.~Visconti,
  ``Digital contact tracing solutions: Promises, pitfalls and challenges,''
  \emph{arXiv preprint arXiv:2202.06698}, 2022.

\bibitem{blasimme2021digital}
A.~Blasimme, A.~Ferretti, and E.~Vayena, ``Digital contact tracing against
  {COVID-19} in europe: Current features and ongoing developments,''
  \emph{Frontiers in Digital Health}, vol.~3, p.~61, 2021.

\bibitem{reichert2021survey}
L.~Reichert, S.~Brack, and B.~Scheuermann, ``A survey of automatic contact
  tracing approaches using {Bluetooth Low Energy},'' \emph{ACM HEALTH}, vol.~2,
  no.~2, pp. 1--33, 2021.

\bibitem{ahmed2020survey}
N.~Ahmed, R.~A. Michelin, W.~Xue, S.~Ruj, R.~Malaney, S.~S. Kanhere,
  A.~Seneviratne, W.~Hu, H.~Janicke, and S.~K. Jha, ``A survey of {COVID-19}
  contact tracing apps,'' \emph{IEEE Access}, vol.~8, pp. 134\,577--134\,601,
  2020.

\bibitem{hatamian2021privacy}
M.~Hatamian, S.~Wairimu, N.~Momen, and L.~Fritsch, ``A privacy and security
  analysis of early-deployed covid-19 contact tracing {Android} apps,''
  \emph{Empirical Software Engineering}, vol.~26, no.~3, pp. 1--51, 2021.

\bibitem{alarabi2021traceall}
L.~Alarabi, S.~Basalamah, A.~Hendawi, and M.~Abdalla, ``Traceall: A real-time
  processing for contact tracing using indoor trajectories,''
  \emph{Information}, vol.~12, no.~5, p. 202, 2021.

\bibitem{ho2022clustering}
Y.~A. Ho, C.~K. Tan, and Y.~H. Ng, ``Clustering indoor location data for social
  distancing and human mobility to combat covid-19,'' \emph{Computers,
  Materials and Continua}, vol.~71, no.~1, pp. 907--924, 2022.

\bibitem{xie2014distance}
X.~Xie, H.~Lu, and T.~B. Pedersen, ``Distance-aware join for indoor moving
  objects,'' \emph{IEEE TKDE}, vol.~27, no.~2, pp. 428--442, 2014.

\bibitem{lu2016finding}
H.~Lu, C.~Guo, B.~Yang, and C.~S. Jensen, ``Finding frequently visited indoor
  pois using symbolic indoor tracking data.'' in \emph{EDBT}, 2016, pp.
  449--460.

\bibitem{liu2021towards}
T.~Liu, H.~Li, H.~Lu, M.~A. Cheema, and L.~Shou, ``Towards crowd-aware indoor
  path planning.'' \emph{Proc. VLDB Endow.}, vol.~14, no.~8, pp. 1365--1377,
  2021.

\end{thebibliography}

\clearpage
\appendices
\section{Algorithm of Finding Uncertainty Region}
\label{sec:appendixA}

Given a location $l$ and a distance $\mathit{dist}$, Algorithm~\ref{alg:findUP} returns a set $\mathit{UP}$ of indoor portions that are within $\mathit{dist}$ from $l$ as the indoor uncertainty region.

\begin{algorithm}
    \caption{\textsc{findIUR} (location $l$, distance $\mathit{dist}$)} \label{alg:findUP}
    \begin{algorithmic}[1]
        \State initialize hashtable $\mathit{UP}: V \mapsto portions$;
        \State $v \gets \mathit{host}(l)$; $P_v \gets \{ (l, \mathit{dist}) \}$; $\mathit{UP}.put(v, P_v)$
        \State initialize distance array $\mathit{dist}[]$ for all doors
        \State initialize last-hop partition array $\mathit{prev}[]$ for all doors
        \State initialize a min-heap $H$
        \For {each door $d_i$ in the indoor space}
	        \If {$d_i \in \mathit{P2D}_\sqsubset(v)$}
                \State $\mathit{dist}[d_i] \gets ||l, d||_E$; $\mathit{prev}[d_i] \gets v$
            \Else
                \State $\mathit{dist}[d_i] \gets \infty$; $\mathit{prev}[d_i] \gets$ \textit{null}
	        \EndIf
    	    \State \text{enheap}($H$, $\left \langle d_i, \mathit{dist}[d_i] \right \rangle$);
        \EndFor
        \While {$H$ is not empty}
  	        \State $\left \langle d_i, \mathit{dist}[d_i] \right \rangle \gets$ \text{deheap}($H$)
  	        \If {$\mathit{dist}[d_i] > \mathit{dist}$}
  		        \textbf{break}
  	        \EndIf
  	        \For {each partition $v_i \in \mathit{D2P}_\sqsupset(d_i)$ $\wedge$ $v_i \neq \mathit{prev}[d_i]$}
  		        \State $\mathit{dist}^l \gets \mathit{dist} - \mathit{dist}[d_i]$
                  \If {$\mathit{UP}.hasKey(v_i)$}
                    \State $\mathit{UP}[v_i].add((d_i, \mathit{dist}^l))$
                  \Else
                    ~$P_{v_i} \gets \{ (d_i, \mathit{dist}^l) \}$; $\mathit{UP}.put(v_i, P_{v_i})$
                  \EndIf
                  \State mark $d_i$ as visited
                  \For {each unvisited door $d_j \in \mathit{P2D}_\sqsubset(v_i)$}
  		            \State $v'_i \gets \mathit{D2P}_\sqsupset(d_j) \backslash v_i$
                      \State $\mathit{dist}_j \gets \mathit{dist}[d_i] + f_\text{d2d}(v_i,d_i,d_j)$
  		            \If {$\mathit{dist}_j < \mathit{dist}[d_j]$}
  			            \State $\mathit{dist}[d_j] \gets \mathit{dist}_j$
                          \State \text{enheap}($H$, $\left \langle d_j, \mathit{dist}[d_j] \right \rangle$); \State $\mathit{prev}[d_j] \gets v_i$
  		            \EndIf
  	            \EndFor
  	        \EndFor

        \EndWhile
          \State \textbf{return} $\mathit{UP}$
    \end{algorithmic}
\end{algorithm}
In line~1, $\mathit{UP}$ is initialized as a hashtable with the partition ID as key and a portion set as the value.
Then, the portion inside the current host partition $v$ of $l$ is obtained (line~2).
In particular, a pair $(l, \mathit{dist})$ is added to a new set $P_v$, meaning that the portion inside $v$ is centered at $l$ with a radius $\mathit{dist}$. Note that this portion is not necessarily a circle or a sector because it is bounded by the walls of the partition.
Next, the algorithm expands outwardly from each leavable door $\mathit{P2D}_\sqsubset(v)$ of $v$ in the spirit of Dijkstra's algorithm (lines~3--27).
The expansion is performed over our enhanced graph model (see Section~\ref{ssec:data_organization}).
In particular, arrays $\mathit{dist}[]$ and $\mathit{prev}[]$ are initialized to maintain the distances to $l$ and previous partitions for all doors in expansion (lines~3--4).
Subsequently, each door's initial distance to $l$ and previous partition are obtained (lines~6--10).
All doors and their corresponding initial distances to $l$ are pushed into a min-heap $H$ (line~11).
In the following, doors are expanded in the order of their initial distances to $l$ as controlled by $H$ (lines~12--27).

For each door $d_i$ deheaped from $H$, if its distance to $l$ is larger than $\mathit{dist}$, i.e., it is out of the range of $\mathit{dist}$, the processing will terminate (line~14).
Otherwise, we check each $d_i$'s enterable partition $v_i \in \mathit{D2P}_\sqsupset(d_i)$~\cite{lu2012foundation} and avoid visiting $v_i$ twice consecutively (line~15).
The left distance $\mathit{dist}^l$ is obtained (line~16) and the corresponding portion inside $v_i$ is obtained as $(d_i, \mathit{dist}^l)$ and added/updated to $\mathit{UP}$ (lines~17--19).
Afterwards, door $d_i$ is marked as visited (line~20) and the expansion to the next door is performed (lines~21--27).
\section{Algorithm of Deriving Samples}
\label{sec:appendixB}

Given an object $o$ and a sampling time $t^s$, the algorithm of deriving samples returns a set $S$ of derived samples based on the indoor uncertainty regions obtained in Algorithm~\ref{alg:findUP}. 

\begin{algorithm}
    \caption{\textsc{Derive} (object $o$, sampling time $t^s$)} \label{alg:sampling}
    \begin{algorithmic}[1]
        \State $S \gets \varnothing$; $\mathit{PL} \gets \varnothing$; get $o$'s raw trajectory $\Psi_o$
        \State get the previous record $\psi_{\dashv} = (l_{\dashv}, t_{\dashv}, et_{\dashv})$ prior to $t^s$ from $\Psi_o$
        \State get the next record $\psi_{\vdash} = (l_{\vdash}, t_{\vdash}, et_{\vdash})$ after $t^s$ from $\Psi_o$
        \State $\mathit{dist}_{\dashv} \gets (t^s - et_{\dashv}) \cdot v_{max}$; $\mathit{dist}_{\vdash} \gets (t_{\vdash} - t^s) \cdot v_{max}$
        \State $\mathit{UP}_{\dashv} \gets \textsc{FindIUR}(l_{\dashv}, \mathit{dist}_{\dashv})$;
        $\mathit{UP}_{\vdash} \gets \textsc{FindIUR}(l_{\vdash}, \mathit{dist}_{\vdash})$
        \State $V_o \gets \mathit{UP}_{\dashv}.keys() \cap \mathit{UP}_{\vdash}.keys()$
        \For {each partition $v_i \in V_o$}
            \State generate lattice points in $v_i$ with side length $ll$
            \For {each portion $(d_{\dashv}, \mathit{dist}^l_{\dashv})$ in $\mathit{UP}_{\dashv}[v_i]$}
                \State $\mathit{dist}^m_{\dashv} \gets$ the max distance to $d_{\dashv}$ in $v_i$
                \For {each portion $(d_{\vdash}, \mathit{dist}^l_{\vdash})$ in $\mathit{UP}_{\vdash}[v_i]$}
                    \State $\mathit{dist}^m_{\vdash} \gets$ the max distance to $d_{\vdash}$ in $v_i$
                    \If {$\mathit{dist}^l_{\dashv} + \mathit{dist}^l_{\vdash} < \mathit{dist}(d_{\dashv}, d_{\vdash})$}
                        \textbf{continue}
                    \EndIf
                    \If {$\mathit{dist}^l_{\dashv} \ge \mathit{dist}^m_{\dashv}$ and $\mathit{dist}^l_{\vdash} \ge \mathit{dist}^m_{\vdash}$}
                        \State add all $v_i$'s lattice points to $\mathit{PL}$
                    \Else
                        \State $mbr \gets \mathit{MBR}((d_{\dashv}, \mathit{dist}^l_{\dashv})) \cap \mathit{MBR}((d_{\vdash},\mathit{dist}^l_{\vdash}))$
                        \For {each lattice point $l$ within $mbr$}
                            \If {$|l, d_{\dashv}|_E < \mathit{dist}^l_{\dashv}$ and $|l, d_{\vdash}|_E < \mathit{dist}^l_{\vdash}$}
                                \State add $l$ to $\mathit{PL}$
                            \EndIf
                        \EndFor
                    \EndIf
                \EndFor
            \EndFor
        \EndFor
        \For {each $l \in \mathit{PL}$}
            add $(l, 1/|\mathit{PL}|)$ to $S$
        \EndFor
        \State \textbf{return} $S$
    \end{algorithmic}
\end{algorithm}

First, the sample set $S$ and the possible location set $\mathit{PL}$ are initialized and object $o$'s raw trajectory $\Psi_o$ is retrieved (line~1).
Next, $o$'s previous and next records $\psi_{\dashv}$ and $\psi_{\vdash}$ with respect to $t^s$ are obtained from $\Psi_o$ (lines~2--3).
With respect to $\psi_{\dashv}$ and $\psi_{\vdash}$, we compute two distance bounds that object $o$ can move (line~4).
Specifically, $o$'s maximum movement distance between the previous expiration timestamp (i.e., $\psi_{\dashv}.et_{\dashv}$) and $t^s$ is obtained as $\mathit{dist}_{\dashv}$; likewise, $o$'s maximum movement distance between $t^s$ and the next positioning time is $\mathit{dist}_{\vdash}$.
As a result, their corresponding uncertainty regions are computed by calling Algorithm~\ref{alg:findUP} (line~5) and the common partitions of the uncertainty regions are obtained in a set $V_o$ (line~6).

The algorithm then iterates through each such common partition $v_i$ (lines~7--22) as follows.
The lattice points associated to $v_i$ is first generated (line~8).
Specifically, $v_i$ is evenly divided into lattices with side length $ll$, and each lattice point is regarded as a location sample of $v_i$.
The effect of varying $\mathit{ll}$ on query processing is studied in experiments section.
In the outer loop, each portion $(d_{\dashv}, \mathit{dist}^l_{\dashv})$ in $\mathit{UP}_{\dashv}[v_i]$ is obtained (line~9).
Also, the maximum distance from any point in $v_i$ to the center $d_{\dashv}$ is obtained as $\mathit{dist}^m_{\dashv}$ (line 10).
The same procedure is applied to the inner loop of each portion $(d_{\vdash}, \mathit{dist}^l_{\vdash})$ in $\mathit{UP}_{\vdash}[v_i]$ (lines~11--12).

The algorithm then determines the geometrical relationship between portion $(d_{\dashv}, \mathit{dist}^l_{\dashv})$ and portion $(d_{\vdash}, \mathit{dist}^l_{\vdash})$.
If the sum of the two radii $\mathit{dist}^l_{\dashv}$ and $\mathit{dist}^l_{\vdash}$ is less than the distance between $d_{\dashv}$ and $d_{\vdash}$, it means that the two portions are disjoint. Therefore, the current pair of portions is skipped (line~13).
If $\mathit{dist}^l_{\dashv}$ equals or exceeds the maximum distance $\mathit{dist}^m_{\vdash}$ and $\mathit{dist}^l_{\vdash}$ equals or exceeds $\mathit{dist}^m_{\dashv}$, it means that the intersection of the two portions fully contains the partition $v_i$.
Therefore, all lattice points are included in $\mathit{PL}$ (lines~14--15).
Otherwise, the two portions are approximated by their MBRs and the intersection $mbr$ of the two MBRs is quickly determined (line~17).
For each lattice point $l$ inside $mbr$, it is added to $\mathit{PL}$ if its distances to $d_{\vdash}$ and $d_{\dashv}$ are both less than the corresponding radius (line~18--20).

Finally, we associate each possible location $l$ in $\mathit{PL}$ with the same probability of $1/|\mathit{PL}|$ (line~21).
Each such sample $(l, 1/|\mathit{PL}|)$ is added to $S$ which in turn is returned as the sample set (line~22).
\section{Sequential Search Algorithm}
\label{sec:appendixC}

Similar to the constrained search method formalized in Algorithm~\ref{alg:constrained_search} in Section~\ref{ssec:ICQ_processing}, the sequential search method can also be evoked by the overall framework (Algorithm~\ref{alg:ICQ}). 
To do this, line~13 in Algorithm~\ref{alg:ICQ} is modified as ``\textbf{return} \textsc{S-Search}($o$, $[t^s_s, t^s_e]$, $\delta$, $\eta$, $k$)''.
The resultant method is referred to as \SICQ{} and have been included in the experimental comparisons in Section~\ref{sec:experiment}.

The sequential search algorithm \textsc{S-Search} is formalized in Algorithm~\ref{alg:sequential_search}.
First, the result set $\mathit{result}$ is initialized (line~1).
Subsequently, the algorithm processes each sampling time $t^s$ sequentially (lines~2--16).
At each current time $t^s$, it looks ahead $(k-1)\Delta t$ time, to find objects in close contact to $o$ for $k$ consecutive sampling times.
Therefore, the end of $t^s$ is set to $t^s_e - (k-1)\Delta t$ in line~2.

For each current time $t^s$, the sample set $S$ is retrieved from the enhanced graph model and a set $O_\mathit{visited}$ is used to record all visited objects in this iteration (line~3).
For each sample $s (l, \rho)$ in $S$, the host partition $v$ is obtained (lines~4--5).
The partition $v$ will be skipped (line~6) if it has been visited (see line~7).
Next, the algorithm checks each candidate object $o_i$ that has samples in $v$ at $t^s$ (line~8).
If $o_i$ is the query object $o$, or it has been processed, it will be skipped (line~9).
Otherwise, $o_i$ is added to the visited set (line~10). Furthermore, the algorithm $\textsc{IsCloseContact}(o, o_i, t^s, \delta, \eta, k)$, which has been formalized in Algorithm~\ref{alg:contact} in Section~\ref{ssec:close_contact}, is called to determine if $o_i$ has close contact with $o$ for $k$ consecutive times from $t^s$ to $t^s + (k-1)\Delta t$ or not (line~11).
The candidate object $o_i$ is added to $\mathit{result}$ if it has close contact with $o$ (lines~11--12).
Lines~13--15 deal with a boundary condition when the current time $t^s$ is earlier than $t^s_s + k \Delta t$.
In this case, $o_i$'s sample sets within $[t^s_s, t^s]$ may have not been computed for close contact determination.
To avoid missing such $o_i$, \textsc{IsCloseContact} is also executed for $t^s_s$ (lines~14--15).
Finally, $\mathit{result}$ is returned (line~17).

\begin{algorithm}
    \caption{\textsc{S-Search} (query object $o$, time period $[t^s_s, t^s_e]$, distance constraint $\delta$, contact probability threshold $\eta$, contact number $k$)}\label{alg:sequential_search}
    \begin{algorithmic}[1]
        \State $\mathit{result} \gets \varnothing$; $t^s \gets t^s_s$
        \While {$t^s \leq t^s_e - (k-1) \Delta t$}
            \State obtain $(S, t^s)$ from $\Psi^s_o$; $O_\mathit{visited} \gets \varnothing$
            \For {each $s (l, \rho) \in S$}
                \State $v \gets \mathit{host}(l)$
                \If {$v$ has been visited}
                    \textbf{continue}
                \EndIf
                \State marked $v$ as visited
                \For {each $o_i \in \mathit{OT}^v[t^s]$}
                    \If {$o_i = o$ or $o_i \in O_\mathit{visited}$}
                        \textbf{continue}
                    \EndIf
                    \State add $o_i$ to $O_\mathit{visited}$
                    \If {\textsc{IsCloseContact}($o, o_i, t^s, \delta, \eta, k$)}
                      \State add $o_i$ to $\mathit{result}$
                    \EndIf
                    \If{$t^s \leq t^s_s + (k-1)\Delta t$}
                      \If {\textsc{IsCloseContact}($o, o_i, t^s_s, \delta, \eta, k$)}
                        \State add $o_i$ to $\mathit{result}$
                      \EndIf
                    \EndIf
                \EndFor
            \EndFor
            \State $t^s \gets t^s + \Delta t$
        \EndWhile
        \State \textbf{return} $\mathit{result}$
    \end{algorithmic}
\end{algorithm}

It is worth noting that in \SICQ{}, the line~3 of $\textsc{IsCloseContact}$ calls the sequential determination algorithm $\textsc{S-InstantContact}$ (to be presented in Algorithm~\ref{alg:instant_contact}) instead of $\textsc{C-InstantContact}$, as \SICQ{} does not consider the three speed-up strategies presented in Section~\ref{ssec:close_contact}.

\textsc{S-InstantContact} is formalized in Algorithm~\ref{alg:instant_contact}.
It checks if $o$ is instant contact with $o'$ and is called by \textsc{IsCloseContact}. 
First, it uses a caching mechanism such that two objects' instant contact result is directly fetched if the result has been computed and cached (line~1).
Then, it prepares the candidate object $o'$'s sample set $S'$ by either calling \textsc{Derive} or retrieving from the enhanced graph model (lines~2--3).
Afterwards, it iterates through each pair of samples from $S$ and $S'$ to compute the contact probability according to Equation~\ref{equation:contact_probability} (lines~4--8).
If the computed contact probability $\mathsf{P}$ is no less than the threshold $\eta$, $\mathit{true}$ is returned to indicate the instant contact (line~9).
Otherwise, $\mathit{false}$ is returned (line~10).
Before returning, the result is cached.

\begin{algorithm}
    \caption{\textsc{S-InstantContact} (object $o$, object $o'$, current time $t^s$, distance constraint $\delta$, contact probability threshold $\eta$)} \label{alg:instant_contact}
    \begin{algorithmic}[1]
        \State return the cached result if it exists
        \If {$t^s$ is not seen in $\Psi^s_{o'}$}
                $S' \gets$ \textsc{Derive}$(o', t^s)$
            \Else
                ~obtain $(S', t^s)$ from $\Psi^s_{o'}$
        \EndIf
        \State $\mathsf{P} \gets 0$; obtain $(S, t^s)$ from $\Psi^s_o$
        \For {each sample $(l,\rho)$ in $S$}
            \For {each sample $(l', \rho')$ in $S'$}
                \If {$\mathit{dist}(l, l') > \delta$}
                    \State $\mathsf{P} \gets \mathsf{P} + \rho \cdot \rho'$
                \EndIf
            \EndFor
        \EndFor
        \If {$\mathsf{P} \geq \eta$}
            cache the result and return $\mathit{true}$
        \EndIf
        \State cache the result and return $\mathit{false}$
    \end{algorithmic}
\end{algorithm}




%








\end{document}